\documentclass[a4paper,12pt,times,numbered,print,index,biblatex,openany]{book}
\usepackage[numbers]{natbib} 
\usepackage{thesis}
\usepackage{ifthen}
\usepackage{datetime}
\usepackage{amsmath,amssymb}
\usepackage[dvipsnames]{xcolor}
\usepackage[T1]{fontenc}
\usepackage{ae,aecompl}
\usepackage{graphicx}	
\usepackage{amssymb}
\usepackage[utf8]{inputenc} 
\usepackage[T1]{fontenc}    
\usepackage{rotating}
\usepackage{pdfpages}
\usepackage{indentfirst}
\usepackage{cases}
\usepackage{babel}
\usepackage{subcaption}
\usepackage{url}            
\usepackage{hyperref}       

\usepackage{fancyhdr}

\fancypagestyle{plain}{
  \fancyhf{}
  \fancyhead[LE]{\normalfont\leftmark}
  \fancyhead[RO]{\normalfont\leftmark}
  \fancyhead[LE,RO]{\thepage}
}
\usepackage{datetime}
\newdateformat{myformat}{\THEDAY\ \monthname[\THEMONTH] \THEYEAR}
\usepackage{array}

\newcommand{\be}{\begin{equation}}
\newcommand{\ee}{\end{equation}}
\newcommand{\lb}[1]{\label{#1}}

\newcommand{\doctitle}{} 
\newcommand{\doctype}{The Gompertz-Zipf city size distribution and urban allometry}

% ...

\newcommand{\authorName}{Sharon Teles de Azevedo Chaves}

\newcommand{\supervisor}{Dr. Marcelo Byrro Ribeiro \\ {\small Orcid: 0000-0002-6919-2624}}

\newcommand{\cosupervisor}{Dr. Amanda Reis Lopes \\ {\small Orcid: 0000-0002-6164-5051}}

\newcommand{\Year}{\the\year{}}

\newcommand{\keywords}{Important, comma, separated, keywords, applicable, to, your, thesis}

\hypersetup
{
    pdfauthor={\authorName},
    pdfsubject={\doctitle},
    pdfkeywords={\keywords}
}

\begin{document}


\begin{titlepage}
\headheight = 0pt

{%
\fontsize{9pt}{9pt}\selectfont%
\renewcommand\arraystretch{2}

%
}

\vspace*{1 cm}
\begin{center}

{\fontsize{33pt}{33pt}\selectfont%
\textsc{\textbf{\doctype}}}

\vspace*{1 cm}
\begin{center}
Graduate dissertation submitted to the Physics Institute, Center for Natural and Mathematical Sciences, Universidade Federal do Rio de Janeiro (UFRJ), in partial fulfillment of the requirements for the degree of Master in Applied Physics\\   
\end{center}

\vspace*{1.5 cm}
\begin{Large}
\textbf{\authorName}
\end{Large}
\\
Orcid: 0000-0003-4497-9161
\vspace*{1.5 cm}


\begin{tabular}{ p{0.6\textwidth} p{0.3\textwidth} }
\parbox[t]{\linewidth}{\supervisor} 
& \parbox[t]{\linewidth}{\centering 1st Adviser} \\
\\

\ifthenelse{\equal{\cosupervisor}{Mr. Co-Supervisor's Name}}{}{%
\parbox[t]{\linewidth}{\cosupervisor} 
& \parbox[t]{\linewidth}{\centering 2nd Adviser} \\}
\end{tabular}

\vspace*{1.5 cm}

\includegraphics[width=0.18\textwidth]{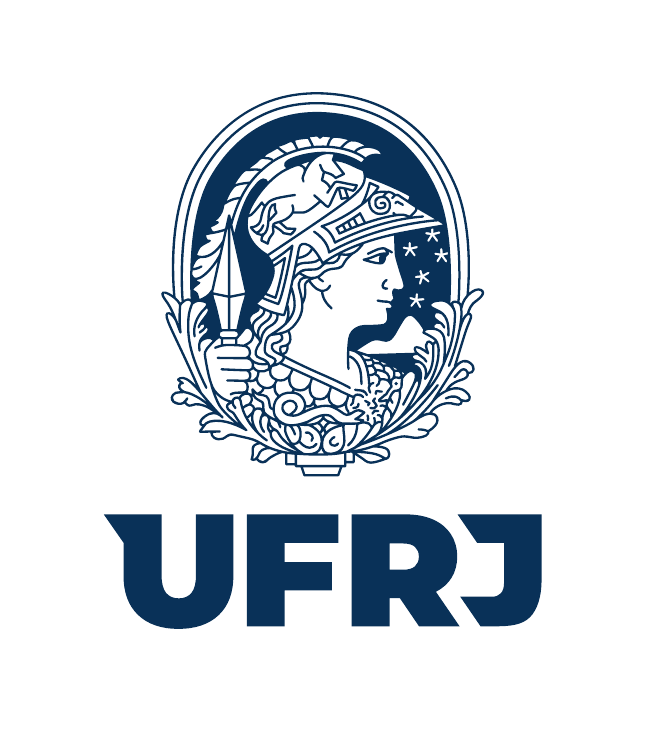}

\myformat\today

\vspace{1cm}

\end{center}
\newpage
\end{titlepage}

\setcounter{page}{0}
\pagenumbering{arabic}   
\addtocounter{page}{1}
\normalsize






\pagebreak

\phantomsection
\setcounter{secnumdepth}{3} 
\setcounter{tocdepth}{1} 
\renewcommand{\contentsname}{Table of Contents}
\tableofcontents

\clearpage \phantomsection
\setcounter{figure}{0}
\addcontentsline{toc}{chapter}{\listfigurename}
\fancyhead[LE,RO]{\normalfont\leftmark}
\markboth{LIST OF FIGURES}{LIST OF FIGURES}
\fancyhead[LE,RO]{\normalfont LIST OF FIGURES}

\listoffigures

\clearpage \phantomsection
\addcontentsline{toc}{chapter}{\listtablename}
\setcounter{table}{0}
\listoftables



\clearpage
\markboth{}{} 
\pagestyle{fancy}
\fancyhf{}
\fancyhead[LE,RO]{\thepage}
\fancyhead[RE]{\leftmark}
\fancyhead[LO]{\rightmark}


\chapter*{\centerline{Acknowledgments}}
\addcontentsline{toc}{chapter}{Acknowledgments}
\label{chapter:acknowledgments} 

This dissertation benefited greatly from the guidance, discussions, corrections, and careful reading provided by my advisers, Professor Marcelo Byrro Ribeiro and Dr. Amanda Reis Lopes.
I thank Professor Marcelo Byrro Ribeiro for teaching me the theoretical content explored throughout this work, for encouraging precise explanations, and for providing detailed recommendations that strengthened the rigor of the dissertation.
I thank Dr. Amanda Reis Lopes for contributing to the organization of the text and for suggestions related to the data analysis that helped clarify the interpretation of the results.
Their many helpful contributions shared in meetings during my master's years were fundamental to the advancement of this project.
I also thank my family for their support throughout this period and for everything they did to help make the completion of this dissertation possible.
Finally, I thank Universidade Federal do Rio de Janeiro and Coordenação de Aperfeiçoamento de Pessoal de Nível Superior (CAPES) for the financial support through a master's scholarship.



\chapter*{\centerline{Agradecimentos}}
\addcontentsline{toc}{chapter}{Agradecimentos}
\label{chapter:agradecimentos} 

Esta dissertação se beneficiou enormemente da orientação, das discussões, das correções e da leitura cuidadosa realizadas por meus orientadores, Professor Marcelo Byrro Ribeiro e Dra. Amanda Reis Lopes.
Agradeço ao Professor Marcelo Byrro Ribeiro por ter me ensinado o conteúdo teórico explorado ao longo deste trabalho, por incentivar explicações precisas e por fornecer recomendações detalhadas que fortaleceram o rigor da dissertação.
Agradeço à Dra. Amanda Reis Lopes por contribuir para a organização do texto e pelas sugestões relacionadas à análise de dados que ajudaram a esclarecer a interpretação dos resultados.
As muitas contribuições compartilhadas por ambos em reuniões durante meu mestrado foram fundamentais para o desenvolvimento deste projeto.
Agradeço também à minha família pelo apoio ao longo deste período e por tudo o que fizeram para ajudar a tornar possível a conclusão desta dissertação.
Por fim, agradeço à Universidade Federal do Rio de Janeiro e à Coordenação de Aperfeiçoamento de Pessoal de Nível Superior (CAPES) pelo apoio financeiro por meio de uma bolsa de mestrado.

\chapter*{\centerline{Abstract}}
\addcontentsline{toc}{chapter}{Abstract}
\label{chapter:abstract}

This dissertation discusses the distribution of city sizes using data from twelve censuses of the Brazilian Institute for Geography and Statistics (IBGE) spanning 150 years, from 1872 to 2022, as well as the allometric scaling relation with the 2022 census data.
The study also explores the applicability of the Gompertz and Zipf distributions to model city populations, suggesting a combined Gompertz-Zipf approach to overcome the limitations of a single power-law model whose lower bound has been often treated as arbitrary. 
The Gompertz distribution is fitted to the population size of less populated Brazilian cities, while the Zipf power law is used to represent the cities having larger populations. 
The threshold separating these two segments varies between 20,216 and 102,857 inhabitants for the entire span of 150 years, which reinforces the need to determine for each dataset where the Zipf power law begins to apply.
The results indicate that this combined distribution offers a coherent and complete representation of the Brazilian city population distribution in all censuses analyzed. 
The data analysis concerning censuses in the period between 1872 and 2022 shows that the Gompertz-Zipf distribution can describe the data across all censuses, with the Gompertz parameter $B$ oscillating between $-2.39\times10^{-5}$ and $-1.49\times10^{-5}$, indicating periods with faster decreases and periods with slower decreases in the number of cities within the less populated segment as population increases.
The Zipf exponent presents a mean value of $a = 2.63 \pm 0.53$ throughout all censuses, ranging from $a = 3.61 \pm 0.09$ in 1872 to $a = 2.28 \pm 0.02$ in 2022, values that are consistent with previous findings in the period 1970-2000 as reported by Moura and Ribeiro (Physica A 367 (2006) 441).
The Gini coefficient $G$ and the Zanardi index $Z$ are also used to quantify and characterize the inequality in the city population distribution, whose obtained values are $G = 0.439$ and $Z = -0.002$ for the 1872 census and $G = 0.733$ and $Z = 0.018$ for the 2022 census dataset, while the Pietra index $p$ and the Kolkata index $k$, associated with the most populated cities, are $p = 31.893\%$ and $k = 65.894\%$ for 1872, and $p = 57.395\%$ and $k = 78.342\%$ for 2022.
These results reveal a high concentration of population in larger cities, with the Gini coefficient increasing from 1872 to 2022 and the Zanardi index becoming positive after 1900, indicating a greater concentration of population in the group of more populated cities than in the group of less populated cities. 
Finally, urban allometry is also examined, generating an exponent $b = 0.835$ for the 2022 data, showing that the scaling relation between area and population holds for urbanized area but not for total administrative area, indicating that scaling is associated with spatial occupation.

\chapter*{\centerline{Resumo}}\label{chapter:resumo}
\addcontentsline{toc}{chapter}{Resumo}

Esta dissertação discute a distribuição dos tamanhos das cidades utilizando dados de doze censos do Instituto Brasileiro de Geografia e Estatística (IBGE) ao longo de 150 anos, de 1872 a 2022, bem como a relação de escala alométrica com os dados do censo de 2022.
O estudo também investiga a aplicabilidade das distribuições de Gompertz e de Zipf para modelar as populações das cidades, propondo uma abordagem combinada Gompertz–Zipf para contornar as limitações de um único modelo de lei de potência, cujo limite inferior tem sido frequentemente tratado como arbitrário.
A distribuição de Gompertz é ajustada ao tamanho populacional das cidades brasileiras menos povoadas, enquanto a lei de potência de Zipf é utilizada para representar as cidades com maiores populações.
O limiar que separa esses dois segmentos varia entre 20.216 e 102.857 habitantes ao longo de todo o período de 150 anos, o que reforça a necessidade de determinar para cada conjunto de dados onde a lei de potência de Zipf começa a se aplicar.
Os resultados indicam que essa distribuição combinada oferece uma representação coerente e completa da distribuição populacional de cidades brasileiras em todos os censos analisados.
A análise dos dados referentes aos censos no período de 1872 a 2022 mostra que a distribuição Gompertz–Zipf consegue descrever os dados em todos os censos, com o parâmetro de Gompertz $B$ oscilando entre $-2{,}39\times10^{-5}$ e $-1{,}49\times10^{-5}$, indicando períodos com reduções mais rápidas e períodos com reduções mais lentas no número de cidades dentro do segmento menos populoso à medida que a população aumenta.
O expoente de Zipf apresenta um valor médio de $a = 2{,}63 \pm 0{,}53$ em todos os censos, variando de $a = 3{,}61 \pm 0{,}09$ em 1872 a $a = 2{,}28 \pm 0{,}02$ em 2022, valores que são consistentes com resultados anteriores para o período de 1970–2000 reportados por Moura e Ribeiro (Physica A 367 (2006) 441).
O coeficiente de Gini $G$ e o índice de Zanardi $Z$ também são utilizados para quantificar e caracterizar a desigualdade na distribuição de população das cidades, cujos valores obtidos são $G = 0{,}439$ e $Z = -0{,}002$ para o censo de 1872 e $G = 0{,}733$ e $Z = 0{,}018$ para o conjunto de dados do censo de 2022, enquanto o índice de Pietra $p$ e o índice de Kolkata $k$, associados às cidades mais populosas, são $p = 31{,}893\%$ e $k = 65{,}894\%$ para 1872, e $p = 57{,}395\%$ e $k = 78{,}342\%$ para 2022.
Estes resultados revelam uma elevada concentração da população em cidades maiores, com o coeficiente de Gini aumentando de 1872 a 2022 e o índice de Zanardi tornando-se positivo após 1900, indicando uma maior concentração da população no grupo das cidades mais populosas do que no grupo das cidades menos populosas.
Por fim, a alometria urbana também é examinada, resultando em um expoente $b = 0{,}835$ para os dados de 2022, mostrando que a relação de escala entre área e população é válida para área urbanizada, mas não para a área administrativa total, indicando que a escala está associada à ocupação espacial.

\chapter*{Introduction}\lb{intro}
\markboth{INTRODUCTION}{INTRODUCTION}
\addcontentsline{toc}{chapter}{Introduction}

Many scientific measurements are centered-peaked, this is, for instance, the case of human body temperature with a value of (37.0 $\pm$ 0.5)°C, the adult resting heart rate (80 $\pm$ 20) bpm, as well as many other quantities. 
However, not all distributions are concentrated around the mean; for instance, the number of people in each city of a country varies over a broad range. The largest city in Brazil is São Paulo in the state of São Paulo, which has 11,451,999 inhabitants, while the smallest is Serra da Saudade in the state of Minas Gerais, populated by only 833 people, according to the 2022 census \cite{IBGE2022census}. 
This disparity illustrates that population distributions span several orders of magnitude and cannot be adequately characterized by traditional statistical measures such as the mean and standard deviation. In other words, there are distributions weakly described by their mean and standard deviation, as exemplified by city populations and income, which are better represented by power laws \cite{newman2005}.

Felix Auerbach (1856-1933) was the first to observe that the way urban aggregates are distributed follows a power-law distribution \cite{auerbach1913}. However, George Kingsley Zipf (1902-1950) is frequently credited with this observation due to his extensive population data study of various countries and the corresponding empirical systematization demonstrating that cities are demographically distributed as a power law with an exponent close to one. This empirical evidence became known as the Zipf law \cite{zipf}.
The importance of this result lies in its apparent universality, as similar scaling patterns have been identified across different countries and historical periods, suggesting that common underlying mechanisms may drive the formation and evolution of city size distributions regardless of geographical or cultural context \cite{newman2005,cristelli2012}.
Moreover, Vilfredo Pareto (1848-1923) found a  power law in the income distribution tails which encompasses the wealthiest segment of the population, known as Pareto law \cite{chami2011}. 
Zipf also studied income distribution besides city population, but from a terminological point of view, a power law with any exponent value is referred to as Pareto law, whereas the name Zipf law is restricted to the power law found in city populations.
Some authors even refer to the Zipf law only when the exponent is approximately equal to one, that is, when the so-called pure Zipf law occurs. While other authors use the term Zipf law more broadly to denote any power law observed in city size distribution regardless of the exponent value, a convention adopted in this dissertation.
In addition, there are those who define the Zipf law using the probability density function (PDF) and those who do so using a cumulative distribution function (CDF), or even the complementary cumulative distribution function (CCDF). Furthermore, there are those who utilize $1/a$ instead of $a$ as the power law exponent. These mathematical differences will be detailed in Section 1.1. 

In a general rank-size formulation of a power-law distribution, size is a measurable quantity such as the number of times a letter appears in a text, the income of certain groups, and in this case it is the number of people in a city, while rank is the position of that quantity in a descending ordering. 
The expression $s(r) \propto r^{-a}$ describes how the size $s$ depends on the rank $r$, where the value $a = 1$ corresponds to the pure Zipf law and suggests a specific kind of hierarchical structure where the demographic size of the largest city $s(r=1)$ is twice the demographic size of the second city $s(r=2)$, three times the size of the third city $s(r=3)$, and so on.
A higher exponent suggests a more equally distributed city population system, which might demonstrate a different pattern of population dynamics compared to a system with a smaller exponent that signals a more uneven distribution where a few large cities concentrate most of the population \cite{arshad2018}.

\begin{figure}[h]
    \centering
    \includegraphics[width=0.7\linewidth]{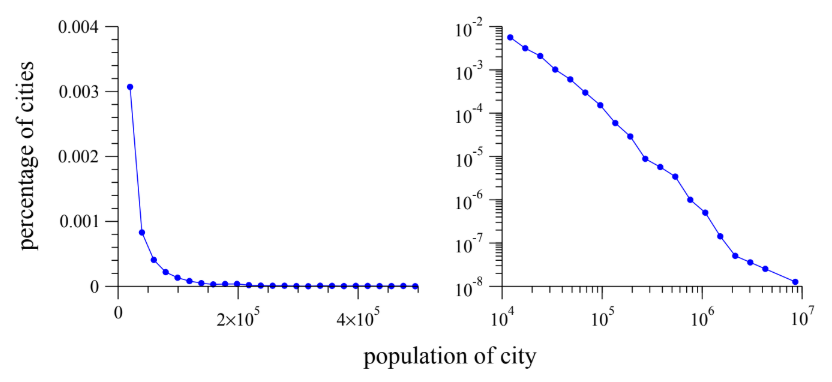}
    \caption[US census: 2000]{Data from the 2000 US Census. Left: histogram of the populations of all US cities with population of 10,000 or more. Right: another histogram of the same data, but plotted on logarithmic scales. The approximate straight line form of the histogram on the right panel suggests that the distribution can be represented by a power law. Figure 2 of Ref. \cite{newman2005}.}
    \label{fig:newman1}
\end{figure}

The Zipf law suggests that the frequency of an event is inversely proportional to its rank, where rank denotes the position of a city in a descending ordering by population size, that is, a categorical index such that the largest city has rank 1, the second largest rank 2, and so on. 
In this context, its general interpretation is that the $n^{\text{th}}$ largest city has a population that is approximately $1/n$ of the population of the largest city \cite{cristelli2012}.
However, many empirical studies examining the Zipf law select cities above a particular population size threshold chosen differently depending on the study's authors because the initial value of the distribution, i.e., less populous cities, does not fit well the power-law behavior \cite{arshad2018}. 
In order to avoid such arbitrariness, one could approach this problem differently by fitting the smaller cities with the Gompertz distribution, which has the same axes as the Zipf law, just on another scale. 
While the Zipf law involves a plot of log-log transformed variables, the Gompertz distribution, which is a double exponential, is represented using a double natural logarithm on one variable and a linear scale on the other, allowing then for a more comprehensive and continuous description of the full range of city sizes.

A city population histogram that forms an approximate straight line when both population size and rank are log-transformed suggests a power-law distribution, as seen on the right of Fig. \ref{fig:newman1}. This means that the probability of a city having a population greater than a certain value $x$ follows a power-law decay, indicating that large cities are statistically more frequent than would be expected under exponential or Gaussian distributions \cite{newman2005}. This feature is characteristic of systems exhibiting scale invariance where no single scale dominates the behavior of the system, and similar patterns can be observed across different levels of aggregation \cite{newman2005}. In this context, the distribution of city sizes can be interpreted as a manifestation of a complex, self-organizing system driven by the interactions of many individual components with finite resources \cite{cristelli2012}.

A collective behavior is defined as emergent in a complex system when the interaction of many parts results in behavior greater than the sum of the individual ones.
From this perspective, the dynamics of city creation and growth can be thought of as being an emergent phenomenon of human cooperation and collective action. 
In this way, the demographic distribution of human beings on Earth's surface has peaks of population concentration, namely the cities, alternating with large areas where the population density is much lower. These peaks follow a power law typical of the non-linear dynamics of a complex system \cite{newman2005, cristelli2012}.
Changes in a country's city population distribution over time can be connected to economic transformations. For example, Ref. \cite{arshad2018} notes that the Pareto exponent in China showed a turning point in 1996, illustrating the country's transition from a planned economy to a more market-oriented economy. 
Another example is the study in Ref. \cite{zhu2018} that discusses how the Zipf law holds approximately for the distribution of city size in each European country but fails for the aggregated European Union data, which suggests that the level of economic integration is a factor in shaping city population distributions. 
These examples highlight that city size distributions are not static but evolve in response to institutional, economic and technological changes.
Consequently, understanding the city population distribution might indicate information on the economic and social properties of a country or region under study.

In this context, this dissertation will discuss Brazil’s city population through the Gompertz-Zipf law, which combines the Zipf law with the Gompertz distribution for the smaller cities. The analysis will be based on data collected by the Brazilian Institute for Geography and Statistics (IBGE) over a 150-year span, from the 1872 census to the 2022 one. This long timescale is particularly fitting, as over time, Brazil has experienced shifts in population distribution associated with internal migration flows and the consolidation of large metropolitan regions, all of which have contributed to shaping Brazil’s current distribution of city sizes \cite{IBGEbook}.

A second objective of this study is to examine possible allometry, specifically the scaling relation between urban area and population in Brazil.
In many complex systems, allometric laws describe how one quantitative measure grows proportionately to another, often following power-law patterns \cite{newman2005}. 
Originally developed in biology, allometry was used to study how morphological, physiological and ecological traits vary with organism size, such as the relationship between brain size and body size, metabolic rate and body mass or wing size and flight performance \cite{shingleton2010}. 
More generally, allometry investigates how characteristics or processes change proportionally with size and how these scaling relations are associated with functional organization and evolution \cite{shingleton2010}. 
In the urban context, this type of analysis helps determine whether larger cities expand in area at the same rate as they grow in population. 
This allometric assessment provides complementary insight into how Brazilian cities organize in terms of infrastructure and land use efficiency.
If larger cities grow more slowly in area relative to population, this suggests increasing population density and more compact land use, whereas faster spatial expansion indicates more dispersed urban development \cite{chen2012}.
Therefore, the allometric analysis alongside the study of population distributions can provide a more comprehensive understanding of Brazilian cities, both in terms of their demographic structure and their spatial organization.

This dissertation is organized as follows: Chapter 1 presents the theoretical foundation of the Zipf power law, the Gompertz distribution, inequality indices and hierarchy in urban systems. 
Chapter 2 describes the Brazilian city population datasets from the censuses between 1872 and 2022, focusing on the 2022 census to analyze different choices of logarithmic binning in order to improve the statistics by generating more data points. 
Chapter 3 presents the results of the Gompertz-Zipf fit for the Brazilian city size distribution, as well as the results of the allometric analysis relating urban area to population restricted to the 2022 data census.

\chapter{Scaling laws and inequality measures}\label{chapter:first_chapter}



\label{sec:visao_geral}


This chapter presents the theoretical concepts and main mathematical formulations used throughout this work, including the Zipf law, the Gompertz distribution, inequality measures such as the Gini and Zanardi coefficients, and hierarchy in urban systems.


\section{The Zipf power law}

The study of rank-size relations has a long history in statistics, linguistics, economics and physics. 
One of the earliest descriptions was given by Auerbach in 1913, who observed that the population of cities in the German Empire approximately followed an inverse relationship between rank and size \cite{auerbach1913}. Later, Zipf studied the frequency of words in several languages and reported a similar regularity in his book \textit{Human Behavior and the Principle of Least Effort} where he analyzed a vast range of languages and dialects including English, Latin, Attic Greek, Gothic, Sanskrit, Chinese, Palestinian Hebrew, Japanese, Icelandic, Dutch, Russian, Italian, Spanish, Portuguese, French and others \cite{zipf}. 
Zipf observed that when words are classified by frequency in a text, the most frequent word appears approximately twice as often as the second most frequent word, three times as often as the third, and so on. This relation can be written as
\begin{equation}
f(r) \propto \frac{1}{r},
\end{equation}
\noindent where $f(r)$ denotes the frequency of the element with rank $r$. 
This empirical pattern is known as the \textit{Zipf law}. In the context of word frequencies, Zipf proposed that the distribution of words reflects a balance between the efforts of speakers and listeners, with language evolving to minimize the total effort required for communication \cite{zipf}.

Although originally formulated in linguistics, similar rank–size relations were later documented in other systems. Examples include the distribution of city populations, wealth of the richest people, intensity of solar flares, earthquake magnitudes and diameter of moon craters \cite{newman2005}. 
In these cases, the probability of observing a value follows a power-law decay, implying that small values are common and large values are rare but still occur with non-negligible probability \cite{newman2005}.
In other words, those power-law distributions were said to have \textit{long tails} because large values have small, but still significant proportions or probabilities. 
Due to this realization, the empirical study of power laws gained attention in several disciplines during the second half of the twentieth century. 
Particularly in city size distributions, it was shown that the largest cities in a country approximately follow the inverse rank-size rule, meaning that the size of an element is inversely proportional to its rank, so that higher-ranked cities have systematically smaller populations.
This regularity has been observed in the United States, European countries, Japan and Brazil \cite{arshad2018,moura2006}.

Power-law behavior also appears in complex systems, extending the rank–size empirical regularities observed in city size distributions discussed above to interaction-based structures.
Unlike static random graph models, real-world networks such as the Internet, social networks and citation networks expand through the constant addition of new vertices. 
Preferential attachment dictates that these new vertices are more likely to link to existing elements that already possess a high number of connections. 
This ``rich-gets-richer'' phenomenon ensures that initial differences in connectivity are amplified over time, as older or more visible vertices acquire new connections at a higher rate. 
As a result, in these systems a small number of elements have many connections while most elements have only a few. 
This self-organizing process serves as a mechanism for the development of heavy-tailed distributions, meaning that the number of connections per element follows a distribution where most values are small while a few are extremely large, consistent with a power-law form.
Thus, it directly gives rise to unequal connectivity patterns, consistent with the disparities often measured in competitive systems. These are systems governed by two fundamental mechanisms: continuous growth and preferential attachment, where new elements connect preferentially to already highly connected ones, leading to cumulative advantage and unequal connectivity \cite{barabasi1999}.

Despite the large number of empirical studies reporting power-law behavior, several authors have noted that estimating power-law exponents is statistically challenging. 
These methodological issues are particularly relevant in the analysis of city size distributions. 
Empirical studies often restrict the sample to cities above a population threshold in order to reduce deviations from the power-law form. 
However, this truncation introduces uncertainty regarding the range over which the power law actually holds \cite{clauset2009}.
Cristelli, Batty and Pietronero \cite{cristelli2012} showed that the rank–size relation depends on the size of the dataset and the coherence of the distribution. 
Their analysis suggests that the simple Zipf relation may not fully capture the statistical structure of city systems.
Therefore, since the early formulations of the Zipf law a recurrent methodological choice in the literature has been the adoption of an arbitrary cutoff defining the lower bound from which the power-law behavior is assumed to hold. 
This cutoff is often chosen arbitrarily, typically based on population thresholds or rank truncation, rather than on a systematic data driven procedure.
As a result, the estimated Zipf exponent becomes conditional on the researcher’s subjective choice of threshold.
In this sense, much of the empirical literature reports Zipf exponents that are not strictly comparable, as they are derived from different truncation criteria. 
Table~\ref{tab:zipfcuts} summarizes some examples of the Zipf power law exponents and the cutoffs adopted in the literature \cite{newman2005,moura2006, Zanette1997}.
In an attempt to bypass such arbitrariness for the lower threshold, in this work the Zipf law is used to model the tail of the city population distribution, while the Gompertz distribution is used for smaller cities \cite{cristelli2012}.


\begin{table}[h!]
\centering
\begin{tabular}{llll}
\hline
The Zipf exponent & $x_{\min}$ & Location and date & Reference \\
\hline
$2.22 \pm 0.34$ & $30{,}000$ & Brazilian cities: 1970-1980 & \cite{moura2006} \\

$2.36$ & $30{,}000$ & Brazilian cities: 1980--2000 & \cite{moura2006}\\

$2.26 \pm 0.11$ & $30{,}000$ & Brazilian cities: 1991-2000 & \cite{moura2006} \\

$2.0 \pm 0.1$ &  $1{,}000$ & Swiss municipalities & \cite{Zanette1997} \\

$2.1 \pm 0.1$ &  $10{,}000$ & USA cities & \cite{Zanette1997} \\

$2.03 \pm 0.05$ & $100{,}000$ & World cities & \cite{Zanette1997} \\

$2.30 \pm 0.05$ & $40{,}000$ & USA cities: 2000 & \cite{newman2005}\\

\hline
\end{tabular}
\caption[The Zipf power law exponents: Literature]{The Zipf power law exponents for city population distributions reported in Refs. \cite{newman2005}, \cite{moura2006} and \cite{Zanette1997}, where $x_{min}$ is the population cutoff defining the lower bound from which the power-law behavior is assumed to hold.}
\label{tab:zipfcuts}
\end{table}

The Zipf law as used to describe the distribution of city population sizes is a special case of the Pareto distribution \cite{arshad2018, zhu2018}. Both distributions exemplify power laws, where the probability of a value is inversely proportional to a power of that value. As mentioned above, the power-law exponent is viewed differently depending on the choice of function, the three most common being the PDF, CDF and CCDF. Each function offers a distinct perspective on the distribution, affecting the exponent's value and its interpretation \cite{moura2006, clauset2009}.
The PDF, denoted here as $p(x)$, describes the likelihood of a variable $x$ taking on a specific value. For a power-law distribution, the PDF is expressed as 
\be
p(x) = C x^{-a} \:,
\ee
where $a$ and $C$ are constants \cite{moura2006}.
Applying the logarithm to both sides, we get:
\be
\log[p(x)] = \log(C) - a \: \log(x) \:.
\ee
This equation represents a linear relationship on a log-log scale, where the slope of the graph is $-a$. Thus, in this form, the exponent $-a$ directly reflects the rate at which the probability decreases as $x$ increases, which corresponds to the slope of the PDF when plotted on logarithmic axes. 

The CDF, expressed as $P^*(x)$ here, represents the probability that the variable $x$ is less than or equal to a certain value. For a power-law distribution, such as the Zipf law, the CDF is,
\be
P^*(x) = \frac{C}{a-1} x^{-(a-1)} \:.
\label{eq:p*x}
\ee
Taking the logarithm of both sides, we get:
\be
\log[P^*(x)] = \log\left(\frac{C}{a-1}\right) - \: (a-1) \: \log(x) \:.
\ee
This equation also represents a linear relationship on a log-log scale, where the slope $\alpha$ of the graph is $-(a-1)$, or $1-a$. The CDF form is associated with the Zipf law or Pareto distribution \cite{newman2005,moura2006}.

The CCDF, here denoted as $P(x)$, characterizes the probability that a variable $x$ is greater than or equal to a certain value. It may be written as follows,
\be
P(x) = 100 - P^*(x) \:,
\ee
\noindent and substituting $P^*(x)$ from Eq. (\ref{eq:p*x}) we have:
\be
P(x) = 100 - \frac{C}{a-1} x^{-(a-1)} \:.
\ee
\noindent Simplifying:
\be
P(x) = \beta x^{1-a} = \beta x^{\alpha} \:,
\label{eq:ccdf}
\ee
\noindent where $\beta = 100 - \frac{C}{a-1}$. Taking the logarithm of both sides of Eq. (\ref{eq:ccdf}), we get:

\be
\log[P(x)] = \log(\beta) + (1-a) \: \log(x) = \log(\beta) + \alpha \: \log(x) \:.
\ee

\noindent Here, the maximum probability is normalized to 100 rather than 1, and the PDF relates to the CDF and the CCDF through the following derivatives:

\be
\frac{dP^*(x)}{dx} = p(x), \qquad \frac{dP(x)}{dx} = -p(x).
\ee

\begin{figure}[h!]
    \centering
    \includegraphics[width=0.8\textwidth]{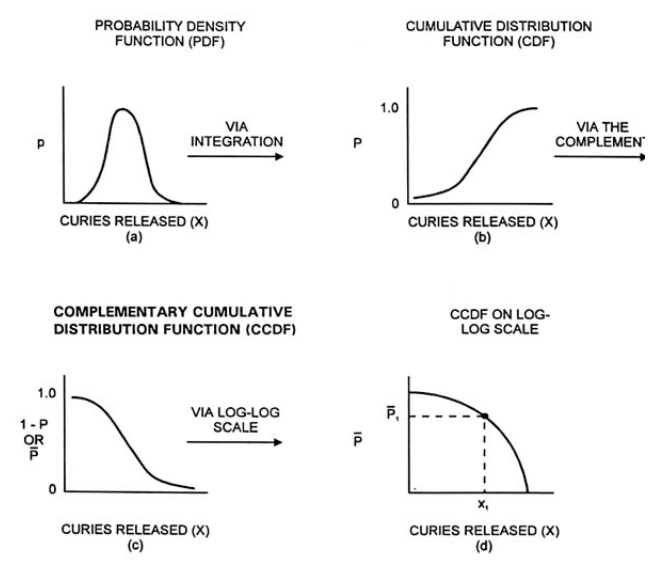}
    \caption[PDF, CDF and CCDF representations]{Illustration of the PDF in panel (a), CDF in panel (b), and CCDF representations for the same probability distribution in panels (c) and (d). The CCDF representation expresses the probability of observing values greater than or equal to a given threshold, providing a smoother and more stable visualization of the tail behavior when compared to the PDF and CDF. Figure B.1 of Ref. \cite[p.~111]{nationalacademies1996}.}
    \label{fig:cdf_ccdf_pdf}
\end{figure}

CCDF plots are more visually stable than PDF and CDF plots, particularly at the tail of the distribution, because the impact of statistical fluctuations from sample analyses is reduced \cite{clauset2009}. 
Figure \ref{fig:cdf_ccdf_pdf} illustrates the differences between the PDF, CDF and CCDF representations for the same distribution, highlighting how the CCDF representation provides a smoother and more stable description of the tail behavior when compared to the PDF and CDF representations.
Similar to the CDF, the CCDF is also associated with a slope of $1-a$ when plotted on a log-log scale, since it follows the same power-law form. 
The Zipf frequency-rank plots typically show $x$ on the horizontal axis and $P^*(x)$ on the vertical axis, and appear as straight lines on log-log plots \cite{newman2005, clauset2009}. In summary, the functional choice, PDF, CDF, or CCDF, affects the exponent's value and interpretation of the graph's slope in data analysis. For the PDF, the slope is equal to the exponent $-a$, while for both CDF and CCDF, the slope is $1-a$. In this study of Brazilian cities, the CCDF will be adopted. 

The Zipf law describes a notable pattern in the distribution of city sizes, in this case size refers to the number of people, suggesting that in a system, the largest city, ranked as one, is approximately twice the size of the second largest, ranked as two, three times the size of the third largest, ranked as three, and so on \cite{arshad2018}. 
The Zipf law, also known as the rank-size rule, is a principle describing the relationship between the rank and size of cities, such as population. 
The CCDF for the Zipf law is represented through a log-log graph, where the axes show the population of cities and the cumulative percentage of cities with equal or greater population than a given population value $x$ \cite{newman2005}. At the minimum value of $x$ the probability that $x$ is greater than or equal to that minimum value is 100\%. As $x$ increases, the probability of observing values greater than or equal to $x$ decreases, resulting in a decaying slope in a graph.

\begin{figure}[h!]
    \centering
    \includegraphics[width=0.8\textwidth]{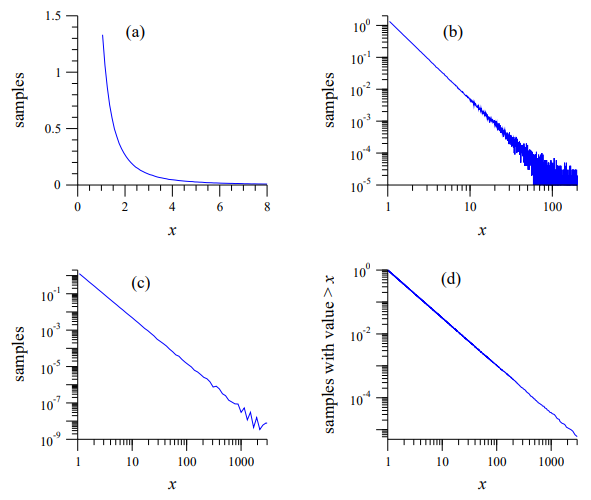}
    \caption[Graphical representations of a power-law distributions]{Different graphical representations of a power-law distribution. 
    Panel (a) shows the histogram in linear scale, while panel (b) presents the same data on logarithmic axes, where the scaling behavior becomes more evident but fluctuations increase in the tail. 
    Panel (c) illustrates the use of logarithmic binning (see main text) to reduce statistical noise, and panel (d) shows the cumulative representation, which provides a smoother and more stable description of the distribution. 
    Figure 3 of Ref. \cite{newman2005}.}
    \label{fig:powerlaw_newman}
\end{figure}

Figure~\ref{fig:powerlaw_newman} shows the same power-law distributed data under different graphical representations in Ref. \cite{newman2005}. 
Panel (a) presents the distribution in the form of a standard histogram of a synthetic dataset of 1 million random numbers generated to follow a power-law distribution. 
In panel (b), the same data plotted on logarithmic scales, making the power-law behavior more apparent through an approximately linear trend, although the tail becomes increasingly noisy due to the small number of observations in that region. 
Panel (c) shows the use of logarithmic binning, which reduces these fluctuations by grouping data into bins that grow geometrically, resulting in a smoother representation of the distribution. 
Finally, panel (d) presents the cumulative form of the distribution, where the data appear significantly more stable, since each point incorporates contributions from all larger values. 
In this representation, the power-law behavior is preserved, but with a different exponent reflecting the cumulative transformation. 
Together, these panels illustrate how the choice of representation affects the visualization of scaling behavior and why cumulative approaches such as the CCDF provide a more reliable description of the distribution, which is further supported by Ref. \cite{Islas2025}, where the CCDF representations are shown to improve the stability of empirical income distributions, particularly in the tail region.

While the Zipf law holds in many instances, empirical results have been criticized due to sample selection biases regarding the choice of the lower bound from which the power-law behavior starts, as well as data limitations related to incomplete coverage of the full city size distribution \cite{cristelli2012,arshad2018}. 
In this work, the data range is assessed as a whole, avoiding the introduction of arbitrary cutoffs that may affect the estimation of the scaling exponent. 
While the end of the distribution, with the most populated cities, is fitted into the Zipf law, the beginning, related to cities with smaller populations, is fitted into a Gompertz distribution. 
Ref. \cite{cristelli2012} supports combining distributions, such as Gompertz-Zipf, lessening the single power-law model limitations and achieving a better description for scaling systems.

\section{The Gompertz distribution}

The Gompertz distribution originates from demographic studies concerned with human mortality. 
In his original work, Benjamin Gompertz (1779-1865) introduced what is now known as the Gompertz law of mortality while analyzing the age dependence of death rates in adult populations. He observed that beyond early life stages, mortality increases approximately exponentially with age, a pattern that could be consistently identified across different datasets available at the time. 
This empirical regularity led to the formulation of the Gompertz function, which provided a mathematical representation of mortality dynamics and later became a tool in survival analysis with direct applications in life insurance price estimation based on survival probabilities, where the CCDF is referred to as the survival function in medical contexts and as the reliability function in engineering, both interpretations describing the probability that a patient survives or a device remains functional beyond a certain time \cite{Islas2025}.
Over time, the flexibility of this functional form enabled its application beyond its original demographic context, extending to multiple domains where growth and decay processes exhibit non-linear temporal behavior \cite{gompertz1825}.

The Gompertz curve $y(t)$ can be written in the form
\begin{equation}
y(t) = a \exp\left(-b {\mathrm{e}}^{-c x}\right),
\label{eq:winsor}
\end{equation}
\noindent corresponding to Eq. (2) in Ref.~\cite{winsor1932}, which was originally expressed as $y = k {\mathrm{e}}^{-{\mathrm{e}}^{a-bx}}$ whit $k,b>0$. The expression above is obtained from that formulation through a redefinition of parameters, in which the inner exponential structure is preserved but written in a more compact form for convenience. In this representation, the parameter $a$ corresponds to the upper limiting value of the curve and is denoted by $k$ in the original formulation, while $b$ and $c$ are the parameters of Eq. (\ref{eq:winsor}) that control the displacement and the rate of growth of the curve, respectively.
This functional form is characterized by an initial phase of slow growth, followed by a period of acceleration, and finally a saturation phase in which the variable approaches an upper bound asymptotically. Such asymmetric behavior distinguishes the Gompertz curve from other growth models, as the logistic curve, which exhibits a symmetric growth pattern around its inflection point. Due to these properties, the Gompertz function has been applied in fields such as biology, demography, epidemiology and economics, where systems evolve under limiting conditions \cite{gompertz1825, winsor1932}.

In the context of population dynamics, the Gompertz function has been employed to describe growth processes in which the rate of increase declines as the system evolves, reflecting the presence of limiting factors. Originally introduced to model age-dependent mortality, the function captures a progressively decreasing growth rate, which can be interpreted as the effect of constraints acting more strongly as the system size increases \cite{gompertz1825}. This characteristic makes the Gompertz curve suitable for representing situations in which initial growth is relatively rapid but gradually slows over time, producing a saturation behavior consistent with bounded systems.
Such decelerating growth patterns are also consistent with observations in complex socioeconomic systems, where aggregate quantities do not grow indefinitely but instead approach limiting regimes. These resource limitations can play a central role in shaping macroscopic behavior \cite{yakovenko2009}. 
The resulting dynamics differ from purely exponential growth because of this intrinsic slowdown in large-scale distributions evolving under constraints.

Another important application of the Gompertz distribution arises in the study of income distributions. Empirical evidence suggests that the majority of income data, corresponding to the lower and middle income classes, can often be approximated by a Gompertz function, while the upper tail follows a Pareto distribution. This two-regime structure reflects the coexistence of different underlying economic mechanisms. Detailed analyses of this phenomenon were provided by F. Chami Figueira et al. \cite{chami2011} and later by Marcelo Byrro Ribeiro in his book \cite{ribeiro2020}, who introduced and empirically validated the Gompertz-Pareto distribution as a model capable of capturing the full range of income data.
The Gompertz function contains parameters that determine how rapidly the probability decreases with increasing values of the variable, making it suitable for describing distributions that decay progressively rather than following a scale-free power law.
This behavior contrasts with power-law distributions, which are scale-invariant, meaning that no preferred scale exists and the same mathematical relation holds across different orders of magnitude.
In empirical socioeconomic data, such as income distributions, the majority of the population typically does not follow a power law, and instead exhibits a faster decay that can be represented by alternative functions. In this context, the Gompertz function has been employed to model the lower and intermediate regions of distributions, where the majority of observations are concentrated \cite{chami2011,ribeiro2020}.

From a statistical perspective, the Gompertz distribution can be expressed through its cumulative distribution function as
\begin{equation}
F(x) = 1 - \exp[-b({\mathrm{e}}^{cx}-1)],
\end{equation}
\noindent where $b$ and $c$ are positive parameters controlling the shape of the distribution. This formulation produces a rapidly decaying tail, reflecting the decreasing probability of large values, and provides a representation capable of fitting empirical data in the region where deviations from power-law behavior are observed \cite{winsor1932}.
Applications of this approach can be found in the analysis of socioeconomic systems, where different mechanisms govern different parts of the distribution. 
In studies of income and wealth, it has been observed that the majority of the population follows a distribution that decays more rapidly than a power law, while the upper tail is better described by a power law associated with multiplicative processes and accumulation effects \cite{yakovenko2009}.
Similarly, empirical analyses of firm sizes and city populations show that large entities often follow power-law behavior, whereas smaller ones deviate from this pattern and require alternative descriptions \cite{axtell2001}.
To account for this heterogeneity, hybrid models combining Gompertz and Pareto components have been proposed. In such models, the Gompertz function describes the bulk of the distribution, while the Pareto tail captures the extreme values. This approach provides a unified structure for representing empirical distributions in their entire range, reflecting the coexistence of different underlying mechanisms and improving the overall fit to observed data \cite{chami2011,ribeiro2020}.


Hence, the Gompertz curve, known to be a good descriptor of population dynamics, mortality rate, and growth processes in biology \cite{chami2011}, can also be applied to city size distributions. 
The Gompertz CCDF is suitable for modeling income distributions, particularly for lower-income populations, while the Pareto distribution models the income of the wealthiest individuals \cite{chami2011, ribeiro2020, pinto2012}. 
The Gompertz distribution will be used here to study less populated cities, instead of income. 
For this reason, a novel contribution to the literature is made in this work with the first use of the Gompertz CCDF to model city population sizes, following the expression:

\be 
F(x) = {\mathrm{e}}^{{\mathrm{e}}^{(A+Bx)}} \:,
\ee

\noindent where $x$ in the present approach represents the population size of a city, $A$ and $B$ are parameters that determine the shape of the curve. These parameters are constants fitted to the data.

When plotted on a linear scale for $x$ on the horizontal axis and a $\ln(\ln)$ scale for $f(x)$ in the vertical axis, the straight line will be $(A+Bx)$ with slope equal to the parameter $B$ and with $A = \ln[\ln 100]$, a value which determines the boundary conditions.
Consequently, $B$ controls how rapidly the CCDF changes with population size.
In this empirical study, the parameters $A$ and $B$ are estimated by fitting the Gompertz curve to the observed data using Python's linear data fitting via the Least Squares Method, as described in the \texttt{scipy.stats.linregress} documentation \cite{scipy_linregress}.
By using the Gompertz distribution, city population size can be effectively modeled, providing information on the dynamics and structure of smaller cities in the context of population size.
Finally, the two-component \textit{Gompertz-Zipf population distribution function} $F_{GZD}$ can be written as in Eq. (\ref{eq:chaves}), similarly to the Gompertz-Pareto income distribution given by Eq. (2.36) in Ref. \cite{ribeiro2020},
\be
F_{GZD}(x) =
\begin{cases} 
    {\mathrm{e}}^{{\mathrm{e}}^{(A+Bx)}}  ,  \,\,\,\, (0 \leq x < x_t) \,, \\ 
    \beta x^{1-a}  ,  \,\,\,\, (x_t \leq x \leq \infty) \, , 
\end{cases}
\label{eq:chaves}
\ee

\noindent where $x_t$ stands for the limit between the Gompertz and Zipf segments of the distribution, as it will be detailed in Chapter 2.

\section[Measures of inequality]{Measures of inequality: Gini, Kolkata, Pietra and Zanardi indices}

The measurement of inequality is a central topic in economics and statistics, where the goal is to quantify how a given resource, such as income, wealth or population is distributed in a set of units. Over time, several quantitative indicators have been developed to summarize the dispersion and concentration present in empirical distributions. Among these measures, the Gini coefficient stands out as one of the most used due to its intuitive geometric interpretation and its applicability in different domains \cite{gini1921}.

The Gini coefficient was introduced by Corrado Gini (1884-1965) in 1912 as a measure of statistical dispersion based on the Lorenz curve representation of a distribution as proposed by Max Otto Lorenz (1876-1959) \cite{lorenz1905}. 
The Lorenz curve describes the cumulative share of a variable as a function of the cumulative share of the population, where individuals or units are ordered from the smallest to the largest value. This construction allows the visualization of how far a given distribution deviates from perfect equality, providing a graphical and analytical foundation for the study of inequality \cite{lorenz1905}.
In the case of perfect equality, the Lorenz curve coincides with the diagonal line, commonly referred to as the line of equality. Any deviation from this diagonal indicates the presence of inequality, with the magnitude of the deviation reflecting the degree of concentration in the distribution. 

The Gini coefficient is defined as the normalized area between the Lorenz curve and the line of equality, taking values in the range between zero and one or between zero and one hundred, depending on the adopted maximum percentage share, or probability, normalization. A value of zero corresponds to perfect equality, while a value of 1 or 100 corresponds to the extreme case in which all quantities are concentrated in a single unit \cite{gini1921}.
Due to its general formulation, the Gini coefficient has been extensively applied not only to income and wealth distributions but also to a wide range of other empirical contexts. These include the analysis of concentration in land ownership, firm sizes, scientific citations and city population distributions \cite{sen1997}.

\begin{figure}[h!]
    \centering
    \includegraphics[width=0.4\textwidth]{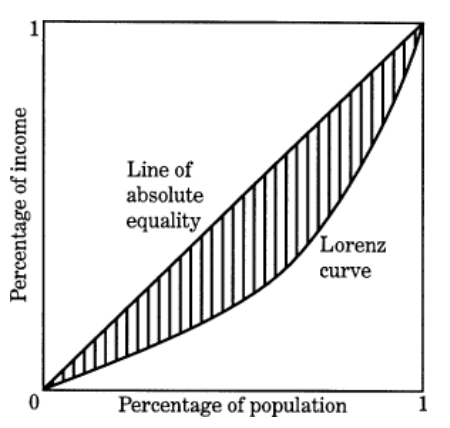}
    \caption[Lorenz curve with equality line]{Lorenz curve and graphical representation of the Gini coefficient. The diagonal line represents perfect equality, while the Lorenz curve describes the cumulative distribution of the variable. The shaded area between the two curves corresponds to the measure used to compute the Gini coefficient. Diagram 2.3 from Ref.~\cite[p.~30]{sen1997}.}
    \label{fig:lorenz73sen}
\end{figure}

\begin{figure}[h!]
    \centering
    \includegraphics[width=0.4\textwidth]{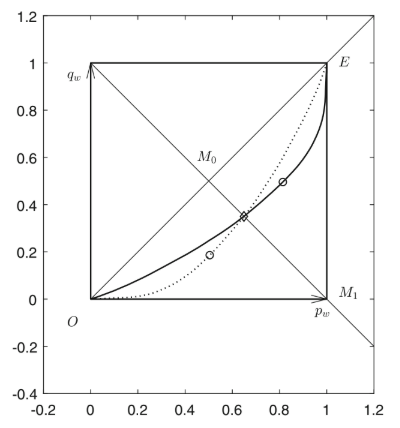}
    \caption[Two Lorenz curves with identical Gini but different shapes]{Two Lorenz curves with identical Gini coefficients but different shapes. 
    Although both distributions exhibit the same overall level of inequality, their internal structures differ, reflecting distinct patterns of concentration across the population. 
    This illustrates a limitation of the Gini coefficient, which does not capture asymmetries in the distribution. Figure 2 from Ref. \cite{clementi2019}.}
    \label{fig:lorenz_asymmetry}
\end{figure}

Figures~\ref{fig:lorenz73sen} and \ref{fig:lorenz_asymmetry} illustrate important aspects of the Lorenz curve representation and the interpretation of the Gini coefficient. 
Figure~\ref{fig:lorenz73sen} shows the standard construction of the Lorenz curve in relation to the line of perfect equality, represented by the diagonal. 
The shaded area between these two curves corresponds to the measure used in the definition of the Gini coefficient, highlighting how deviations from equality translate into quantitative measures of inequality. 
As the distribution becomes more unequal, the Lorenz curve moves further away from the diagonal, increasing the relative internal shaded area and, consequently, the value of the Gini coefficient. 
In contrast, Figure~\ref{fig:lorenz_asymmetry} presents two distinct Lorenz curves that share the same Gini coefficient despite exhibiting different shapes. 
This result emphasizes that the Gini index captures only the overall level of inequality, without providing information about how this inequality is distributed across the population. 
In particular, different distributions may present concentration either among poorer or richer segments while having the same Gini value, indicating that additional measures may be required for a more detailed characterization of inequality structures.
This limitation has motivated the development of complementary indices, such as the Kolkata index and the Zanardi coefficient, that provide additional information on concentration and the relative contribution of different segments of the population \cite{ribeiro2020,clementi2019,ghosh2014,chakrabarti2025,clementi2025}.
For more detailed information on the Kolkata index, see Ref.~\cite[p.~62]{ribeiro2020}. 
A discussion of inequality measures, including the Gini and Zanardi coefficients, can be found in Section 2 of Ref.~\cite{clementi2025}.

\begin{figure}[h]
    \centering
    \includegraphics[width=0.5\linewidth]{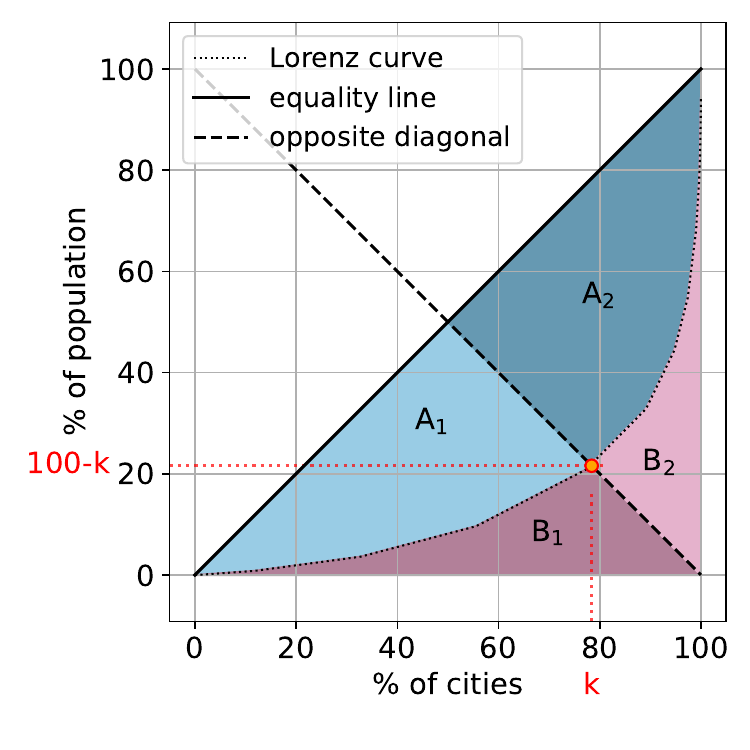}
    \caption[Hypothetical Lorenz curve]{Hypothetical Lorenz curve for Gini and Zanardi indices where $A = A_1 + A_2$ and $B = B_1 + B_2$. The partial Gini coefficients are defined as $G_1 = A_1/(A_1+B_1)$ and $G_2 = A_2/(A_2+B_2)$.}
    \label{fig:lorenz}
\end{figure}

The Gini coefficient or Gini index, here denoted as $G$, and the Kolkata index, also known as the $k$-index, are both measures of inequality, often used to describe income distribution, although applicable to any unequally distributed quantity such as city population. 
The Gini coefficient is derived from the Lorenz curve, which quantifies statistical dispersion, reflecting the inequality of the distribution, such as that of income or wealth \cite{ribeiro2020}. 
In the context of city demographic size, the Lorenz curve displays the cumulative proportion of the total population against the cumulative proportion of cities when the cities are ranked by population. 
The Gini index is then defined as the ratio of the area $A$ between the egalitarian line and the Lorenz curve to the total area $A + B$, where $B$ is the area beneath the Lorenz curve. As seen in Fig. \ref{fig:lorenz}, $A = A_1 + A_2$ and $B = B_1 + B_2$. The defining equation is: 
\begin{equation}
    G \equiv \frac{A}{A + B} \:.
\end{equation}

\noindent In a scenario of perfect equality, the Lorenz curve aligns with the egalitarian line and G = 0, indicating that all cities have equally distributed populations among them, that is, for example, 30\% of the cities hold 30\% of the population, 70\% of the cities hold 70\% of the population, and so on.
In contrast, under perfect inequality where G = 1, one city holds the entire population of the country or region being analyzed. 

The $k$-index is defined as the fraction $k$ of cities that account for $100 - k$ percent of the total population. Graphically, $k$ corresponds to the $x$-axis coordinate on the Lorenz curve where a diagonal orthogonal to the egalitarian line intersects the Lorenz curve, as shown in Fig. \ref{fig:lorenz}. 
Unlike the Gini coefficient, the $k$-index lacks a direct expression. 
Originally defined in the context of income, the $k$-index was later found to suggest an empirical linear relationship with $G$, given by $k = 50 + k_1 G$, with $k_1 = 36.5 \pm 0.5$, after suitable normalization \cite[p.~62]{ribeiro2020}.



\begin{figure}[h!]
    \centering
    \includegraphics[width=0.4\textwidth]{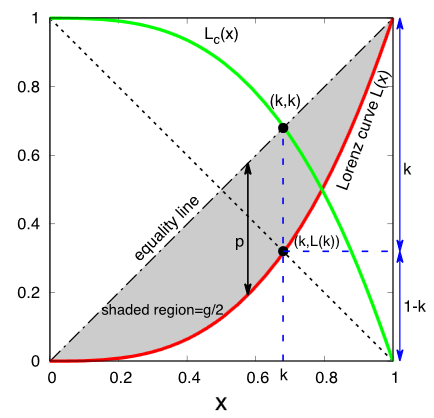}
    \caption[Geometric interpretation of the Pietra inequality measure]{Lorenz curve $L(x)$ representing the cumulative share of a quantity held by a fraction $x$ of elements ordered in ascending value, under the normalization adopted by the authors where the total quantity equals 1, not 100. The green curve represents the complementary Lorenz curve $L_c(x)=1-L(x)$. The Pietra index $p$ is given by the maximum vertical distance between the Lorenz curve and the equality line. Figure 1 from Ref. \cite{chakrabarti2025}.}    
    \label{fig:pietra_geom}
\end{figure}

Another inequality measure associated with the Lorenz curve is the Pietra index, also known as the Robin Hood or Hoover index \cite{chakrabarti2025}.
The Pietra index, here denoted by $p$, is defined as the maximum vertical distance between the Lorenz curve and the equality line, as illustrated in Fig.~\ref{fig:pietra_geom}, and it quantifies the maximum deviation between the observed distribution and perfect equality. 
Perfect equality corresponds to a Lorenz curve coinciding with the equality line, where $x\%$ of the cities contain $x\%$ of the population for every value of $x$, but the Pietra index does not define a redistribution process nor mathematically determine how the distribution would change after redistribution.


Ref.~\cite{chakrabarti2025} considered a quadratic approximation for the Lorenz curve, written as $L(F)=DF+EF^2$, satisfying the conditions $L(0)=0$ and $L(100)=100$.
Under this approximation, analytical relations between the inequality indices were obtained, namely $p \approx \frac{3}{4}G$ and $k \approx \frac{1}{2}+\frac{3}{8}G$ for variables normalized between 0 and 1, or equivalently $p \approx 100 \times \frac{3}{4}G$ and $k \approx \frac{100}{2}+\frac{300}{8}G$ in the 0 to 100 scale.
For convenience, the approximated quantities obtained from the Gini index are here denoted by $p_G$ and $k_G$, respectively,
\begin{equation}
p_G = 100 \times \frac{3}{4}G ,
\label{eq:pG}
\end{equation}
and
\begin{equation}
k_G = \frac{100}{2}+\frac{300}{8}G .
\label{eq:kG}
\end{equation}
However, these relations remain approximations with an estimated error of about 5\%, since they depend on the quadratic representation of the Lorenz curve.
More accurate values of the Pietra index can instead be obtained numerically from the maximum distance between the Lorenz curve and the equality line using higher-order polynomial representations \cite{chakrabarti2025}.

The Zanardi coefficient is an inequality index that, similarly to the Gini index, is based on the Lorenz curve \cite{clementi2019}. 
However, instead of using the two regions $A$ and $B$ as in the Gini index, it divides the area below and above the Lorenz curve into four regions: $A_1$, $A_2$, $B_1$ and $B_2$. 
The Zanardi index Z can then be expressed as
\begin{equation}
Z = \frac{k(100-k) (G_2 - G_1)}{10^4G},
\end{equation}
\noindent where $G_2$ and $G_1$ are the Gini coefficients associated with the upper and lower portions of the Lorenz curve shown in Fig.~\ref{fig:lorenz}, respectively, with $G_1 = A_1/(A_1+B_1)$, $G_2 = A_2/(A_2+B_2)$, while $10^4$ is the normalization factor used when the axes are expressed in percentages from 0 to 100 rather than proportions from 0 to 1.

If $G_2 > G_1$, then $Z > 0$, indicating that the upper part of the distribution is more concentrated; if $G_2 < G_1$, then $Z < 0$, indicating higher concentration in the lower part. Therefore, the Zanardi index captures the asymmetry of the Lorenz curve and indicates whether the asymmetry favors the upper or the lower part of the distribution.
When used in conjunction with the Gini index, it provides a more complete description of inequality.
When normalized to a $100 \times 100$ box, the area $A + B = \frac{100 \times 100}{2} = 5000$ corresponds to the colored bottom triangle in Fig. \ref{fig:lorenz}. 
With $G = A/5000$, $G_2 = A_2/2500$ and $G_1 = A_1/2500$, the dimensionless Zanardi index takes values in the range $-1 \leq Z \leq +1$.
\begin{equation}
Z = \frac{k(100-k)}{10^4(A/5000)} \left( \frac{A_2-A_1}{2500}\right) = \frac{2k(100 - k)}{10^4} \left( \frac{A_2 - A_1}{A}\right),
\end{equation}
\noindent and in terms of the areas $B$, $B_1$ and $B_2$ under the Lorenz curve:
\begin{equation}
Z = \frac{2k(100 - k)}{10^4} \left[ \frac{(2500-B_2) - (2500-B_1)}{5000-B}\right] = \frac{2k(100 - k)}{10^4} \left( \frac{B_1-B_2}{5000-B}\right).
\end{equation}
\noindent In the event of perfect equality, the Lorenz curve coincides with the equality line, so that $A_1 = A_2 = 0$ and $B_1 = B_2 = 2500$. In this case, $G_1 = G_2 = G = 0$ and the Zanardi index is indeterminate, since $Z = 0/0$, reflecting the absence of both concentration and asymmetry.
In the opposite extreme, perfect inequality occurs when the Lorenz curve collapses onto the axes, implying $B_1 = B_2 = 0$ and $A_1 = A_2 = 2500$. Then $G_1 = G_2 = G = 1$, and $Z = 0$, since there is maximum concentration but no asymmetry between the two parts.
More generally, for any symmetric distribution, the two parts of the Lorenz curve exhibit the same degree of concentration, so that $G_2 = G_1$, which implies $Z = 0$. 
In terms of areas, this corresponds to $B_1 = B_2$ and $A_1 = A_2$, meaning that the two regions defined by the $k$-point are balanced.
When $B_1 > B_2$, it follows that $A_1 < A_2$, hence $G_2 > G_1$ and therefore $Z > 0$. 
This indicates that the segment corresponding to larger values of the variable is more concentrated ($G_2 > G_1$), even though the geometric area under the Lorenz curve is smaller in that region ($B_1 > B_2$).
Conversely, when $B_1 < B_2$, then $A_1 > A_2$, hence $G_2 < G_1$ and thus $Z < 0$, indicating greater concentration in the lower part of the distribution.

While $G$ summarizes the overall level of concentration, 
the $k$-index offers a more direct interpretation in terms of the fraction of cities accounting for the majority of the population and $Z$ captures the asymmetry of this distribution. 
These complementary measures allow for a detailed understanding of the structure of city systems. 
As stated above, the Gini and Zanardi indices were originally developed to quantify income and wealth inequality but can be applied to any form of inequality. In this study, they are used to quantify the city population distribution using data from the demographic Brazilian censuses from the years of 1872 to 2022.


\section{Hierarchy in urban areas}

The concept of hierarchy in urban systems has been studied in geography and urban economics in relation to scaling laws and self-organization. 
Scaling approaches show that urban systems exhibit self-similar patterns in spatial and temporal scales \cite{chen2012}. 
In this context, the Zipf law can be interpreted as a representation of hierarchical organization, where cities are distributed across levels following scaling relations \cite{chen2014}. 
These patterns are associated with urban growth, spatial interaction and resource allocation, contributing to the emergence of power-law distributions in city sizes. 
Empirical studies show that the relation between population, urban area and number of cities can be described through scaling and allometric relations \cite{chen2012}. 
The Zipf law can be used to describe the distribution of city sizes within a region and this distribution reflects how urban area is filled based on a rank-size rule \cite{chen2014}.
The hierarchical structure of cities can be mathematically described using exponential equations that relate urban area and rank \cite{chen2012}, expressed as:
\be N(m) = N_1 \cdot r_N^{-(m-1)} \:, \ee
\be P(m) = P_1 \cdot r_P^{-(m-1)} \:, \ee
\be A(m) = A_1 \cdot r_A^{-(m-1)} \:. \ee
Where $m$ denotes the rank (ordinal number from bottom to top), $N(m)$, $P(m)$ and $A(m)$ represent the number of cities, the number of people and urban area of rank $m$, respectively.
$N_1$ $P_1$, and $A_1$ represent the number of cities, the number of people and urban area at the bottom order, respectively, with additional condition $ N_1 >> 1 $.
The quantities $r_N$, $r_P$, and $r_A$ are defined as the following ratios:
\be r_N= \frac{N_{m+1}}{N_m} \:, \ee
is the class number ratio of cities,
\be r_P= \frac{P_m}{P_{m+1}} \:, \ee
is the population size ratio, and
\be r_A= \frac{A_m}{A_{m+1}} \:, \ee
is the urban area ratio. These equations model the cascade structure of urban hierarchies. 

The relationship between urban area and the number of cities can be described by a power-law scaling relation of the form $N \propto A^{-d}$, or equivalently, $\ln N = -d \ln A + \text{constant}$, where $d = \ln r_N / \ln r_A$ can be interpreted as the fractal dimension associated with urban area. In double logarithmic plots this relation appears as a straight line, indicating a fractal scaling within urban hierarchies \cite{chen2012}.
Thus, the Zipf law indicates hierarchical structures within urban systems. These structures can be modeled through exponential equations and power-law relationships, providing insight into the scaling and self-similarity observed in urban area distributions \cite{chen2012, chen2014}.
In addition, the relationship between urban area and population can be mathematically expressed with:
\be A = aP^b \: ,
\label{eq:allometry}
\ee 
\noindent where $A$ is the urban area, $P$ is the population, and $a$ and $b$ are constants \cite{chen2012}.

Allometry refers to the study of how different quantities scale with each other \cite{chen2012}.
In this context, the scaling exponent $b$ in Eq. (\ref{eq:allometry}) quantifies how urban land area grows in relation to population growth and reflects different types of urban land use, whether intensive or extensive, concentrated or dispersed.
If $b > 1$, it indicates positive allometry, meaning urban land area grows at a faster rate than population. This suggests that per capita land area increases as a city becomes larger.
If $b < 1$, it indicates negative allometry, meaning urban land area grows at a slower rate than population. This implies that per capita land area decreases as a city becomes larger.
If $b = 1$, it indicates isometric scaling, meaning urban area and population grow at the same rate, and per capita land area remains constant.
In summary, the allometric exponent $b$ in Eq.~(\ref{eq:allometry}) characterizes how urban area scales with population, distinguishing between intensive land use (negative allometry), extensive land use (positive allometry) and proportional growth (isometric scaling) \cite{chen2012, chen2014}.


\chapter{Brazilian city data}\label{chapter:second_chapter}
\markboth{CHAPTER 2. BRAZILIAN CITY DATA}{CHAPTER 2. BRAZILIAN CITY DATA}
\thispagestyle{plain}

The Brazilian Institute for Geography and Statistics (IBGE) conducts official censuses primarily through national field operations usually every ten years. 
In order to collect detailed data on demographics, education, employment, housing and migration, trained agents visit all households in Brazil, both urban and rural areas, including regions hard to reach. 
The information selected for the present analysis was the number of people per city based on census data from the years 1872, 1900, 1920, 1940, 1950, 1960, 1970, 1980, 1991, 2000, 2010 and 2022.
The methodology used to process and analyze these data is described in this chapter, including the application of logarithmic binning and the identification of the Gompertz and Zipf segments.
In addition, data on urbanized and total areas only available for the 2022 census were assessed to investigate scaling relations between population size and city area.

\section{Population from the 2022 census}

\begin{figure}[H]
    \centering
    \includegraphics[width=0.9\linewidth]{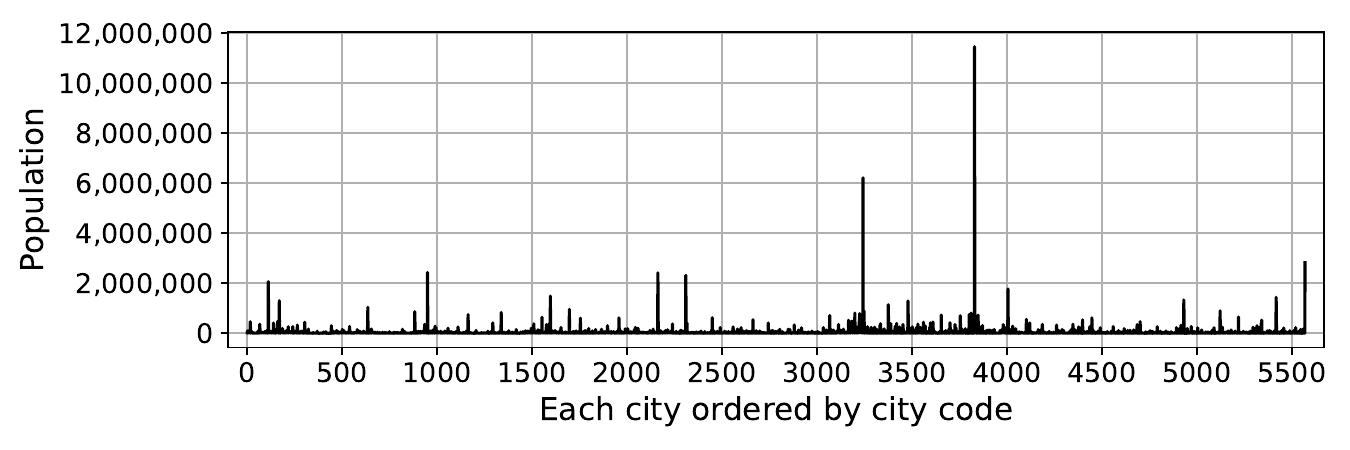}
    \caption[2022 city population]{Each bar in the graph represents the population of one of the 5,570 cities in Brazil according to the 2022 census. The cities are ordered along the x-axis by their official IBGE city codes, which are unique numerical identifiers assigned to each municipality, while the y-axis shows the population, summing up to 203,080,756 people.}
    \label{fig:pop2022}
\end{figure}

According to the 2022 census in Ref. \cite{IBGE2022census}, Brazil has 5,570 cities with 203,080,756 people in total.
Fig. \ref{fig:pop2022} shows the population per city, ordered by its official IBGE city code. 
For analytical purposes, these data were grouped into bins whose sizes grow in accordance with the expression below \cite{moura2006}, in order to lessen statistical fluctuations present at the tail of the CCDF distribution. 
This process was continued up to the 14th bin, corresponding to the point where only one city remains above the lower bound of the bin:
\be 
x_i = 2^{i-1} \cdot x_{min} \: , 
\label{eq:bin2}
\ee
where $x_i$ represents the upper bound of the $i$-th bin and $x_{min}$ is the minimum value of $x$, namely the population of the city with the least people minus 1.
The term $2^{i-1}$ makes the width of each bin double as $i$ increases. This method is called logarithmic binning, where the bin widths increase exponentially. 
Logarithmic binning is used to reduce statistical fluctuations that arise primarily at the tail of a distribution, because in many real-world data sets the number of observations decreases as the variable increases leading to noisy tails in histograms. Logarithmic binning addresses this problem by grouping data into increasingly larger intervals, thus smoothing out the fluctuations, as seen in Fig. \ref{fig:powerlaw_newman}.

The 2022 Brazilian census has $x_{min} = 833 - 1$ that refers to the city of Serra da Saudade in Minas Gerais, whose population was counted as being formed by only 833 people, the smallest in Brazil. Then, the bins are: 
\be x_1 = 2^{1-1} \cdot 832 = 832 \ee
\be x_2 = 2^{2-1} \cdot 832 = 1664 \ee
\be x_3 = 2^{3-1} \cdot 832 = 3328 \ee
and so on. 
The logarithmic binning technique is particularly useful when analyzing power-law distributions or other data where the frequency of observations decreases rapidly with increasing values. 
The first bin refers to all 5,570 cities because every city has at least 832 inhabitants. The second bin expresses that 5,522 cities have at least 1,664 people each, and so on for the following bins. 

\begin{figure}[H]
    \centering
    \begin{minipage}{0.48\textwidth}
        \centering
        \includegraphics[width=\linewidth]{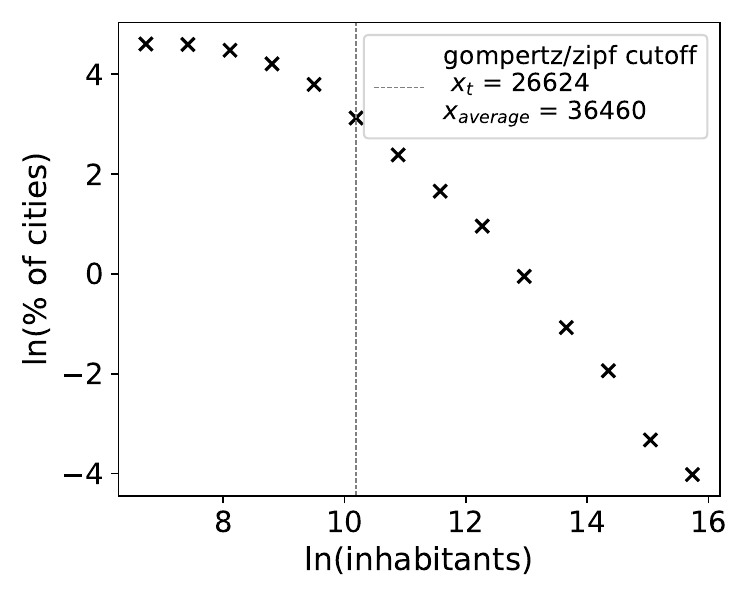}
        \caption[Log-log plot: 2022]{Plot of $\ln$(percentage of cities) versus $\ln$(number of inhabitants), both axes using transformed variables with a linear scale. A linear fit from $x_t$ onward characterizes the Zipf segment. The value $x_{average}$ represents the average city population, obtained by dividing the total population by the number of cities.}
        \label{fig:loglog2022}
    \end{minipage}    
    \hfill
    \begin{minipage}{0.48\textwidth}
        \centering
        \includegraphics[width=\linewidth]{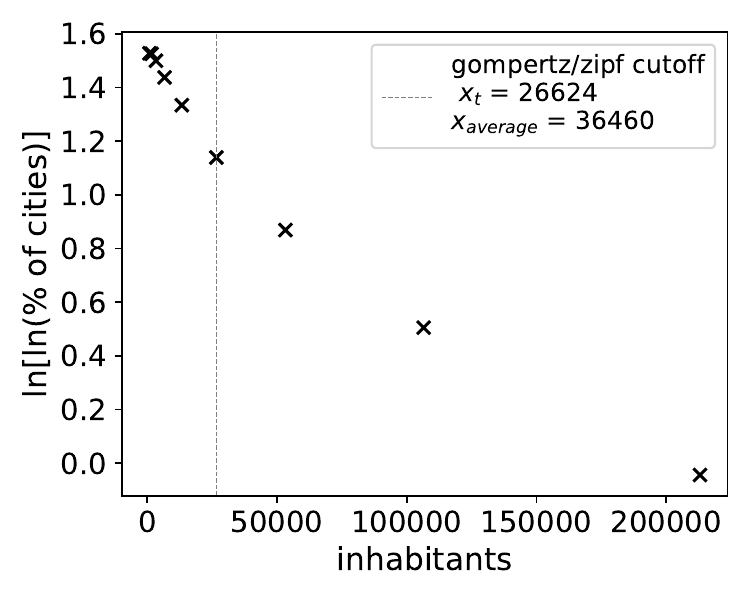}
        \caption[Double log plot: 2022]{Plot with a linear scale on the $x$-axis and a double natural logarithm transformation $\ln$[$\ln$(percentage of cities)] on the $y$-axis. The y-intercept boundary condition reveals the Gompertz segment to extend up to $x_t$. The value $x_{average}$ represents the average city population, obtained by dividing the total population by the number of cities.}
        \label{fig:doublelog2022}
    \end{minipage}
\end{figure}

In a log-log plot of city size distribution, where the axes represent log(rank) versus log(size), a straight line suggests that the distribution follows a power law, which is closely related to the Zipf law \cite{arshad2018}.
However, Fig. \ref{fig:loglog2022}, a plot of $\ln$(percentage of cities) versus $\ln$(inhabitants number), shows that the initial part of the distribution is not exponential, since an exponential function would appear as a straight line on a $\ln$--$\ln$ scale. 
Therefore, a double-exponential form was tested to fit the data, following the same approach used in Gompertz income analysis \cite{chami2011}. 
The Gompertz function captures a higher degree of inequality compared to a simple exponential function.

In order to determine the lower limit of the Zipf law, that is, the point from which a line can be fitted using the Least Squares Method, a separate evaluation was made using a plot of $\ln $[$\ln$(percentage of cities)] versus inhabitants number, as shown in Fig. \ref{fig:doublelog2022}, where Python is used to determine the range over which the data produces the best linear fit, quantified by the maximum coefficient of determination $R^2$, subject to the boundary condition that the y-intercept equals $\ln$[$\ln$(100)] = 1.5272 since the y-axis assumes values strictly above zero up to 100. 
The Least Squares Method minimizes the sum of squared residuals between the observed data and the fitted model. 
This method is used in the analysis of power-law and exponential distributions due to its simplicity \cite{newman2005}. 
The quality of the fit is evaluated through the coefficient of determination, which measures the proportion of variance of the model. 
Values of $R^2$ closer to 1 indicate a better agreement between the model and the empirical data.

The results indicate that the best fit separates the data at bin number 6, corresponding to the cutoff $x_{t} = 26,624$ people and $R^2 = 0.99846$. 
That means that every city with at least 26,624 inhabitants will be fitted into the Zipf law in the second segment, and every city with less than that will be fitted into a Gompertz distribution in the first segment as shown in Figs. \ref{fig:loglog2022} and \ref{fig:doublelog2022}.
The value $x_{average} = 36{,}460$ displayed in the figures corresponds to the average city population, obtained by dividing the total population by the number of cities, and differs from the cutoff $x_t = 26{,}624$, which marks the transition between the Gompertz and Zipf segments.

\subsection{Logarithmic binning}

\begin{figure}[H]
    \centering
    \includegraphics[width=0.7\linewidth]{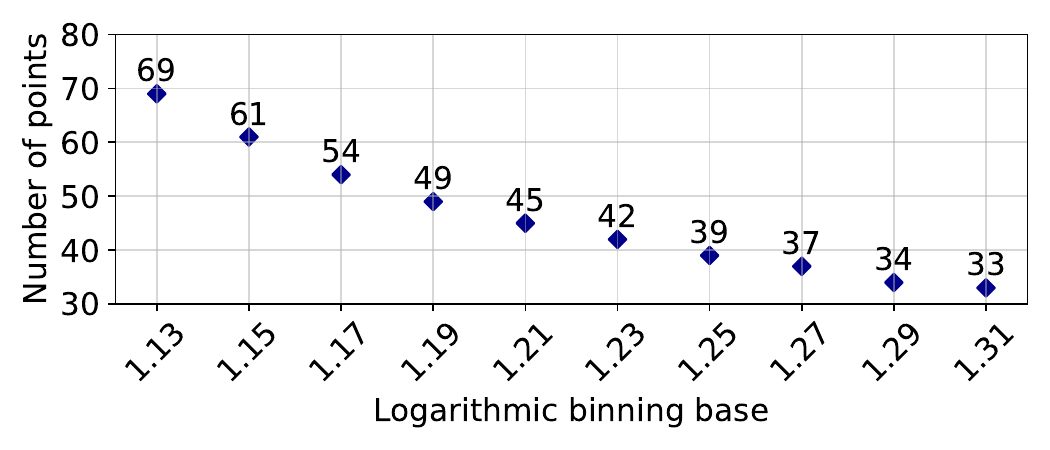}
    \caption[Number of points: 2022]{Number of points generated for the 10 logarithmic base options using the 2022 census data.}
    \label{fig:numberofpoints2022}
\end{figure}

As seen in Section 2.1, Eq. (\ref{eq:bin2}) groups the 2022 census data into 14 bins, which may be viewed as a small number providing few data points. When the base in $x_i = \mathrm{base}^{i-1} \cdot x_{min}$ is too close to 1, the resulting lacunae regions add no new points to the CCDF, causing adjacent points to become indistinguishable. 
Conversely, a larger base produces fewer points. In this context, the objective is to identify a binning configuration that provides a sufficient number of data points for a reliable determination of the cutoff $x_t$ between the Gompertz and Zipf segments, and to indicate whether or not the cutoff value is actually independent of binning. 
Therefore, the base was varied within the interval $1.13-1.31$, from which 10 equally spaced candidate values were selected.
These 10 alternative bases were then tested to generate more CCDF points, according to the following set of equations:
\be 
x_i = 1.13^{i-1} \cdot x_{min} \: , 
\ee
\be 
x_i = 1.15^{i-1} \cdot x_{min} \: , 
\label{eq:bins115}
\ee
\be 
x_i = 1.17^{i-1} \cdot x_{min} \: , 
\ee
and so forth, resulting in the amount of data points shown in Fig. \ref{fig:numberofpoints2022}. 

\begin{figure}[H]
    \centering
    \includegraphics[width=0.85\linewidth]{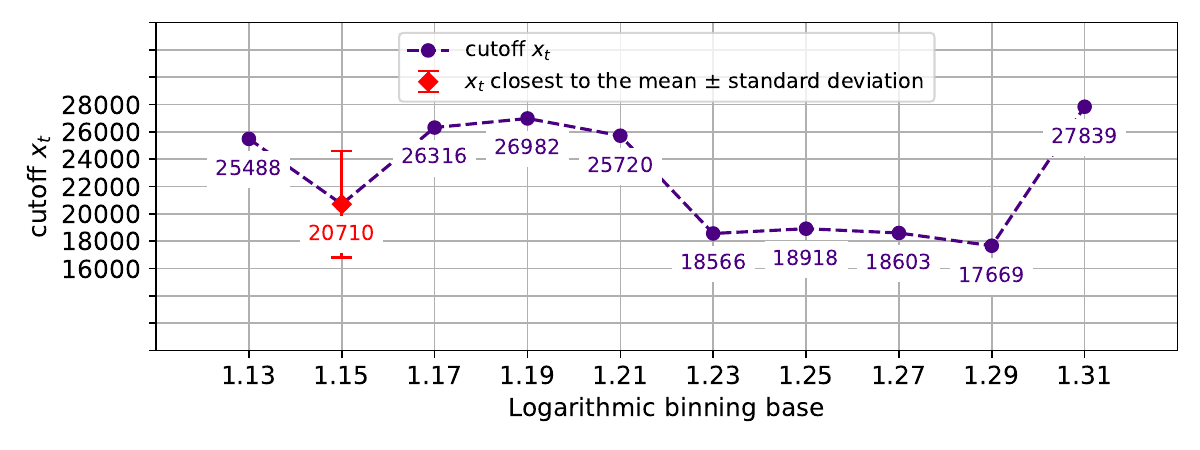}
    \caption[Cutoff $x_t$: 2022]{Cutoff $x_t$ separates the Gompertz segment from the Zipf segment and the best $x_t$ is defined as the one closest to the mean.}
    \label{fig:cut2022bases}
\end{figure}

\begin{figure}[H]
    \centering
    \begin{minipage}{0.48\textwidth}
        \centering
        \includegraphics[width=\linewidth]{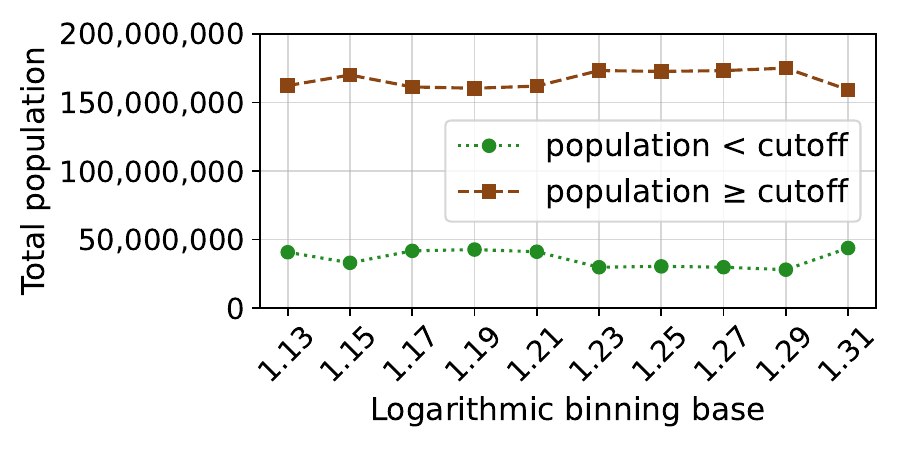}
        \caption[Population in the Gompertz and in the Zipf segments: 2022]{Population below and above the cutoff $x_t$, corresponding to the Gompertz segment and the Zipf segment, respectively, for each option of logarithmic binning base: 2022 census data.}
        \label{fig:pop2022bases}
    \end{minipage}%
    \hfill
    \begin{minipage}{0.48\textwidth}
        \centering
        \includegraphics[width=\linewidth]{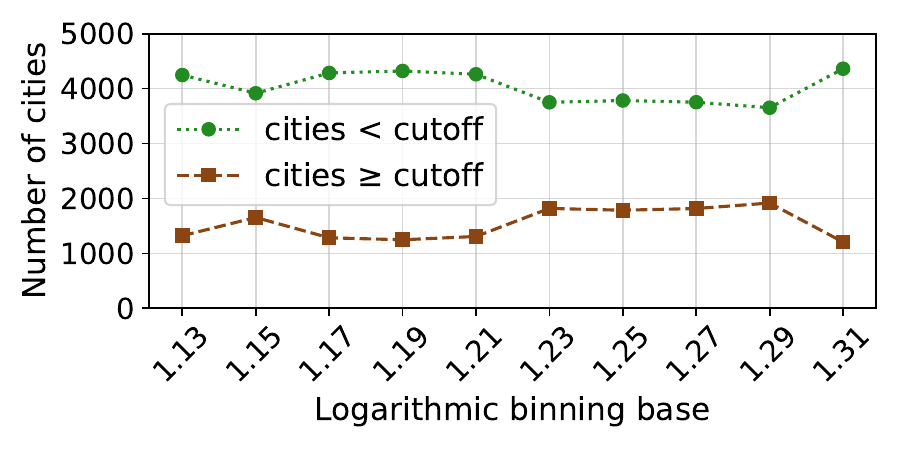}
        \caption[Number of cities in the Gompertz and in the Zipf segments: 2022]{Number of cities below and above the cutoff $x_t$, corresponding to the Gompertz segment and the Zipf segment, respectively, for each option of logarithmic binning base: 2022 census data.}
        \label{fig:cities2022bases}
    \end{minipage}
\end{figure}

When the 2022 census data is grouped using a different logarithmic binning base, the cutoff $x_t$ that separates the Gompertz segment and the Zipf segment changes, as displayed in Fig. \ref{fig:cut2022bases}. 
The base option closest to the mean in Fig. \ref{fig:cut2022bases} along with the number of points in Fig. \ref{fig:numberofpoints2022} justifies the use of the expression $x_i = 1.15^{i-1} \cdot x_{min}$ for all other censuses.
The population and the number of cities below the cutoff $x_t$ for each option of base correspond to the Gompertz segment, whereas the population and the number of cities above the cutoff $x_t$ are related to the Zipf segment, as can be seen in Figs. \ref{fig:pop2022bases} and \ref{fig:cities2022bases}.


\section{150 years of city population}

\begin{figure}[H]
    \centering
    \begin{minipage}{0.48\textwidth}
        \centering
        \includegraphics[width=\linewidth]{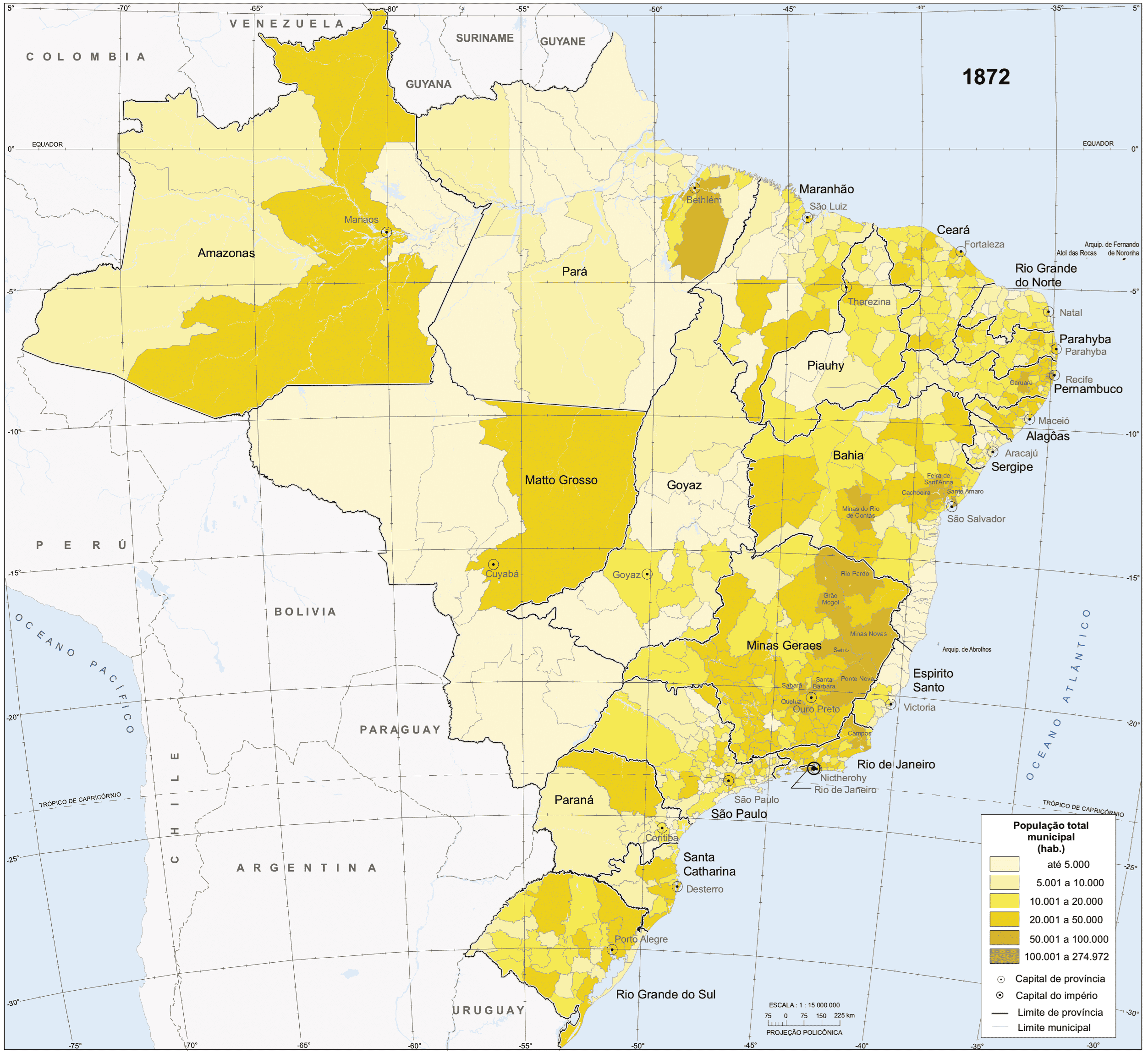}
        \caption[Brazilian city size map: 1872]{Brazilian city size distribution map in 1872. The darker areas on the map represent a city population range of 100,001 to 274,972 people and the lighter areas symbolize a city population up to 5,000 people. Figure taken from Ref. \cite[p.~33]{IBGEbook}.}
        \label{fig:map1872}
    \end{minipage}%
    \hfill
    \begin{minipage}{0.48\textwidth}
        \centering
        \includegraphics[width=\linewidth]{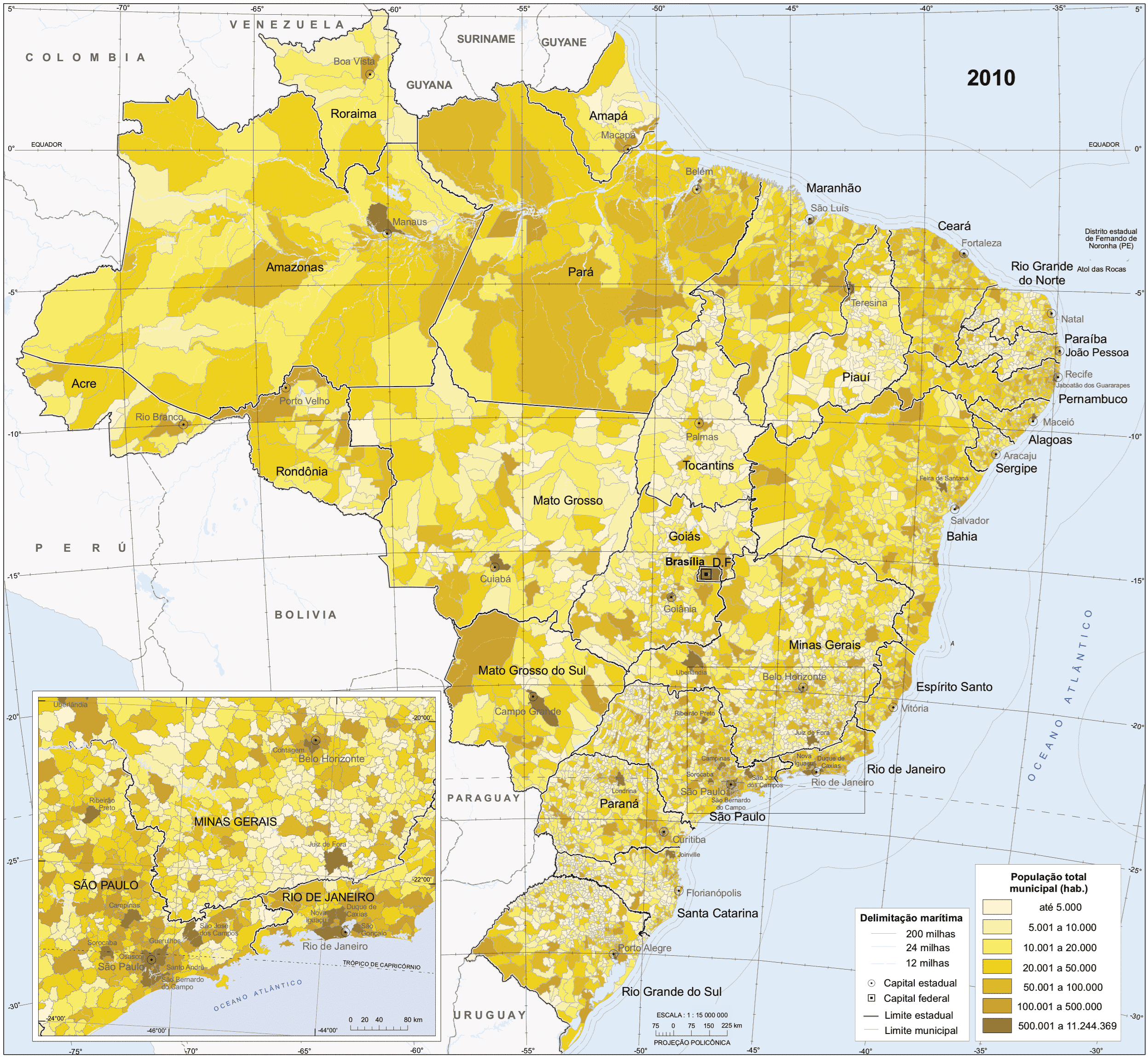}
        \caption[Brazilian city size map: 2010]{Brazilian city size distribution map in 2010. The darker areas on the map represent a city population range of 500,001 to 11,244,369 people and the lighter areas symbolize a city population up to 5,000 people. Figure taken from Ref. \cite[p.~43]{IBGEbook}.}
        \label{fig:map2010}
    \end{minipage}
\end{figure}

The maps in Figs. \ref{fig:map1872} and \ref{fig:map2010} depict the difference in Brazil's total city population in 1872 and 2010, as documented by the IBGE. These maps are part of a series demonstrating populational changes over time.
A comparison between the two maps shows an increase in the number of cities with larger populations, as well as change in the overall area of the country with the acquisition of new territory in the west that formerly belonged to Bolivia. In the earlier period, population was concentrated in fewer locations, while in the later period the distribution becomes more dispersed, although still uneven. This visual evidence suggests an expansion in the number of cities and a change in the distribution profile, which motivates a quantitative analysis of how population is distributed among cities in different censuses.

\begin{figure}[H]
    \centering
    \includegraphics[width=1\linewidth]{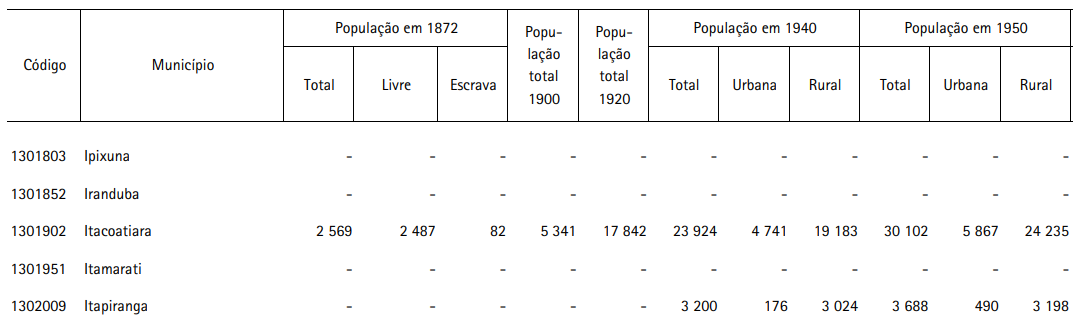}
    \caption[IBGE book: 1872-2010]{Extract from the IBGE book \cite{IBGEbook} showing the data before digitalization.}
    \label{fig:ibgebook}
\end{figure}

\begin{figure}[H]
    \centering
    \includegraphics[width=1\linewidth]{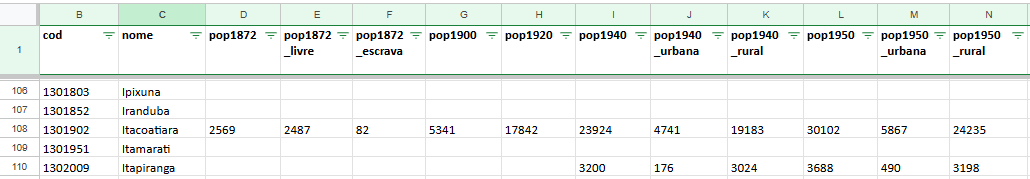}
    \caption[Digitalized IBGE data: 1872-2010]{Extract from the digitalized IBGE data part of the book in Ref. \cite{IBGEbook}.}
    \label{fig:planilhaibge}
\end{figure}

The 2022 IBGE census can be downloaded from Ref. \cite{IBGE2022census} as tabulated data, however, the 1872, 1900, 1920, 1940, 1950, 1960, 1970, 1980, 1991, 2000 and 2010 data in Ref. \cite{IBGEbook} needed a digital treatment before it could be analyzed with the Python computing language, since the data of these years are given in PDF files.
Figure \ref{fig:ibgebook} illustrates an excerpt from the originally printed IBGE publication \cite{IBGEbook}. 
Since city population data for the first censuses were only available in book format, the numerical values were extracted using a PDF digitization process with the online tool \textit{Tabula} available at \href{https://tabula.technology}{tabula.technology}, in which the tables were converted into a structured spreadsheet format, as shown in Fig. \ref{fig:planilhaibge}. 
This procedure allowed the historical population data to be systematically organized and used in the statistical analyses presented in this work.

\begin{figure}[H]
    \centering
    \begin{minipage}{0.48\textwidth}
        \centering
        \includegraphics[width=\linewidth]{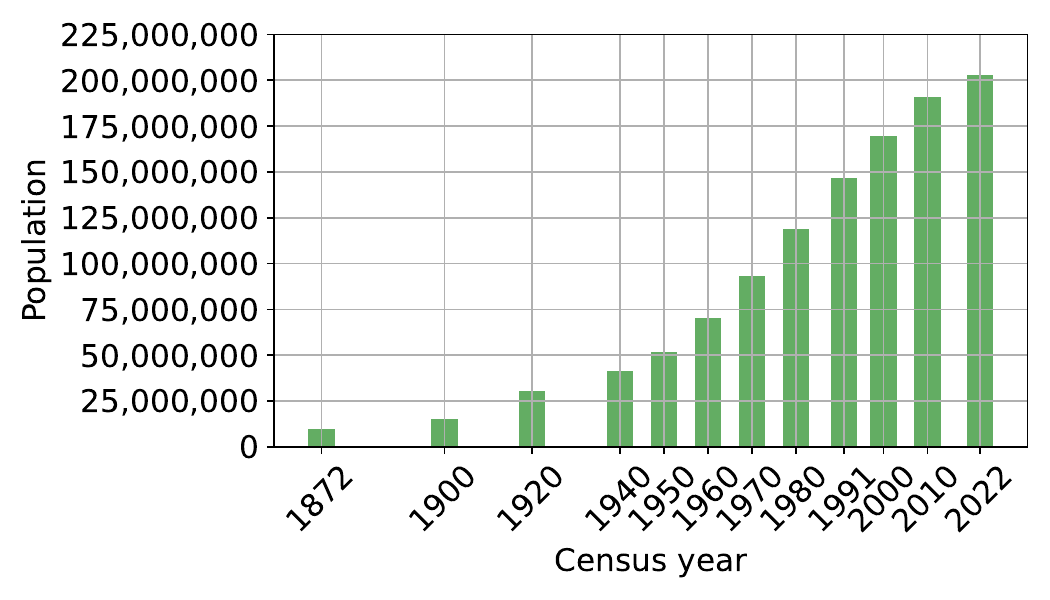}
        \caption[Population from 1872 to 2022]{Brazilian population in every census for the span of 150 years.}
        \label{fig:bar_pop1872to2022}
    \end{minipage}%
    \hfill
    \begin{minipage}{0.48\textwidth}
        \centering
        \includegraphics[width=\linewidth]{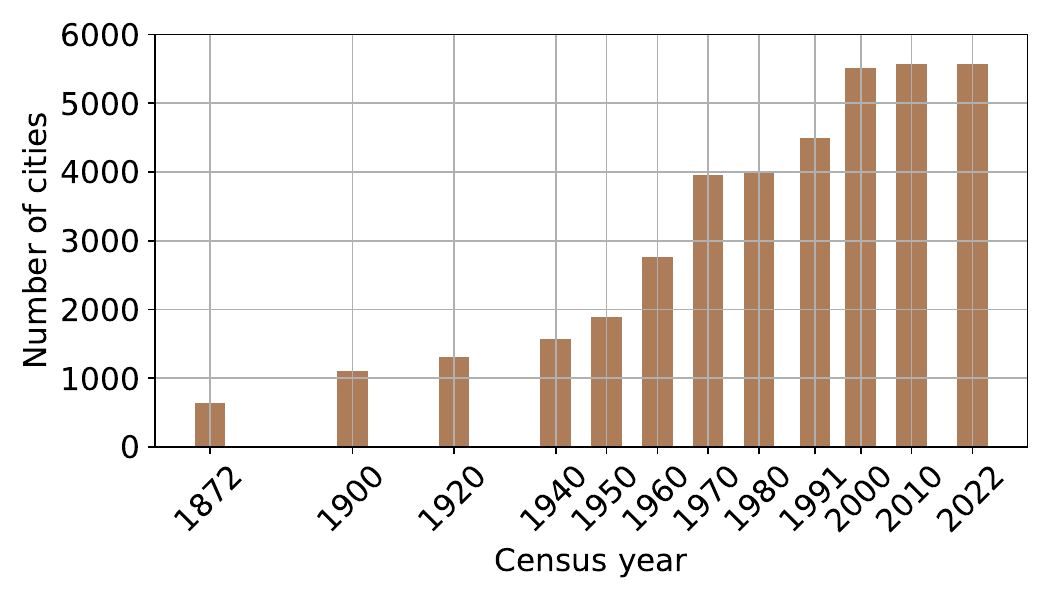}
        \caption[Number of cities from 1872 to 2022]{Number of cities in Brazil for every census within a 150 years span.}
        \label{fig:bar_cities1872to2022}
    \end{minipage}
\end{figure}

\begin{figure}[H]
    \centering
    \includegraphics[width=0.9\linewidth]{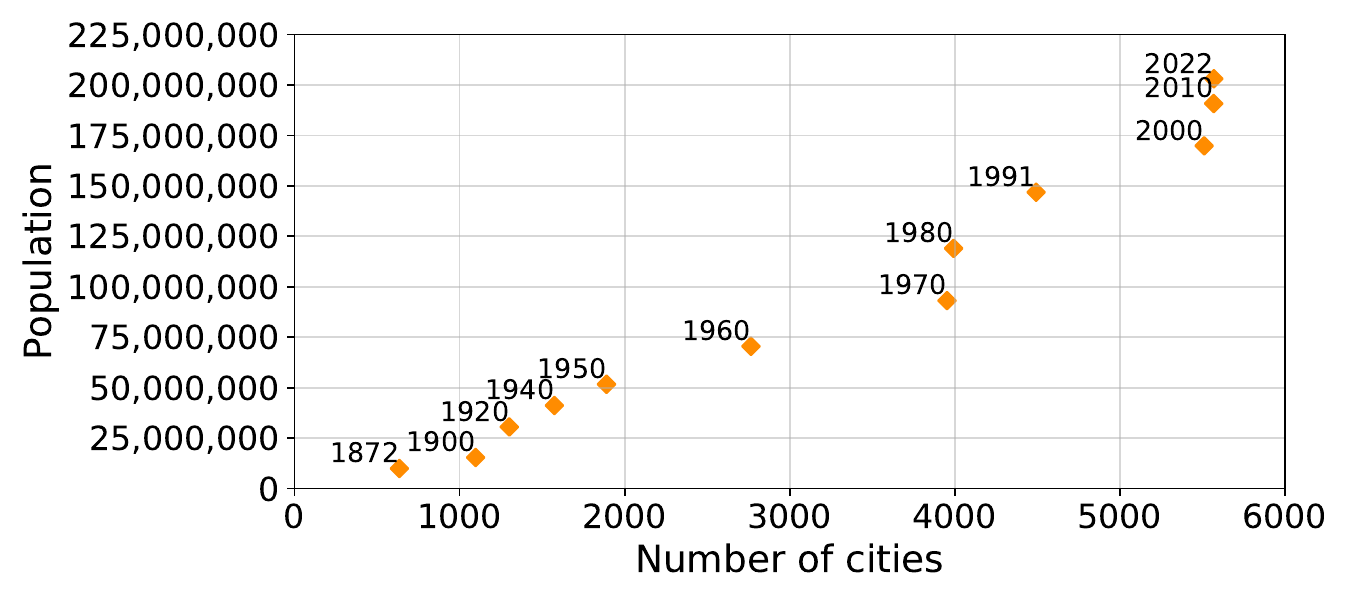}
    \caption[Population versus number of cities: 1872-2022]{Evolution of population versus number of cities in Brazil from 1872 to 2022. Numbers above each bullet refer to the census year.}
    \label{fig:citiespop1872to2022}
\end{figure}

The same data analysis presented in Section 2.1 was executed for the other Brazilian censuses. Figs. \ref{fig:bar_pop1872to2022} and \ref{fig:bar_cities1872to2022} show the population and the number of cities in each census, respectively. 
Fig. \ref{fig:citiespop1872to2022} displays the population on the vertical axis versus the number of cities on the horizontal axis for each census.
The growth rates of the population and of the number of cities do not increase at the same pace. 
There was a burst of growth in city numbers from 1950 to 2000, and another burst in population growth from 2000 to 2022 while the number of cities changed only slightly, as shown in Fig. \ref{fig:citiespop1872to2022}.

In addition to the synthetic inequality indices discussed in Section 1.3, an alternative way to describe concentration can be obtained through the population shares accumulated by the upper percentiles of the city hierarchy, following the approach popularized by Thomas Piketty in the analysis of income and wealth inequality \cite{piketty2014}. 
This approach examines the top decile (10\%), top centile (1\%) and top thousandth centile (0.1\%) separately in order to characterize the concentration structure of a society \cite{ribeiro2020}. 
As discussed by Ribeiro in Ref. \cite{ribeiro2020}, Piketty preferred the use of distribution tables and percentile groups instead of relying exclusively on synthetic indices such as the Gini coefficient, because percentile shares provide a more intuitive description of how strongly income and wealth are concentrated at the top of the hierarchy.

\begin{figure}[H]
    \centering
    \includegraphics[width=1\linewidth]{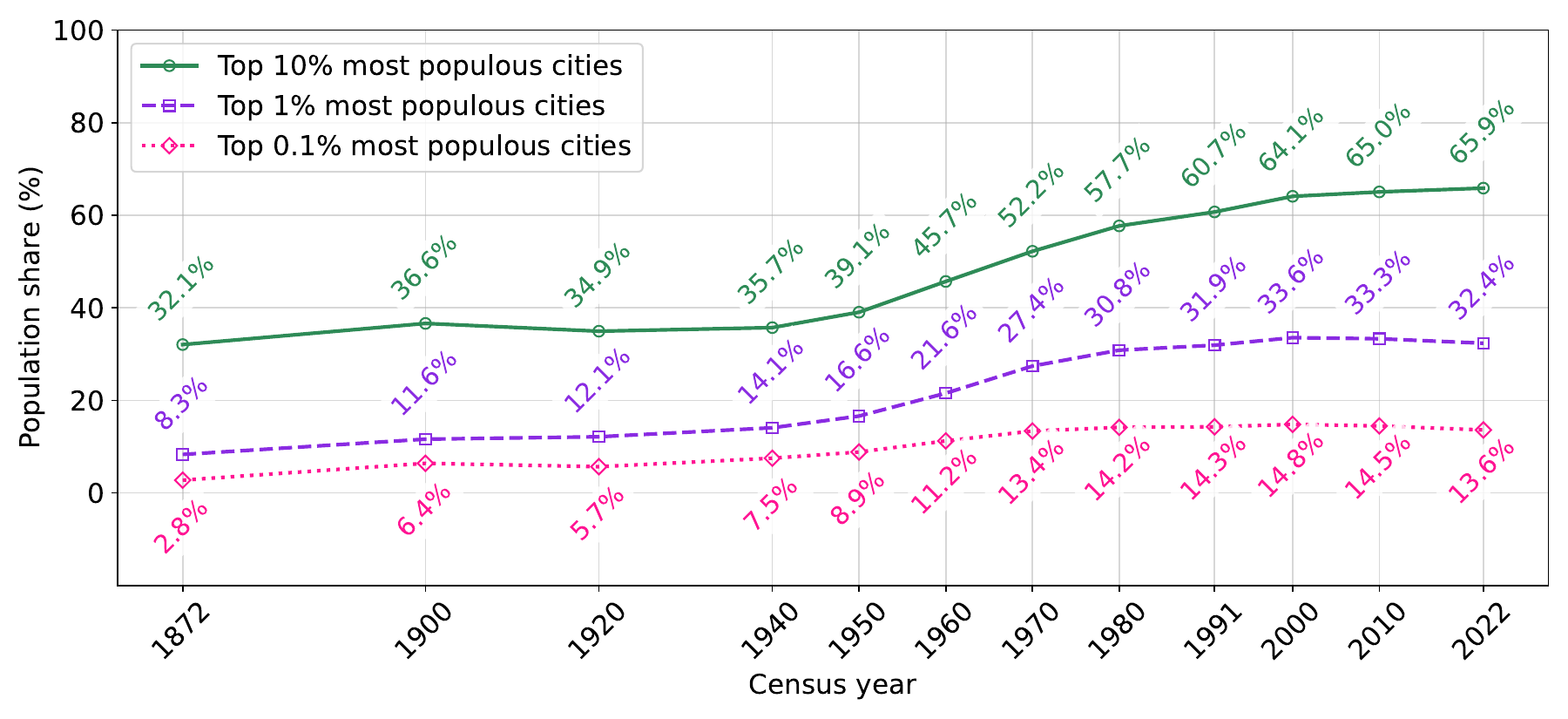}
    \caption[Population concentration in the top cities: 1872-2022]{Share of the Brazilian population concentrated in the top 10\%, top 1\% and top 0.1\% most populous cities for each census between 1872 and 2022. The curves are analogous to the percentile concentration measures used by Piketty in the study of income and wealth inequality, but applied here to the distribution of population among cities.}
    \label{fig:topshares1872to2022}
\end{figure}

In the present work, the same idea is adapted from the context of income distributions to the distribution of population among cities. 
Instead of measuring the share of total income appropriated by the richest individuals, the analysis measures the share of the Brazilian population concentrated in the top 10\%, top 1\% and top 0.1\% most populous cities for each census. 
These quantities are obtained by ordering cities according to their population sizes and computing the cumulative population shares associated with the top percentiles of the city hierarchy, which are equivalent to the information represented by the Lorenz curve. 
They provide an alternative visualization of population concentration over time, as shown in Fig. \ref{fig:topshares1872to2022} and Table \ref{tab:topshares1872to2022}. 
The results indicate a strong historical increase in concentration within the Brazilian city system. 
The share of the total population living in the top 10\% most populous cities increased from 32.06\% in 1872 to 65.85\% in 2022, while the share concentrated in the top 1\% increased from 8.32\% to 32.35\%. 
Similarly, the top 0.1\% most populous cities increased their population share from 2.77\% to 13.61\%. 
The strongest acceleration occurs mainly after 1950, reflecting the expansion of large cities during the process of urbanization and metropolitan growth in Brazil. 
This procedure allows the long-term evolution of concentration in the Brazilian city system to be examined in a manner analogous to the historical analyses of inequality performed by Piketty for income and wealth distributions.

\begin{table}[H]
\centering
\begin{tabular}{c c c c}
\hline
\noalign{\vskip 3pt}
Year & Top 10\% & Top 1\% & Top 0.1\% \\
     & population share (\%) & population share (\%) & population share (\%) \\
\noalign{\vskip 3pt}
\hline
\noalign{\vskip 3pt}
1872 & 32.06 & 8.32  & 2.77  \\
1900 & 36.62 & 11.62 & 6.41  \\
1920 & 34.95 & 12.14 & 5.68  \\
1940 & 35.72 & 14.09 & 7.51  \\
1950 & 39.05 & 16.61 & 8.85  \\
1960 & 45.69 & 21.55 & 11.24 \\
1970 & 52.21 & 27.42 & 13.39 \\
1980 & 57.69 & 30.85 & 14.17 \\
1991 & 60.71 & 31.93 & 14.29 \\
2000 & 64.09 & 33.55 & 14.82 \\
2010 & 65.04 & 33.31 & 14.48 \\
2022 & 65.85 & 32.35 & 13.61 \\
\hline
\end{tabular}
\caption[Population concentration in the top cities: 1872-2022]{Share of the Brazilian population concentrated in the top 10\%, top 1\% and top 0.1\% most populous cities for each census between 1872 and 2022.}
\label{tab:topshares1872to2022}
\end{table}

\begin{figure}[H]
    \centering
    \includegraphics[width=0.8\linewidth]{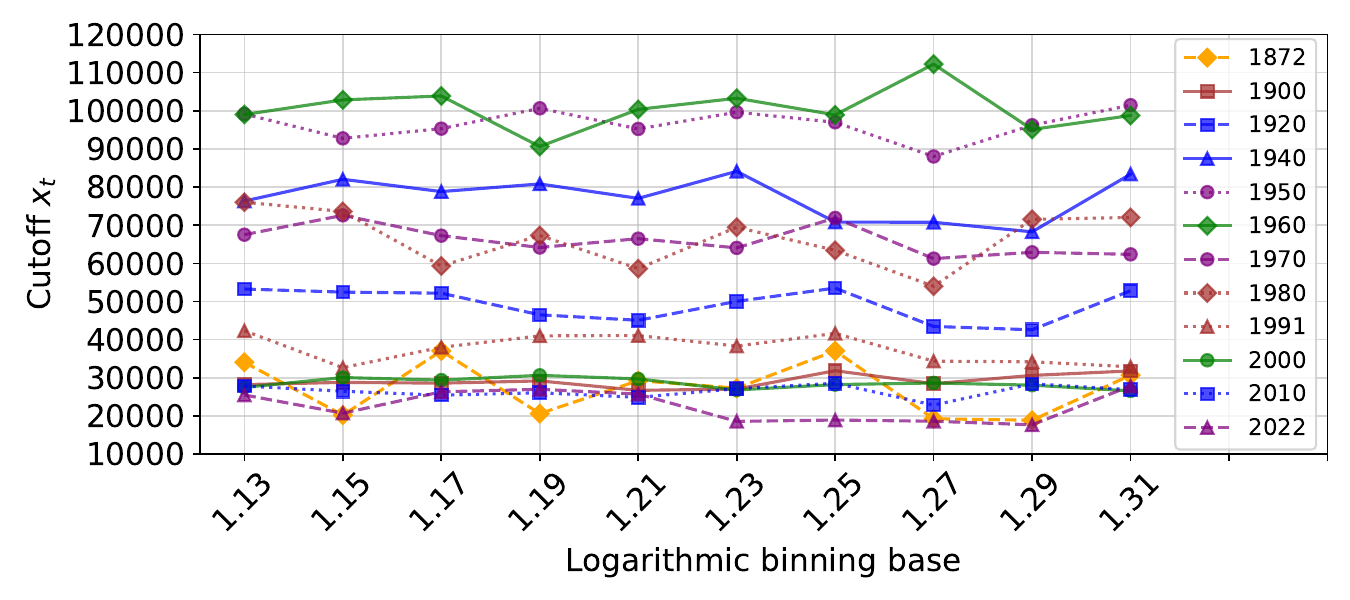}
    \caption[Cutoff $x_t$ for 10 options of logarithmic binning base: 1872-2022]{Cutoff $x_t$ for the 12 censuses studied with 10 options of logarithmic binning base, showing consistent results across binning choices, suggesting that the estimates are robust to the binning selection.}
    \label{fig:cut_xt_bins_1872to2022}
\end{figure}

\begin{figure}[H]
    \centering
    \includegraphics[width=0.8\linewidth]{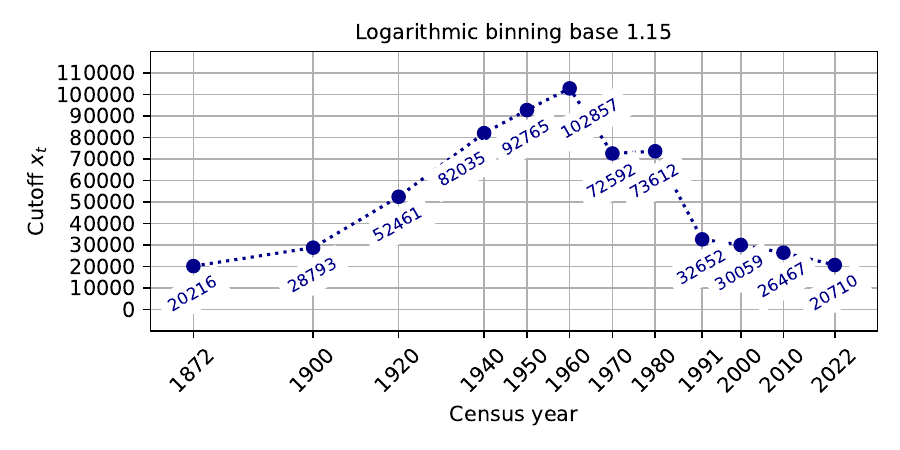}
    \caption[Cutoff $x_t$: 1872-2022]{Cutoff $x_t$ shows the minimum population of a city in the Zipf segment for the 12 censuses studied with logarithmic binning base 1.15.}
    \label{fig:cut1872to2022}
\end{figure}

Fig. \ref{fig:cut_xt_bins_1872to2022} reveals that each census has its own range of cutoff $x_t$ from which the Zipf power law begins to apply.
For the 2022 data, the chosen expression for logarithmic binning is Eq. (\ref{eq:bins115}) because it generates a significant number of points and has cutoff $x_t$ closest to the mean.
Therefore, to maintain consistency, each of the 12 censuses have their data organized according to Eq. (\ref{eq:bins115}) as well. The individual cutoffs $x_t$ are shown in Fig. \ref{fig:cut1872to2022}, with $R^2$ values ranging from 0.980 to 0.999 across all censuses, indicating an excellent fit for the selected thresholds in all periods. 
The censuses remain internally consistent with respect to binning, as can be seen in Fig. \ref{fig:cut_xt_bins_1872to2022} where each census exhibits a well defined range of variation for the cutoff $x_t$. 
These variations may reflect changes over time in the way the IBGE conducts its data collection.

\begin{figure}[H]
    \centering
    \includegraphics[width=0.65\linewidth]{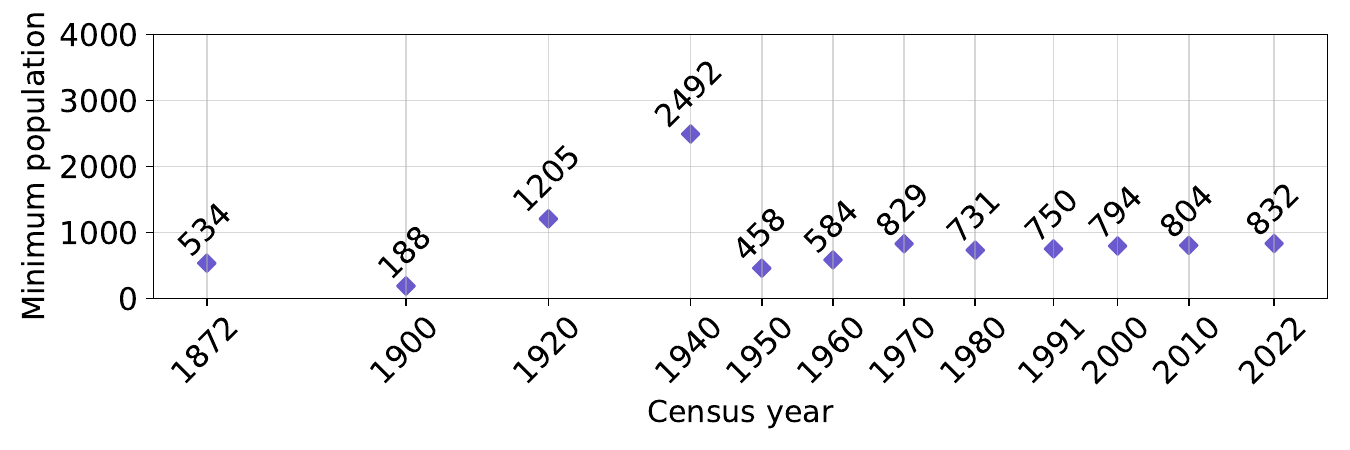}
    \caption[Minimum population versus number of cities: 1872-2022]{Evolution of the minimum population versus the number of cities registered in Brazil from 1872 to 2022.}
    \label{fig:min1872to2022}
\end{figure}

\begin{figure}[H]
    \centering
    \begin{minipage}{0.48\textwidth}
        \centering
        \includegraphics[width=\linewidth]{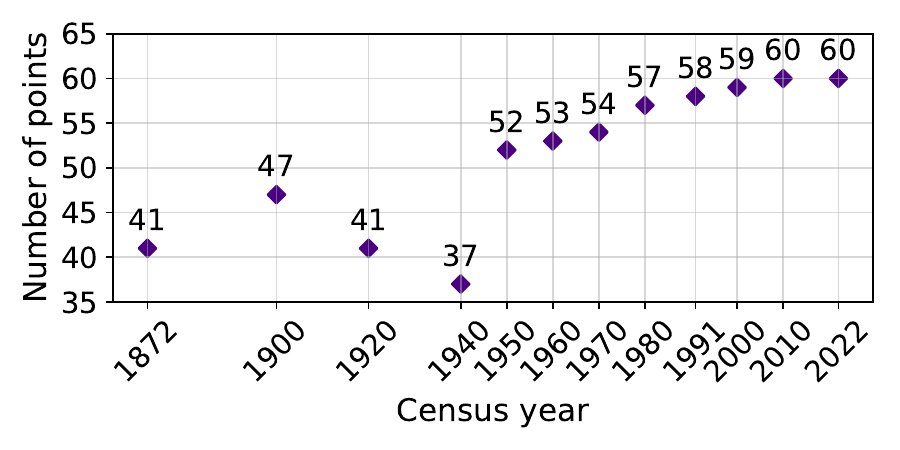}
        \caption[Number of points: 1872-2022]{Number of points when using the logarithmic binning base 1.15 to organize the data of each of the 12 censuses.}
        \label{fig:numberofpoints1872to2022}
    \end{minipage}%
    \hfill
    \begin{minipage}{0.48\textwidth}
        \centering
        \includegraphics[width=\linewidth]{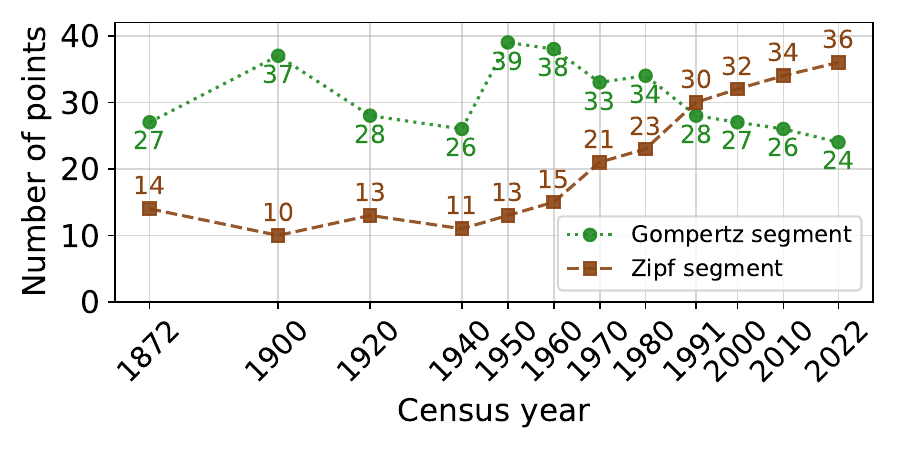}
        \caption[Number of points per segment: 1872-2022]{Number of points in the Gompertz and the Zipf segments when using the logarithmic binning base 1.15 to organize the data of each of the 12 censuses.}
        \label{fig:segment_points_1872to2022}
    \end{minipage}
\end{figure}

\begin{figure}[H]
    \centering
    \begin{minipage}{0.48\textwidth}
        \centering
        \includegraphics[width=\linewidth]{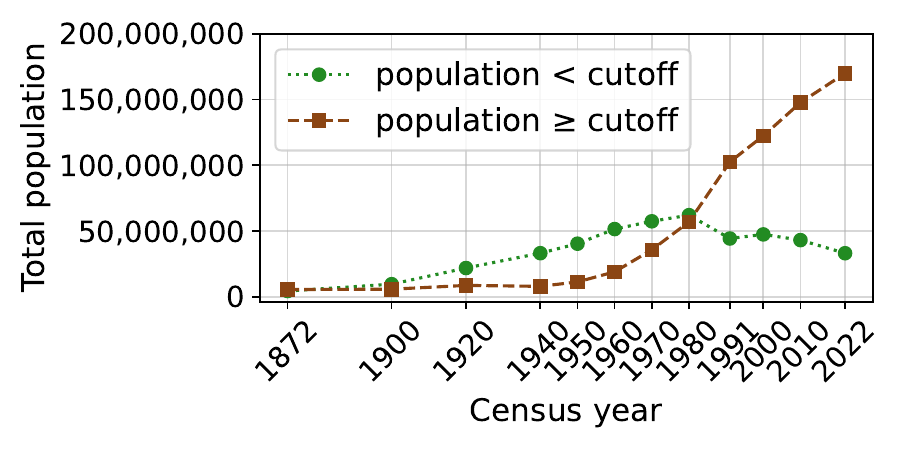}
        \caption[Population per segment: 1872-2022]{Total population in the Gompertz segment ($<$ cutoff) and total population in the Zipf segment ($\geq$ cutoff) for each census.}
        \label{fig:pop1872to2022}
    \end{minipage}%
    \hfill
    \begin{minipage}{0.48\textwidth}
        \centering
        \includegraphics[width=\linewidth]{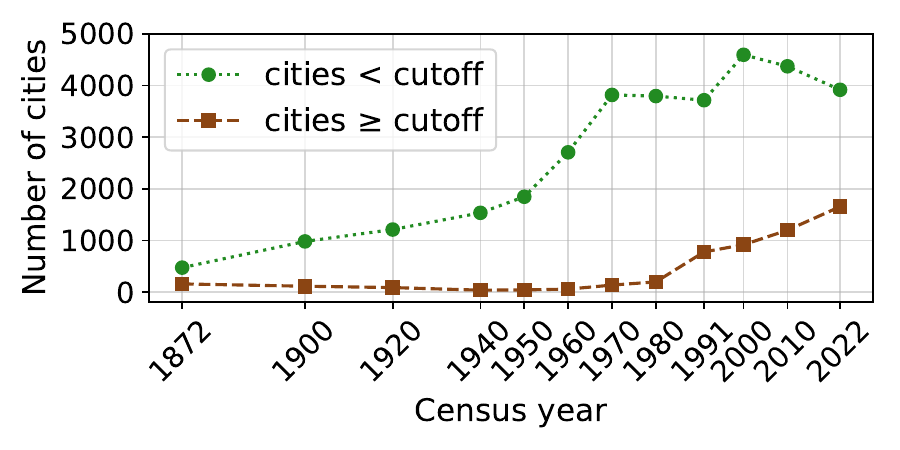}
        \caption[Number of cities per segment: 1872-2022]{Number of cities in the Gompertz segment ($<$ cutoff) and number of cities in the Zipf segment ($\geq$ cutoff) for each census.}
        \label{fig:cities1872to2022}
    \end{minipage}
\end{figure}

As displayed in Fig. \ref{fig:min1872to2022}, the recorded minimum population differs across censuses, with most censuses showing consistent minimum values, except for 1920 and 1940, which present discrepancies.
Figs. \ref{fig:cut1872to2022}, \ref{fig:numberofpoints1872to2022}, \ref{fig:segment_points_1872to2022}, \ref{fig:pop1872to2022} and \ref{fig:cities1872to2022} are related to the logarithmic binning base 1.15 and show, for each census, respectively, the cutoff $x_t$, the total number of points, the number of points per segment, the population and the number of cities in each segment Gompertz-Zipf.
The numbers of points in Fig. \ref{fig:numberofpoints1872to2022} range from 37 to 60, and the numbers of points per segment in Fig. \ref{fig:segment_points_1872to2022} range from 10 to 39, which are enough to perform a linear regression fit.
Furthermore, Fig. \ref{fig:pop1872to2022} exhibits a large population growth in the Zipf segment and a more moderate population growth in the Gompertz segment. In contrast, Fig. \ref{fig:cities1872to2022} shows that the number of cities in the Gompertz segment increases more than the number of cities in the Zipf segment.

\section{Urban area from the 2022 census}

The data used to study urban areas were collected from 01 January 2019 to 20 July 2022 and can be found in Ref. \cite{IBGE2022areas}. Urbanized areas, as defined by the IBGE, were mapped using Sentinel-2 satellite images, classifying them as dense or low-density with subcategories like urban facilities and intra-urban voids. 
These urbanized areas include neighborhoods, streets, residential complexes, urban parks and vacant lots, while excluding rural zones, forests, rivers and agricultural areas \cite{IBGE2022urbanized, IBGE2022metadata}.

Following Ref. \cite{chen2012}, Table \ref{tab:areas2022} illustrates the hierarchical structure and scaling relationships within urban systems, where the class $m$ represents the hierarchical level of cities, $N_m$ indicates the number of cities belonging to the class $m$, $P_m$ shows the average population size of cities in the class $m$, $A_m$ displays the average urban area of cities in the class $m$, and $\bar{A}_m$ represents the average total area of cities in the class $m$.
The cities were grouped into classes following a top-down cascade structure, in which the largest city in terms of population, São Paulo, composes the first class $m=1$, the next two largest cities, Rio de Janeiro and Brasília, form the second class $m=2$, the following four cities, Fortaleza, Salvador, Belo Horizonte and Manaus, form the third class $m=3$, and so on, in a non-cumulative manner, with the number of cities doubling at each level $N_m = 2^{m-1}$. 
The last class contains the remaining cities when the total number of cities is not an exact power of 2, as shown in the final row of the second column of Table \ref{tab:areas2022}.
For each class, the average population $P_m$, the average urban area $A_m$, and the average total area $\bar{A}_m$ were computed from the cities belonging to that class.
The data from Table \ref{tab:areas2022} are organized to describe the hierarchical structure of cities and the associated measures of population and urban area. 
The analysis of the scaling relations derived from these data will be presented in the next chapter.

\begin{table}[H]
\centering
\begin{tabular}{c c c c c}
\hline
\noalign{\vskip 3pt}
Class & Number of cities & Average population & Average urban area & Average total area \\
$m$ & $N_m$ & $P_m$ & $A_m$ (km$^2$) & $\bar{A}_m$ (km$^2$) \\
\noalign{\vskip 3pt}
\hline
\noalign{\vskip 3pt}
1  & 1    & $1.145200 \times 10^{7}$ & 914.564400 & 1521.202000 \\
2  & 2    & $4.514302 \times 10^{6}$ & 615.276550 & 3480.556500 \\
3  & 4    & $2.306409 \times 10^{6}$ & 250.268950 & 3184.560250 \\
4  & 8    & $1.350606 \times 10^{6}$ & 213.865850 & 579.274500 \\
5  & 16   & $7.856289 \times 10^{5}$ & 131.471506 & 1430.638438 \\
6  & 32   & $4.838237 \times 10^{5}$ & 91.272734 & 2175.307375 \\
7  & 64   & $2.989749 \times 10^{5}$ & 65.229417 & 1743.645422 \\
8  & 128  & $1.636583 \times 10^{5}$ & 40.142064 & 2828.483086 \\
9  & 256  & $8.696227 \times 10^{4}$ & 23.837535 & 3316.254426 \\
10 & 512  & $4.428397 \times 10^{4}$ & 12.616479 & 3259.010869 \\
11 & 1024 & $2.306017 \times 10^{4}$ & 6.344016 & 2268.322653 \\
12 & 2048 & $1.008249 \times 10^{4}$ & 2.908502 & 1144.223381 \\
13 & 1475 & $3.550037 \times 10^{3}$ & 1.176278 & 489.241612 \\
\hline
\end{tabular}
\caption[Urban and total areas hierarchy: 2022]{Hierarchy of the 5,570 cities in Brazil organized into classes following a cascade structure $N_m = 2^{m-1}$, where $N_m$ is the number of cities, $P_m$ is the average population, $A_m$ is the average urban area, and $\bar{A}_m$ is the average total area in each class $m$ for 2022.}
\label{tab:areas2022}
\end{table}

\chapter{Modeling the Brazilian city size distribution}\label{chapter:third_chapter}
As previously discussed, this study aims to test the validity of the Gompertz and Zipf distributions in their ability to describe Brazilian cities. The code created for this analysis was developed by this author, and is available at \href{https://github.com/s-teles}{github.com/s-teles}.
The 2022 census results presented in Sec. 3.1 serve as an introductory example, while Sec. 3.1.1 introduces alternative logarithmic binning bases to generate more points for the analysis. Subsequently, Sec. 3.2 contains the final research results for the censuses from 1872 to 2022 using this chosen logarithmic binning base.

\section{Gompertz-Zipf population distribution: 2022 census}

The city size distribution plot in Fig. \ref{fig:loglog2022} shows the relationship between $\ln($percentage of cities$)$ and $\ln($inhabitants number$)$ revealing two distinct segments. For cities having fewer than $26,624$ inhabitants (first segment), the distribution can be represented by a Gompertz function, capturing the gradual decline in the proportion of smaller cities. For cities having at least $26,624$ inhabitants (second segment), the distribution can be described by the Zipf law, as indicated by the linear behavior of a power-law function.

\begin{figure}[h!]
    \centering
    \begin{minipage}{0.48\textwidth}
        \centering
        \includegraphics[width=\linewidth]{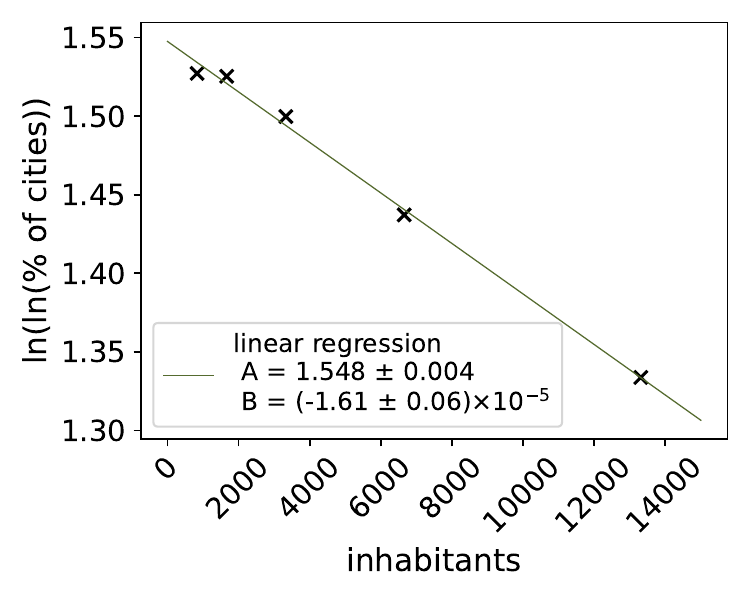}
        \caption[Gompertz linear fit: 2022]{The first segment of the 2022 census data for cities having fewer than 26,624 people can be represented by a Gompertz distribution. The linear regression is $\ln\{\ln[P(x)]\} = A+Bx$ where these five points make $A$ closest to $\ln[\ln(100)] = 1.5272$ and $B$ is the slope of the function. Since the fitted line has negative slope then $B<0$.}
        \label{fig:gompertz2022}
    \end{minipage}%
    \hfill
    \begin{minipage}{0.48\textwidth}
        \centering
        \includegraphics[width=\linewidth]{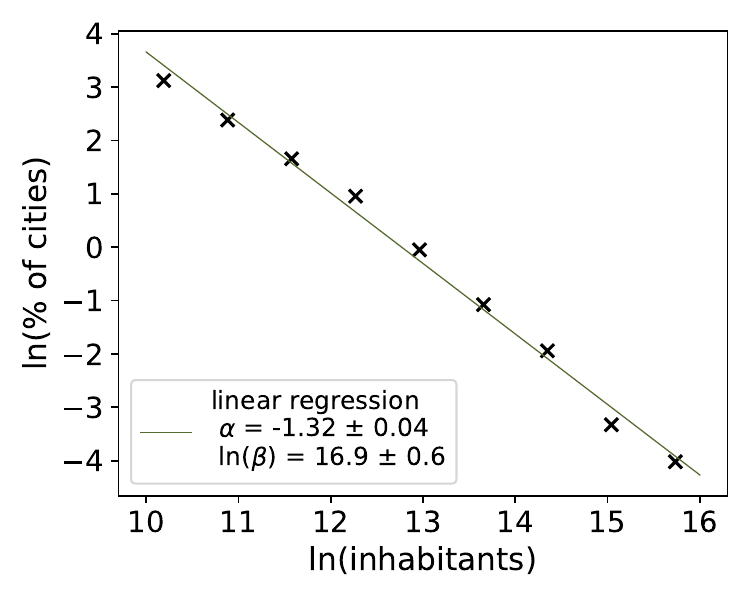}
        \caption[Zipf linear fit: 2022]{The second segment of the 2022 census data for cities having 26,624 people or more can be represented by the Zipf law. The linear regression is $\ln[F(x)] = \alpha \ln(x) + \ln(\beta)$ where $\ln(\beta)$ is the Y intercept. The estimated values are $\beta = 2.8 \pm 0.6$ and the slope $\alpha$ is $1-a$ so the Zipf exponent is $a = 2.32 \pm 0.04$.}
        \label{fig:zipf2022}
    \end{minipage}
\end{figure}

The linear regression graph of the Gompertz CCDF, plotted with the number of inhabitants on the horizontal axis and $\ln[\ln(\text{percentage of cities})]$ on the vertical axis, resulted in a straight line with a negative slope equal to the estimated parameter $B$. 
The fitted model confirms that as city size increases, the proportion of cities exceeding a given population declines at a rate governed by the magnitude of $B$. 
Since $B<0$, larger absolute values of $B$ correspond to a steeper decline, meaning fewer large cities, whereas smaller absolute values correspond to a more gradual decrease. 
The linear fit in Fig. \ref{fig:gompertz2022} supports the Gompertz distribution's applicability in modeling city size distribution, particularly for smaller cities, in this case, cities with fewer than 26,624 people.
To determine the power-law relationship in the Zipf segment of a city size distribution, linear regression using the least squares method was applied to a plot of ln(percentage of cities) versus ln(inhabitants number) \cite{arshad2018, moura2006, hayata2022}. As shown in Fig. \ref{fig:zipf2022}, a straight-line fit was performed for the second segment which encompasses cities with at least $26,624$ inhabitants and was found to follow the Zipf law. To reduce statistical fluctuations, particularly in the tail of the distribution, logarithmic binning was employed, grouping data into increasingly larger intervals as discussed in Sec. 2.1.1 above \cite{newman2005,moura2006}.

\begin{figure}[h]
    \centering
    \includegraphics[width=0.36\linewidth]{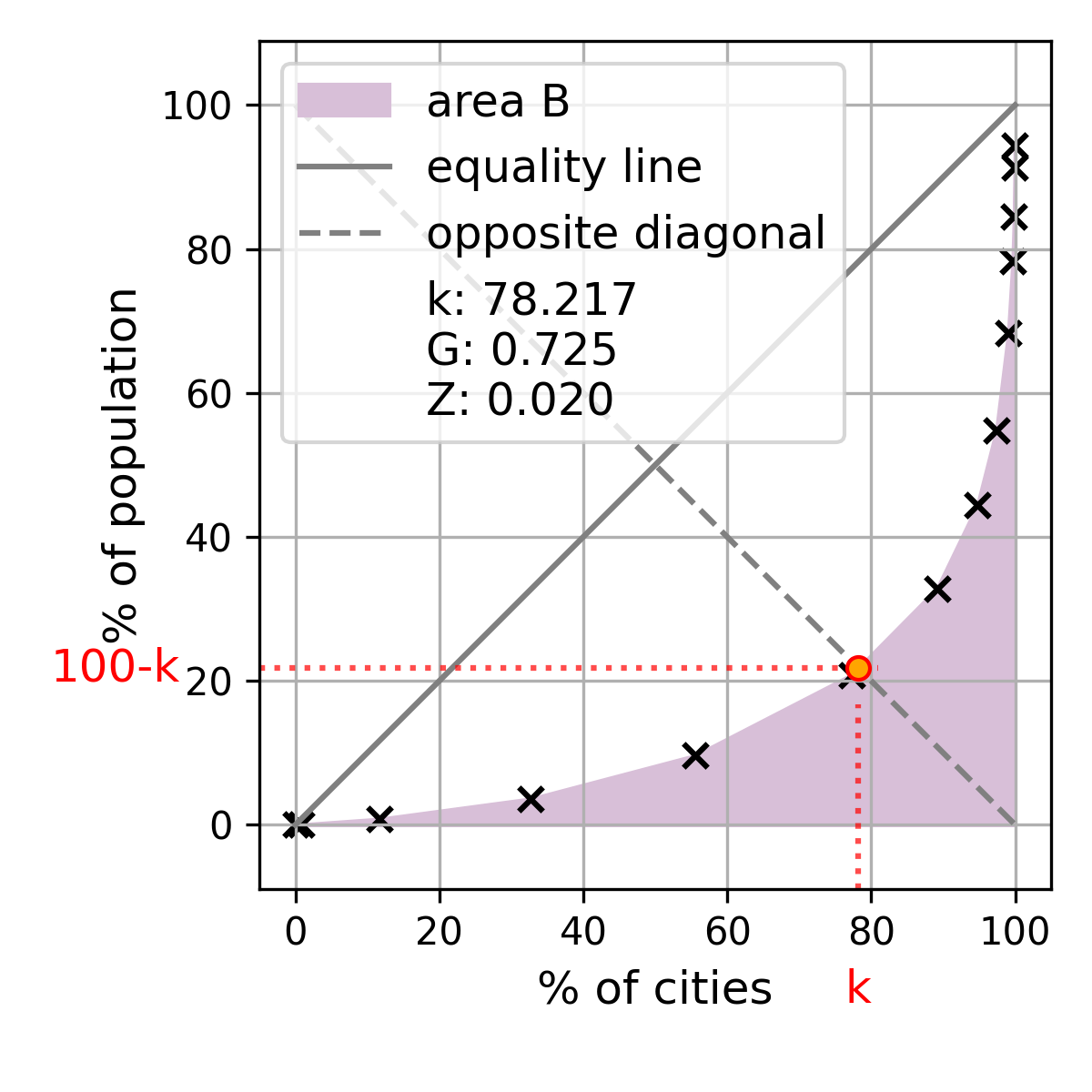}
    \caption[Lorenz curve: 2022]{Lorenz curve with Kolkata, Gini and Zanardi indices for the 2022 census dataset (see Section 1.3). The estimated parameters are related to the logarithmic binning base 2 as in Eq. (\ref{eq:bin2}). These results are presented as an example, since the standard historical comparison throughout the dissertation uses another logarithmic binning base, as will be seen next in Sec. 3.1.1.}
    \label{fig:lorenz2022}
\end{figure}

The Gini, Pietra, Kolkata and Zanardi indices were used to analyze the city population distribution based on the 2022 Brazilian census data.
Fig. \ref{fig:lorenz2022} shows that the Gini coefficient results in $G = 0.725$ and the Zanardi index is $Z = 0.020$, obtained using logarithmic binning base 2, as defined in Eq. (\ref{eq:bin2}), following the procedure adopted by Eq. (7) in Ref. \cite{moura2006}. 
These values indicate a high concentration of population in the entire distribution along with a small asymmetry between the upper and lower segments.
Since $Z > 0$, the asymmetry is oriented towards larger cities, indicating that inequality is slightly more concentrated in the upper tail.
The combination of a high Gini coefficient and a positive Zanardi index portrays a highly concentrated system in which large cities account for a substantial share of the population.
While $G=0.725$ confirms the overall high level of concentration, $Z=0.020$ reveals that the direction of this imbalance favors larger cities, providing a more complete description of the distribution.

The area under the Lorenz curve in Fig. \ref{fig:lorenz2022} was numerically calculated from a monotonic cubic spline interpolation of the empirical data using the trapezoidal rule applied to a dense grid of interpolated points.
Because the Lorenz curve is not represented by a single global polynomial, the interpolation generates a sequence of local cubic polynomials defined over consecutive intervals of the empirical data. 
In the present case, the interpolation produced thirteen cubic polynomials, corresponding to the number of intervals between the fourteen empirical points of the Lorenz curve. 
Each polynomial describes only a restricted portion of the curve and is smoothly connected to neighboring segments, ensuring continuity and preserving the monotonic behavior expected for cumulative distributions.
This piecewise construction provides a stable numerical representation of the Lorenz curve for calculating the Gini, Kolkata and Zanardi indices.


The empirical results can be compared with previous studies on city size distributions and inequality measures. 
For Brazil, Ref. \cite{moura2006} reports a Zipf exponent close to 2, whereas the estimated value in this study is $a = 2.32$, indicating a steeper decay in the tail. 
Deviations from a single power-law behavior are consistent with the findings of Ref. \cite{cristelli2012}, which show that the distribution of city sizes cannot be fully described by a pure Zipf law. 
This supports the use of a combined Gompertz–Zipf distribution, in line with Ref. \cite{chami2011}, where the majority and the tail of the distribution follow different functions. 
For inequality, Ref. \cite{ghosh2014} reports Gini coefficients typically in the range $0.5$ to $0.7$, whereas the result obtained here is $G = 0.725$, indicating a higher level of concentration.
In terms of asymmetry, Ref. \cite{clementi2019} reports Zanardi index values typically close to zero for income distribution, with most observations lying within $|Z| \lesssim 0.1$, whereas the result obtained here is $Z = 0.020$, indicating a comparatively low level of asymmetry, although this comparison should be interpreted with caution since the referenced study concerns income distributions, while the present result is based on city size distributions.

\subsection{Gompertz-Zipf alternative logarithmic binning: 2022 census}

\begin{figure}[H]
    \centering
    \includegraphics[width=0.7\linewidth]{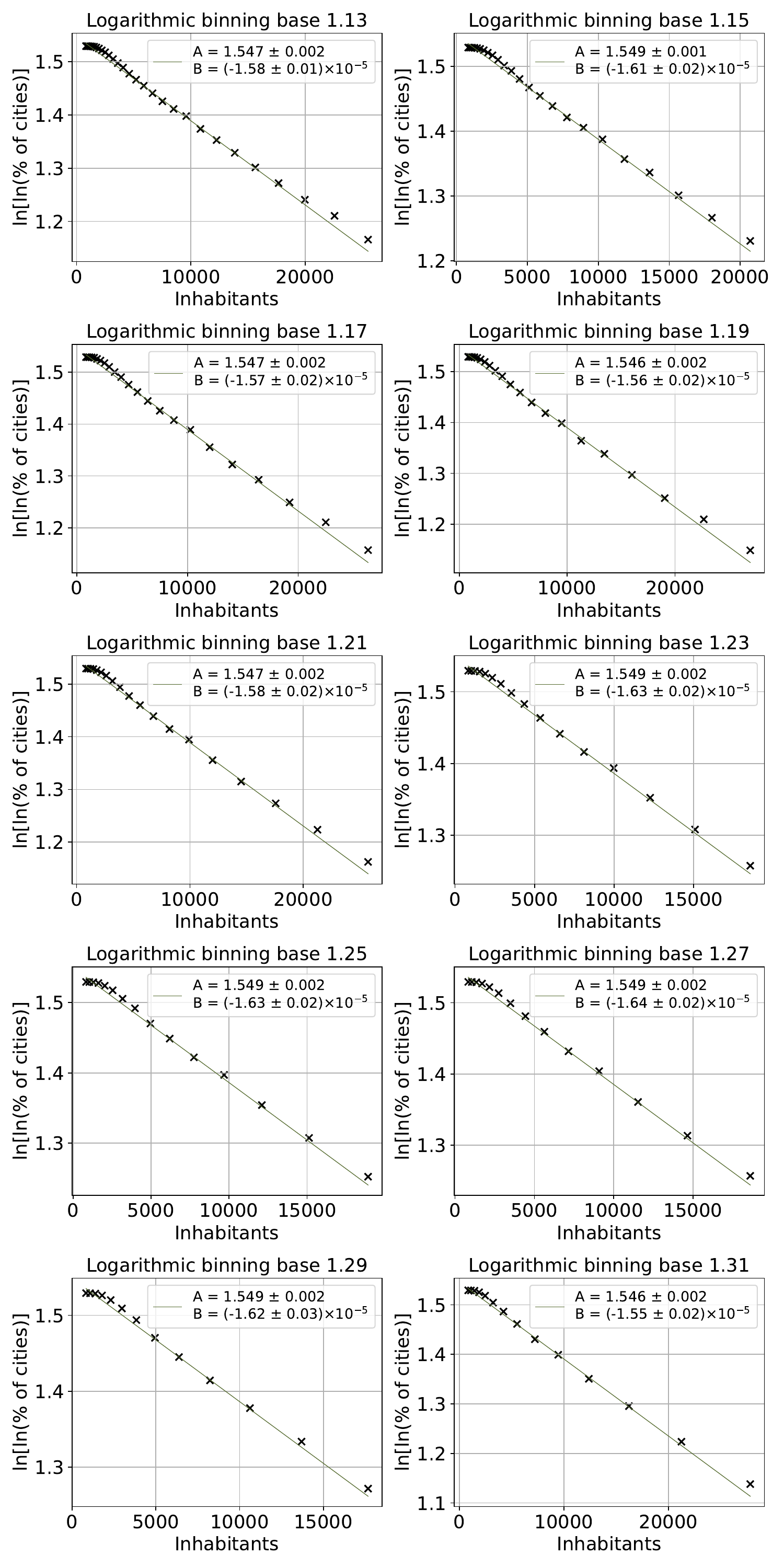}
    \caption[Gompertz linear fits for 10 logarithmic binning base options: 2022]{The linear regression for the 10 logarithmic binning base options of the 2022 data is $\ln\{\ln[P(x)]\} = A+Bx$ where this set of points makes $A$ closest to $\ln[\ln(100)] = 1.5272$ and $B$ is the slope of the function.}
    \label{fig:gompertz2022bases}
\end{figure}

\begin{figure}[H]
    \centering
    \includegraphics[width=0.7\linewidth]{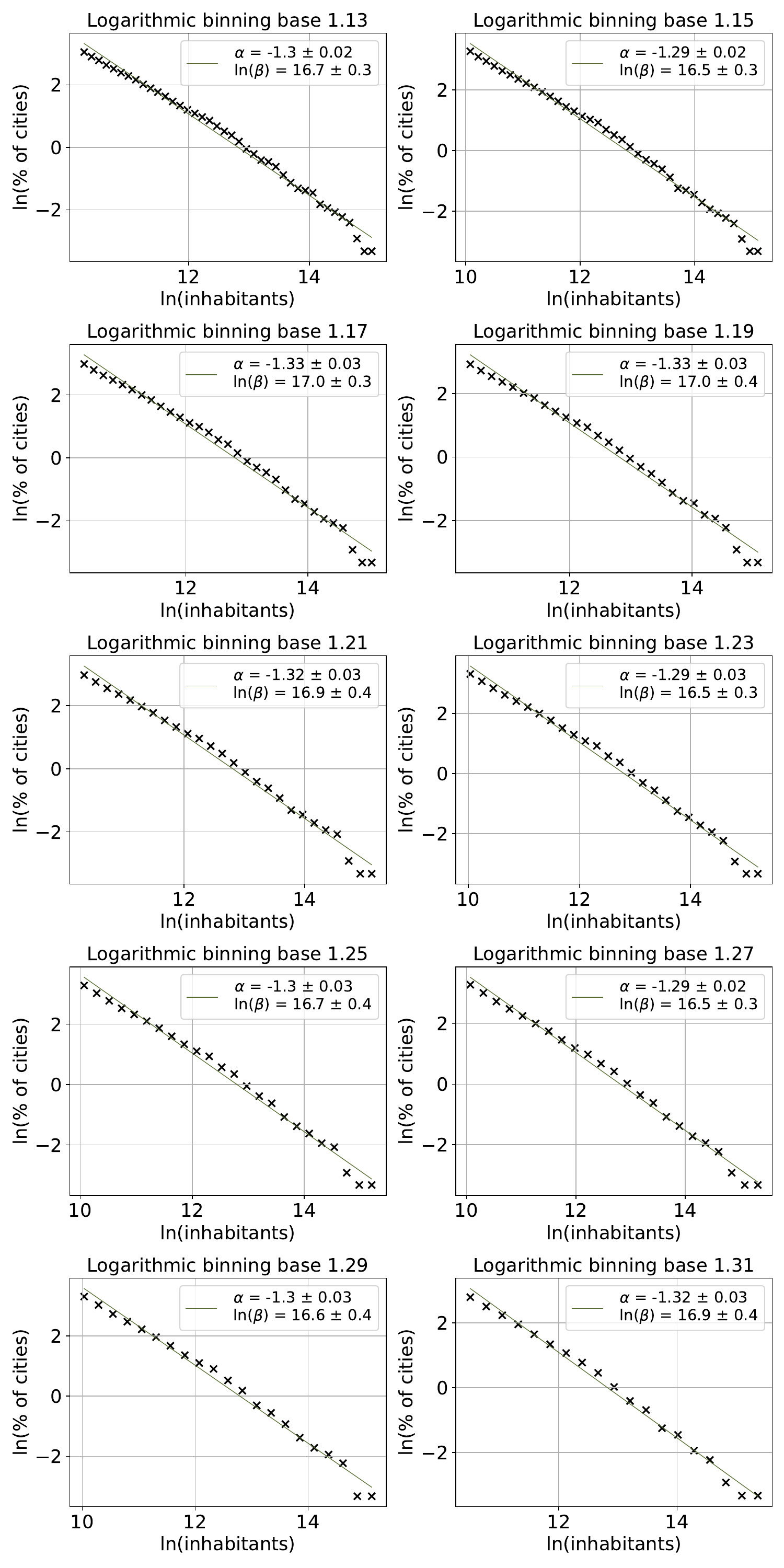}
    \caption[Zipf linear fits for 10 logarithmic binning base options: 2022]{The linear regression for the 10 logarithmic binning options of the 2022 data is $\ln[F(x)] = \alpha \ln(x) + \ln(\beta)$ where $\ln(\beta)$ is the Y intercept. The estimated values for the Zipf exponent are $a = 1 - \alpha$.}
    \label{fig:zipf2022bases}
\end{figure}

\begin{figure}[H]
    \centering
    \includegraphics[width=0.7\linewidth]{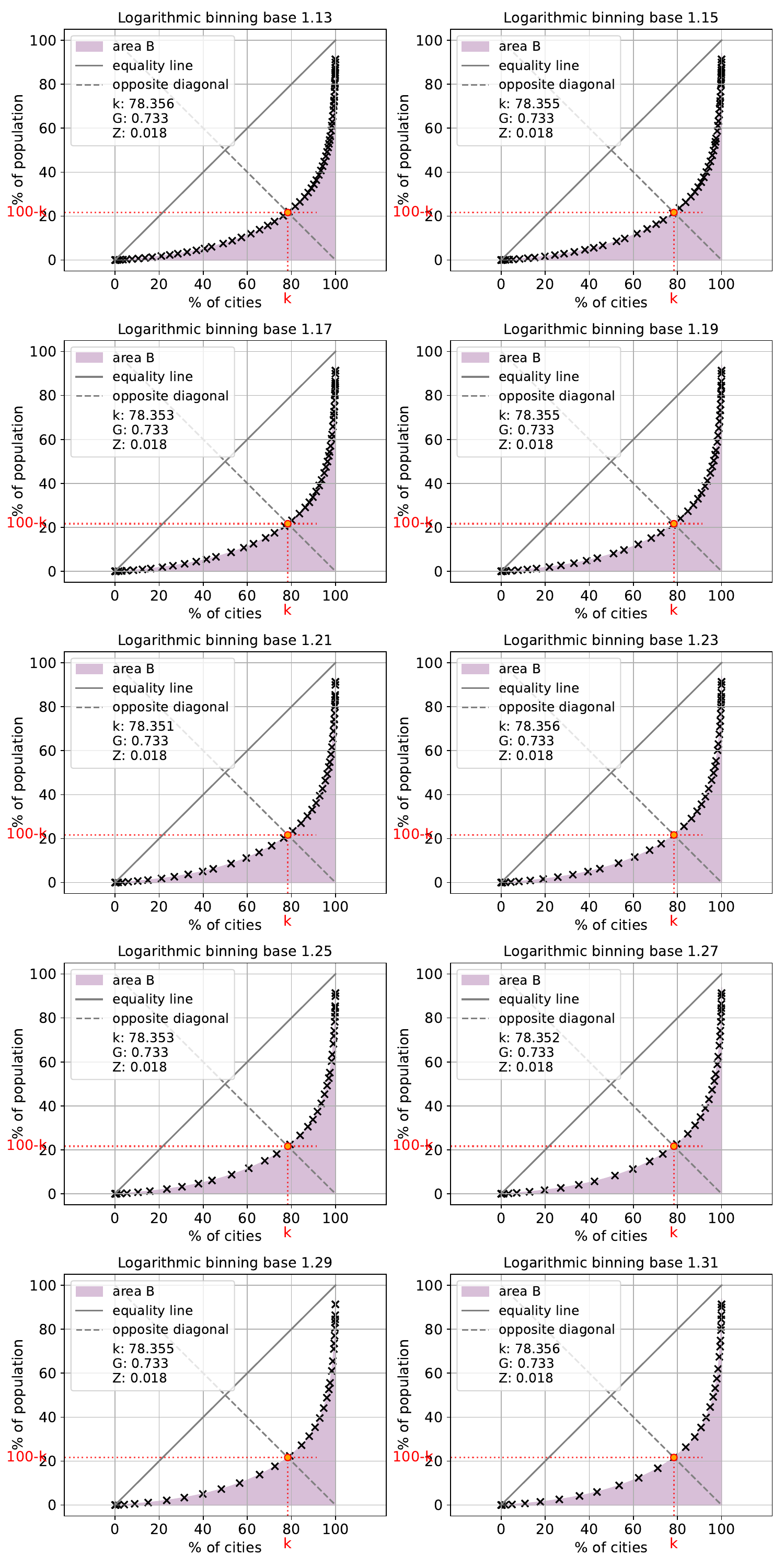}
    \caption[Lorenz curves for 10 logarithmic binning base options: 2022]{Lorenz curves for the 10 logarithmic binning options of the 2022 data. Despite small visual differences caused by the logarithmic binning choices, the corresponding inequality measures remain highly stable across all bases. The relative variations are only $0.006\%$ for the Kolkata index $k$, $0.012\%$ for the total Gini index $G$, and $0.46\%$ for the Zanardi index $Z$.}
    \label{fig:lorenz2022bases}
\end{figure}

To evaluate the sensitivity of the Gompertz–Zipf distribution to different logarithmic binning bases, this section compares the results obtained from 10 base values (1.13, 1.15, 1.17, 1.19, 1.21, 1.23, 1.25, 1.27, 1.29, 1.31) in $x_i = \mathrm{base}^{i-1} \cdot x_{min}$ for the 2022 census data, as described in Section 2.1.1. 
The goal is to assess the robustness of the estimated parameters and the stability of the Gompertz and Zipf regimes with respect to the choice of logarithmic binning base.
Figs. \ref{fig:gompertz2022bases} and \ref{fig:zipf2022bases} show the linear fits of the Gompertz and Zipf segments, while Fig. \ref{fig:lorenz2022bases} displays the corresponding Lorenz curves.

\begin{figure}[H]
    \centering
    \begin{minipage}{0.48\textwidth}
        \centering
        \includegraphics[width=\linewidth]{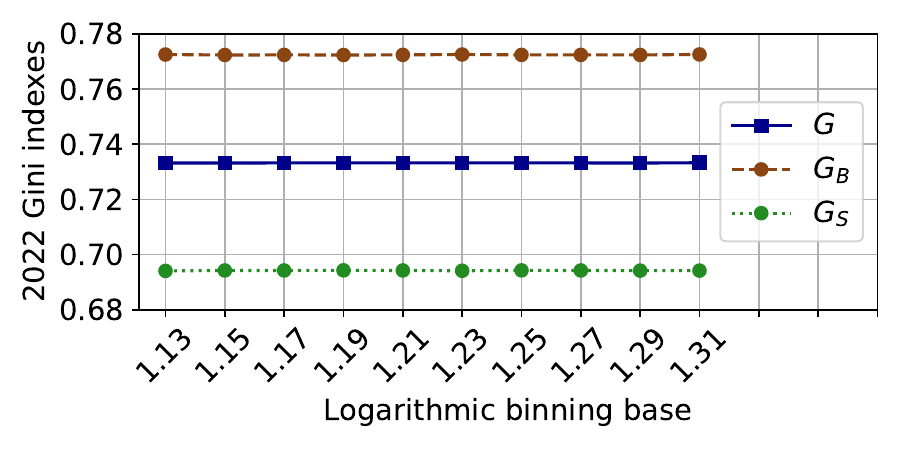}
        \caption[Gini indices: 2022]{$G$ is the Gini index for the 2022 data options of logarithmic binning base. $G_B$ is the Gini index for the upper part of the Lorenz curve related to the bigger cities and $G_S$ is the Gini index for the lower part of the Lorenz curve related to the smaller cities. The relative variation across all logarithmic binning bases is $0.012\%$ for $G$, $0.021\%$ for the small-city Gini index $G_S$ and $0.025\%$ for the big-city Gini index $G_B$, confirming the stability of the inequality measurements despite changes in the logarithmic binning procedure.}
        \label{fig:gini2022bases}
    \end{minipage}%
    \hfill
    \begin{minipage}{0.48\textwidth}
        \centering
        \includegraphics[width=\linewidth]{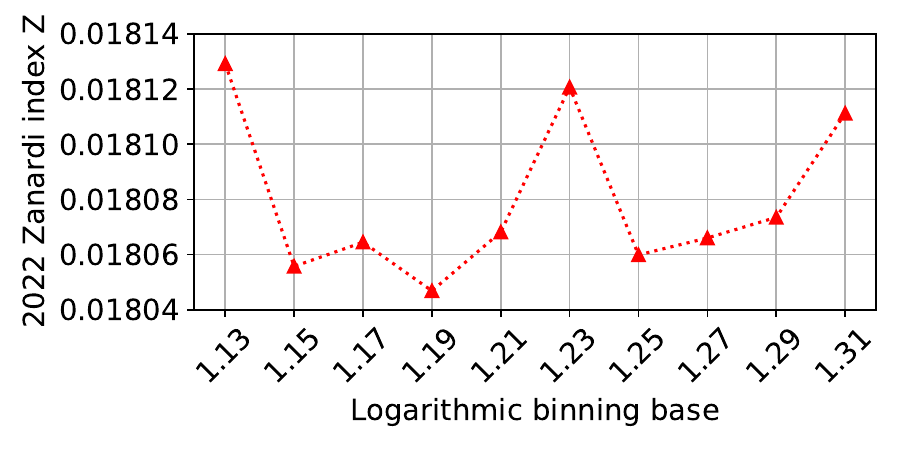}
        \caption[Zanardi indices: 2022]{Zanardi index for the 2022 data options of logarithmic binning base. The relative variation across all logarithmic binning bases is $0.46\%$, indicating that the asymmetry between the big-city and small-city segments remains nearly unchanged for different logarithmic binning choices.}
        \label{fig:zanardi2022bases}
    \end{minipage}
\end{figure}

\begin{figure}[H]
    \centering
    \begin{minipage}{0.48\textwidth}
        \centering
        \includegraphics[width=\linewidth]{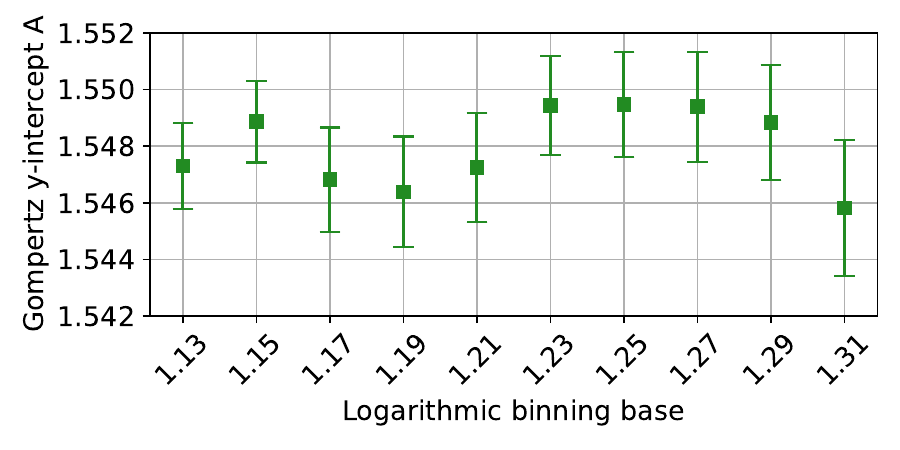}
        \caption[Gompertz y-intercepts $A$: 2022]{Gompertz segment y-intercepts $A$ for each option of logarithmic binning base: 2022. The relative variation of the Gompertz intercepts is only $0.24\%$ across all logarithmic binning bases.}
        \label{fig:gompertz2022intercepts}
    \end{minipage}%
    \hfill
    \begin{minipage}{0.48\textwidth}
        \centering
        \includegraphics[width=\linewidth]{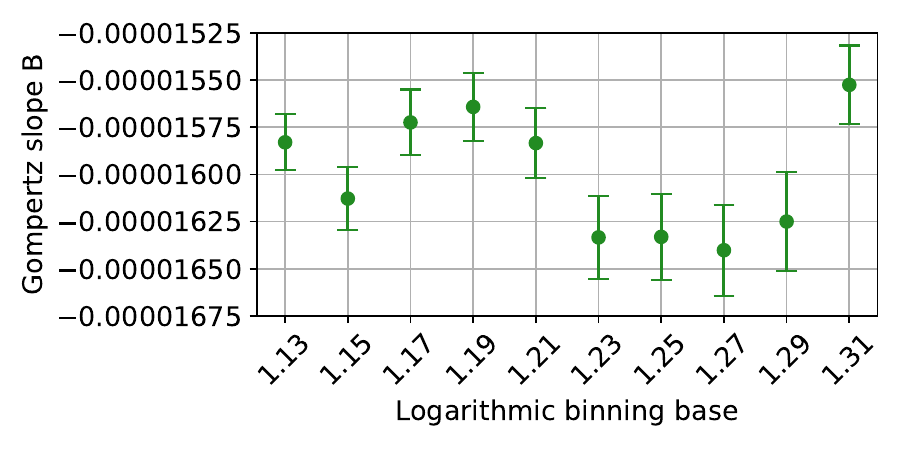}
        \caption[Gompertz slopes $B$: 2022]{Gompertz segment slopes $B$ for each option of logarithmic binning base: 2022. The relative variation of the Gompertz slopes is $5.50\%$ across all logarithmic binning bases.}
        \label{fig:gompertz2022slopes}
    \end{minipage}
\end{figure}

\begin{figure}[H]
    \centering
    \begin{minipage}{0.48\textwidth}
        \centering
        \includegraphics[width=\linewidth]{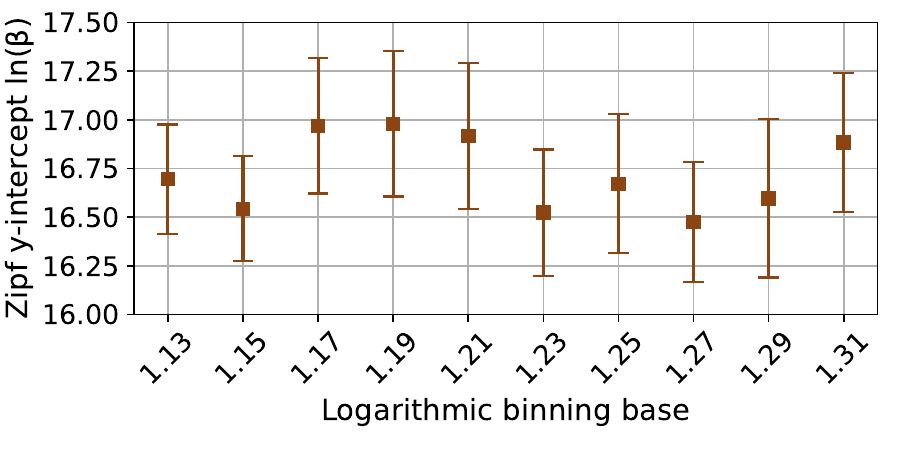}
        \caption[Zipf y-intercepts $\ln(\beta)$: 2022]{Zipf segment y-intercepts $\ln(\beta)$ for each option of logarithmic binning base: 2022. The relative variation of the Zipf intercepts is $3.01\%$ across all logarithmic binning bases.}
        \label{fig:zipf2022intercepts}
        \centering
    \end{minipage}%
    \hfill
    \begin{minipage}{0.48\textwidth}
        \includegraphics[width=\linewidth]{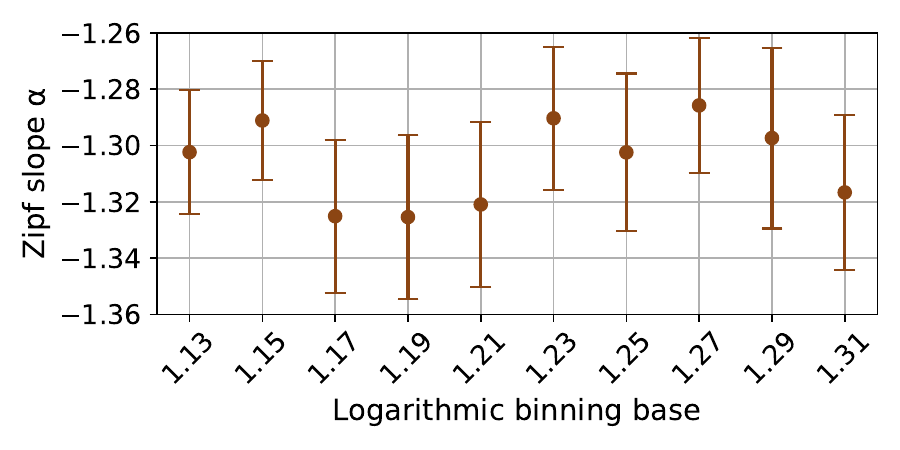}
        \caption[Zipf slopes $\alpha$: 2022]{Zipf segment slopes $\alpha$ for each option of logarithmic binning base: 2022. The relative variation of the Zipf slopes is $3.04\%$ across all logarithmic binning bases.}
        \label{fig:zipf2022slopes}
    \end{minipage}
\end{figure}

\begin{figure}[H]
    \centering
    \includegraphics[width=0.85\linewidth]{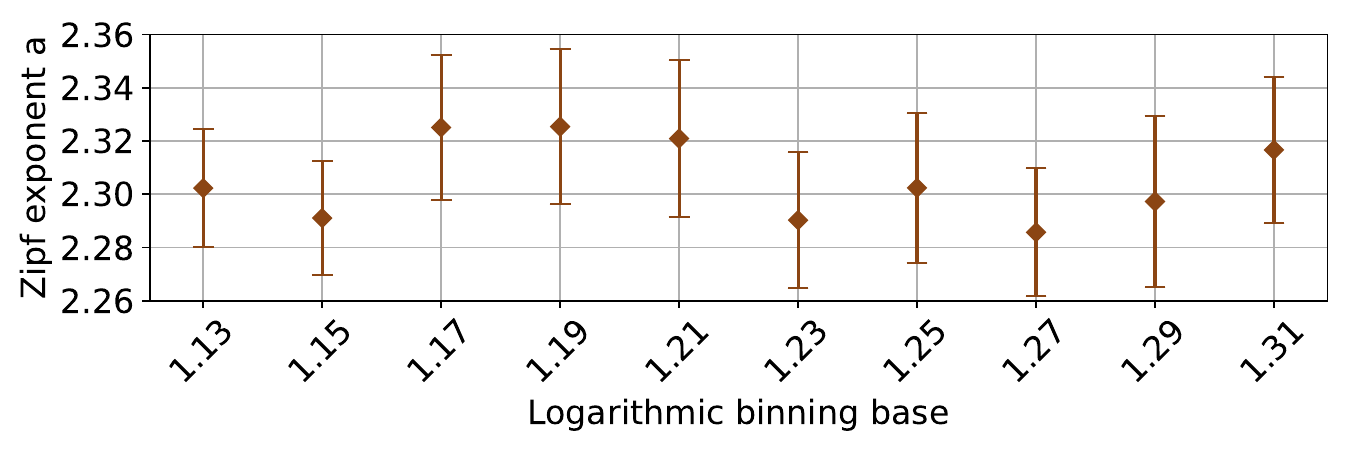}
    \caption[Zipf exponents $a$: 2022]{Zipf exponents $a$ for each option of logarithmic binning base: 2022. The relative variation of the Zipf exponents is only $1.72\%$ across all logarithmic binning bases.}
    \label{fig:zipf2022exponents}
\end{figure}

The inequality measures derived from the Lorenz curves are summarized in Figs. \ref{fig:gini2022bases} and \ref{fig:zanardi2022bases}, which present the Gini and Zanardi indices for each base. 
Throughout all logarithmic binning bases, the results remain consistent. 
The inequality measures in Figs. \ref{fig:gini2022bases} and \ref{fig:zanardi2022bases} further confirm that the overall shape of the Lorenz curves is robust to the choice of logarithmic binning base. 
Across all bases, the relative variation given by $(\max-\min)/\mathrm{mean}\times100\%$ is only $0.012\%$ for the Gini index and $0.46\%$ for the Zanardi index, despite the apparent visual differences caused by the axis scales.
Additionally, the estimated Gompertz intercepts and slopes are shown in Figs. \ref{fig:gompertz2022intercepts} and \ref{fig:gompertz2022slopes}, whereas the Zipf intercepts, slopes, and resulting Zipf power-law exponents are illustrated in Figs. \ref{fig:zipf2022intercepts}, \ref{fig:zipf2022slopes}, and \ref{fig:zipf2022exponents}. 
Figs. \ref{fig:gompertz2022intercepts} and \ref{fig:gompertz2022slopes} show that the Gompertz parameters $A$ and $B$ exhibit only minor fluctuations across binning choices, indicating stable behavior in the small-city segment. 
Using the relative variation, the intercept $A$ changes by only $0.24\%$ across all bases, while the slope $B$ varies by $5.50\%$.
In the Zipf segment, the intercepts and slopes displayed in Figs. \ref{fig:zipf2022intercepts} and \ref{fig:zipf2022slopes} also vary only slightly and the Zipf exponents $a$ in Fig. \ref{fig:zipf2022exponents} remain within a narrow range. 
The relative variation is $3.01\%$ for the intercept $\ln(\beta)$, $3.04\%$ for the slope $\alpha$, and $1.72\%$ for the Zipf exponent $a$.
These findings demonstrate that the conclusions drawn from Section 3.1 are independent of a particular binning choice, but instead reflect the properties of the Brazilian city size distribution for both small and big cities.

\section{Gompertz-Zipf population distribution: 1872-2022 censuses}

\begin{figure}[H]
    \centering
    \includegraphics[width=\linewidth]{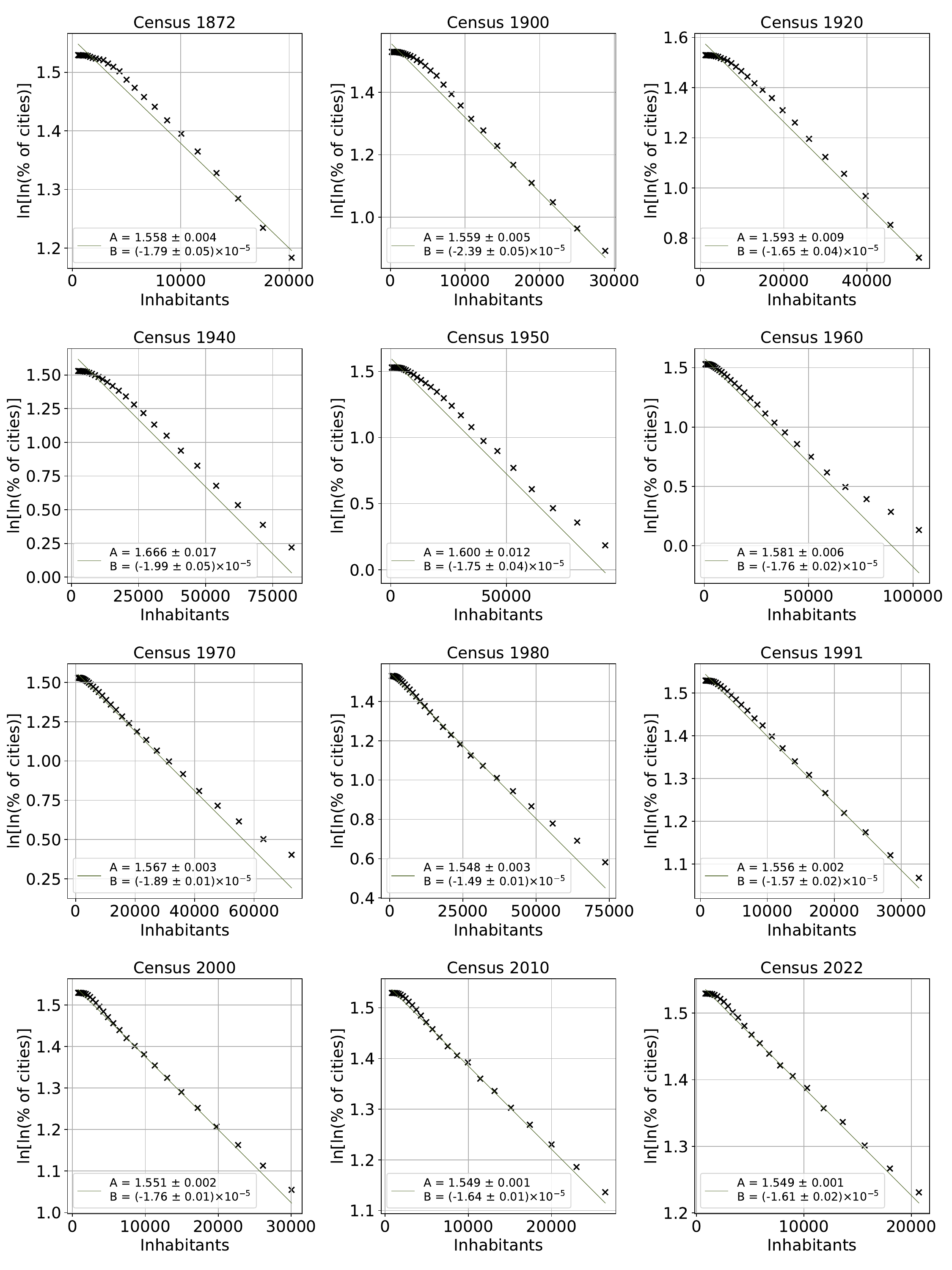}
    \caption[Gompertz linear fits: 1872-2022]{The Gompertz linear regression for the 12 censuses is $\ln\{\ln[P(x)]\} = A+Bx$ where this set of points makes $A$ closest to $\ln[\ln(100)] = 1.5272$ and $B$ is the slope of the function.}
    \label{fig:gompertz1872to2022}
\end{figure}

\begin{figure}[H]
    \centering
    \includegraphics[width=\linewidth]{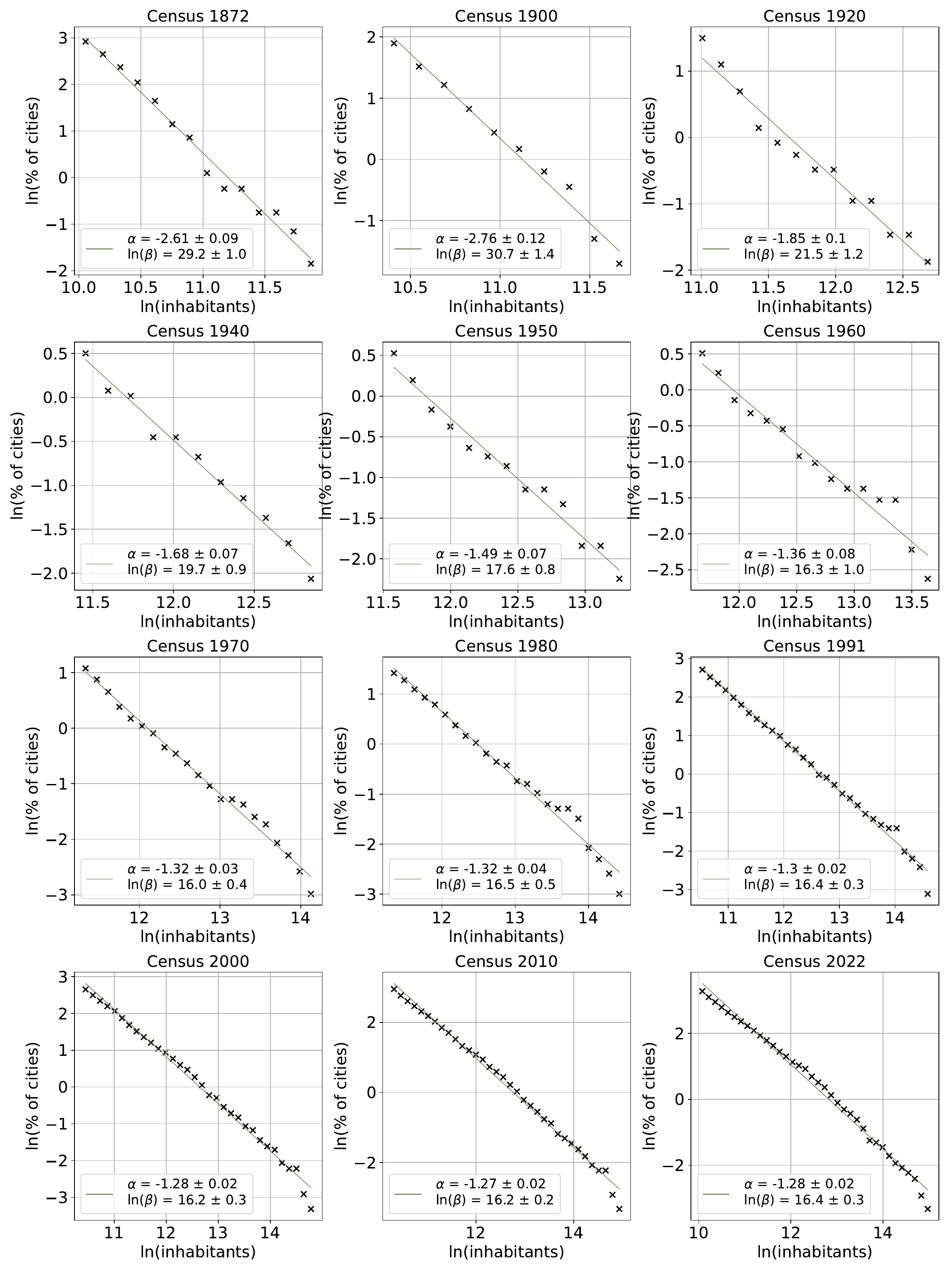}
    \caption[Zipf linear fits: 1872-2022]{The Zipf linear regression for the 12 censuses is $\ln[F(x)] = \alpha \ln(x) + \ln(\beta)$ where $\ln(\beta)$ is the Y intercept. The estimated values for the Zipf exponent are $a = 1 - \alpha$.}
    \label{fig:zipf1872to2022}
\end{figure}

The data from Refs. \cite{IBGE2022census} and \cite{IBGEbook} include censuses for the years of 1872, 1900, 1920, 1940, 1950, 1960, 1970, 1980, 1991, 2000, 2010 and 2022, that is, a data window spanning 150 years.
Using these twelve censuses, the long-term applicability of the Gompertz–Zipf distribution is assessed. 
The Gompertz linear regressions are shown in Fig. \ref{fig:gompertz1872to2022} and the Zipf linear regressions appear in Fig. \ref{fig:zipf1872to2022} showing that the Gompertz–Zipf distribution can describe the Brazilian city-size distribution across all censuses analyzed, with cutoffs $x_t$ between the segments reported in Table \ref{tab:cuts1872_2022}.

\begin{table}[h!]
\centering
\begin{tabular}{c c}
\hline
\noalign{\vskip 3pt}
Census year & Cutoff $x_t$ \\
\hline
1872 & 20,216 \\
1900 & 28,793 \\
1920 & 52,461 \\
1940 & 82,035 \\
1950 & 92,765 \\
1960 & 102,857 \\
1970 & 72,592 \\
1980 & 73,612 \\
1991 & 32,652 \\
2000 & 30,059 \\
2010 & 26,467 \\
2022 & 20,710 \\
\hline
\end{tabular}
\caption[Cutoff $x_t$ values for Brazilian censuses: 1872-2022]{Cutoff $x_t$ values obtained for each of the twelve Brazilian censuses from 1872 to 2022, used to separate the Gompertz and Zipf segments.}
\label{tab:cuts1872_2022}
\end{table}

\begin{figure}[H]
    \centering
    \begin{minipage}{0.48\textwidth}
        \centering
        \includegraphics[width=\linewidth]{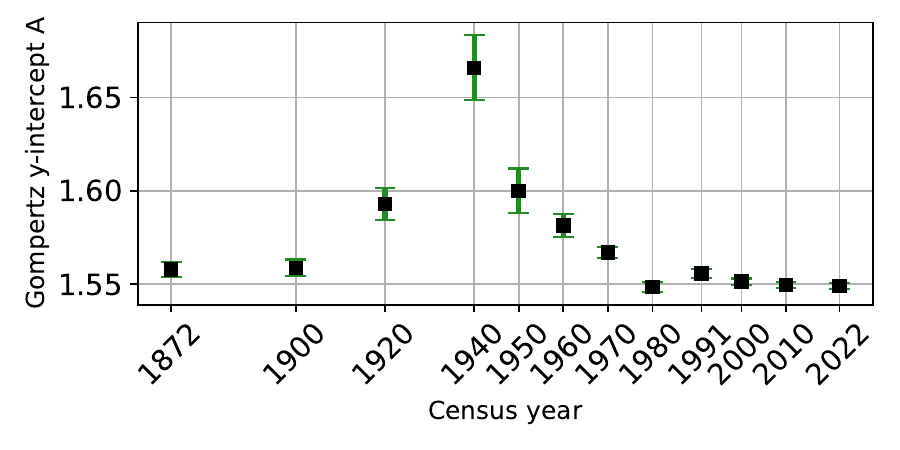}
        \caption[Gompertz y-intercepts $A$: 1872-2022]{Gompertz segment y-intercepts $A$ in the expression $\ln\{\ln[P(x)]\} = A+Bx$ from 1872 to 2022.}
        \label{fig:gompertz1872to2022intercepts}
    \end{minipage}%
    \hfill
    \begin{minipage}{0.48\textwidth}
        \centering
        \includegraphics[width=\linewidth]{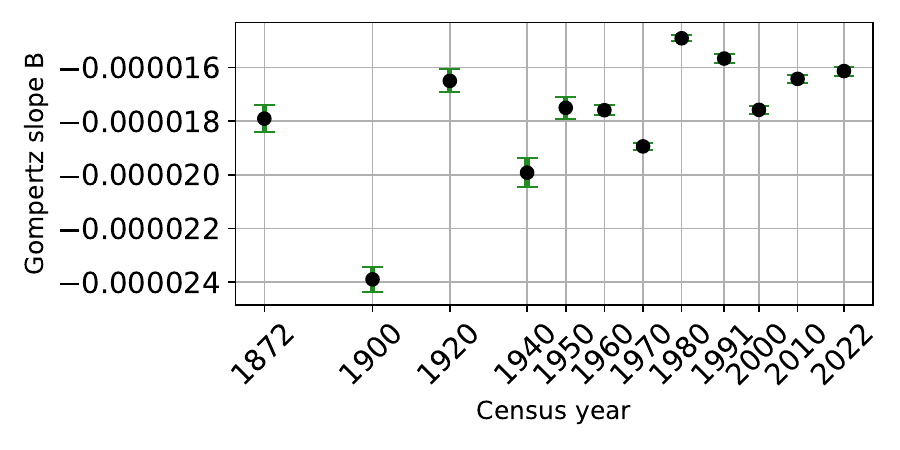}
        \caption[Gompertz slopes $B$: 1872-2022]{Gompertz segment slopes $-B$ in the expression $\ln\{\ln[P(x)]\} = A+Bx$ from 1872 to 2022.}
        \label{fig:gompertz1872to2022slopes}
    \end{minipage}
\end{figure}

The Gompertz parameters $A$ and $B$ are documented in Figs. \ref{fig:gompertz1872to2022intercepts} and \ref{fig:gompertz1872to2022slopes} for the small-city segment. 
As seen by the oscillations on the y-axis of Fig. \ref{fig:gompertz1872to2022slopes} and in Table \ref{tab:gompertz}, the values of the linear coefficient $B$ have relative variation of approximately $46\%$, indicating periods with faster decreases and periods with slower decreases in the number of cities within the less populated segment as population increases.
For the intercept $A$, the relative variation is approximately $7.5\%$. 
Excluding the 1940 census, this variation decreases to approximately $3.3\%$, reinforcing that the 1940 value is noticeably separated from the remaining censuses.
In contrast to the oscillation of $B$, the intercept $A$ shows a noticeable peak in 1940. 
This peak in 1940 may be related to the higher minimum population observed in that census, as shown in Fig. \ref{fig:min1872to2022}, where 1940 presents the largest minimum value among all censuses, indicating a shift in the lower bound of city sizes that can affect the intercept estimation.
Further investigation would be required to confirm this effect.

\begin{figure}[H]
    \centering
    \begin{minipage}{0.48\textwidth}
        \centering
        \includegraphics[width=\linewidth]{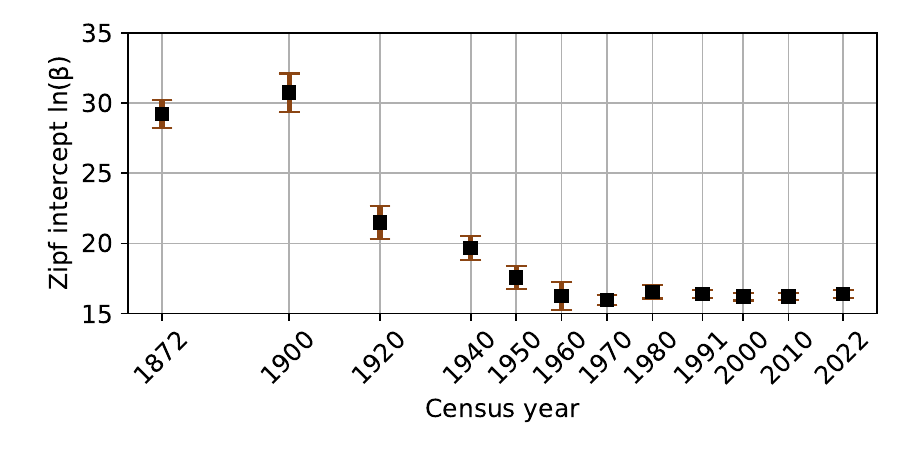}
        \caption[Zipf y-intercepts $\ln (\beta)$: 1872-2022]{Zipf segment y-intercepts $\ln (\beta)$ in the expression $\ln[F(x)] = \alpha \ln(x) + \ln(\beta)$ from 1872 to 2022.}
        \label{fig:zipf1872to2022intercepts}
    \end{minipage}%
    \hfill
    \begin{minipage}{0.48\textwidth}
        \centering
        \includegraphics[width=\linewidth]{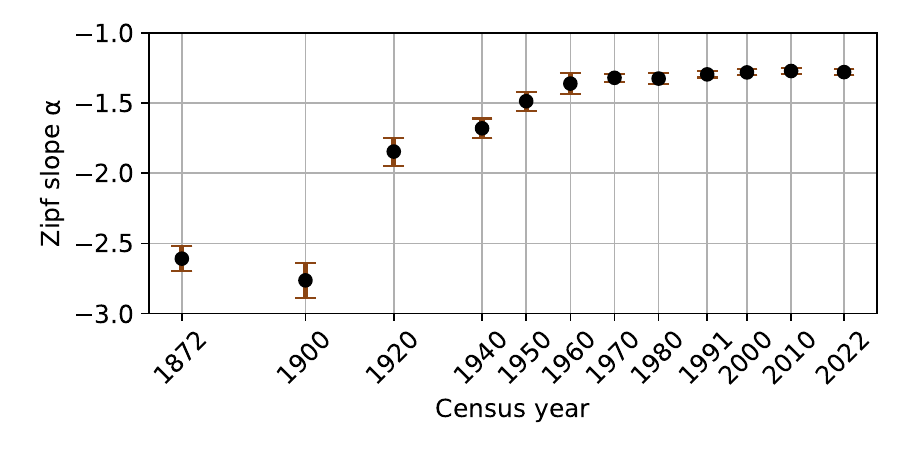}
        \caption[Zipf slopes $\alpha$: 1872-2022]{Zipf segment slopes $\alpha$ in the expression $\ln[F(x)] = \alpha \ln(x) + \ln(\beta)$ from 1872 to 2022.}
        \label{fig:zipf1872to2022slopes}
    \end{minipage}
\end{figure}

\begin{figure}[H]
    \centering
    \includegraphics[width=0.6\linewidth]{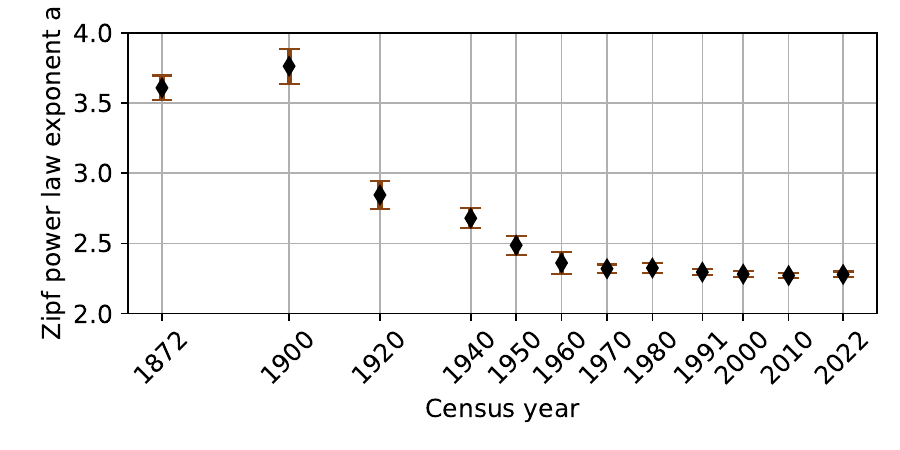}
    \caption[Zipf power-law exponents $a$: 1872-2022]{Zipf power-law exponents $a = 1-\alpha$ in the expression $\ln[F(x)] = \alpha \ln(x) + \ln(\beta)$ from 1872 to 2022.}
    \label{fig:zipf1872to2022exponents}
\end{figure}

Correspondingly, the Zipf intercepts, slopes and power-law exponents are shown in Figs. \ref{fig:zipf1872to2022intercepts}, \ref{fig:zipf1872to2022slopes} and \ref{fig:zipf1872to2022exponents}, respectively. The Zipf exponent decreases from $a=3.61\pm0.09$ in 1872 to $a=2.28\pm0.02$ in 2022, indicating an increasing presence of very populous cities, while the parameters become stable after 1960.
Using the values reported in Table \ref{tab:zipf}, the relative variation across all censuses is approximately $53\%$ for the intercept $\ln(\beta)$, $53\%$ for the slope $\alpha$, and $39\%$ for the Zipf exponent $a$. 
However, after 1960 the values become much more stable, with the relative variation of the exponent remaining below $4\%$.
The numerical values of the fitted parameters for both segments are summarized in Tables \ref{tab:gompertz} and \ref{tab:zipf}. 
These tables allow a direct comparison of the temporal evolution of the Gompertz and Zipf parameters, highlighting the oscillations of the Gompertz parameter $B$ throughout the period and the stabilization of the Zipf parameters after the mid-20th century, when the Zipf exponent becomes nearly constant.

\begin{table}[h]
\centering
\begin{tabular}{ccc}
\hline
Year & A & B \\
\hline
1872 & $1.558 \pm 0.004$ & $(-1.79 \pm 0.05) \times 10^{-5}$ \\
1900 & $1.559 \pm 0.005$ & $(-2.39 \pm 0.05) \times 10^{-5}$ \\
1920 & $1.593 \pm 0.009$ & $(-1.65 \pm 0.04) \times 10^{-5}$ \\
1940 & $1.666 \pm 0.017$ & $(-1.99 \pm 0.05) \times 10^{-5}$ \\
1950 & $1.600 \pm 0.012$ & $(-1.75 \pm 0.04) \times 10^{-5}$ \\
1960 & $1.581 \pm 0.006$ & $(-1.76 \pm 0.02) \times 10^{-5}$ \\
1970 & $1.567 \pm 0.003$ & $(-1.89 \pm 0.01) \times 10^{-5}$ \\
1980 & $1.548 \pm 0.003$ & $(-1.49 \pm 0.01) \times 10^{-5}$ \\
1991 & $1.556 \pm 0.002$ & $(-1.57 \pm 0.02) \times 10^{-5}$ \\
2000 & $1.551 \pm 0.002$ & $(-1.76 \pm 0.01) \times 10^{-5}$ \\
2010 & $1.549 \pm 0.001$ & $(-1.64 \pm 0.01) \times 10^{-5}$ \\
2022 & $1.549 \pm 0.001$ & $(-1.61 \pm 0.02) \times 10^{-5}$ \\
\hline
\end{tabular}
\caption[Gompertz parameters: 1872-2022]{Estimated Gompertz parameters $A$ and $B$ obtained from linear regressions for each Brazilian census from 1872 to 2022. The fits follow $\ln\{\ln[P(x)]\} = A + Bx$, where $A$ is the y-intercept and $B$ is the slope. Reported uncertainties correspond to the standard errors of the linear fits.}
\label{tab:gompertz}
\end{table}

\begin{table}[h]
\centering
\begin{tabular}{cccc}
\hline
Year & $\ln(\beta)$ & $\alpha$ & $a$ \\
\hline
1872 & $29.2 \pm 1.0$ & $-2.61 \pm 0.09$ & $3.61 \pm 0.09$ \\
1900 & $30.7 \pm 1.4$ & $-2.76 \pm 0.12$ & $3.76 \pm 0.12$ \\
1920 & $21.5 \pm 1.2$ & $-1.85 \pm 0.10$ & $2.85 \pm 0.10$ \\
1940 & $19.7 \pm 0.9$ & $-1.68 \pm 0.07$ & $2.68 \pm 0.07$ \\
1950 & $17.6 \pm 0.8$ & $-1.49 \pm 0.07$ & $2.49 \pm 0.07$ \\
1960 & $16.3 \pm 1.0$ & $-1.36 \pm 0.08$ & $2.36 \pm 0.08$ \\
1970 & $16.0 \pm 0.4$ & $-1.32 \pm 0.03$ & $2.32 \pm 0.03$ \\
1980 & $16.5 \pm 0.5$ & $-1.32 \pm 0.04$ & $2.32 \pm 0.04$ \\
1991 & $16.4 \pm 0.3$ & $-1.30 \pm 0.02$ & $2.30 \pm 0.02$ \\
2000 & $16.2 \pm 0.3$ & $-1.28 \pm 0.02$ & $2.28 \pm 0.02$ \\
2010 & $16.2 \pm 0.2$ & $-1.27 \pm 0.02$ & $2.27 \pm 0.02$ \\
2022 & $16.4 \pm 0.3$ & $-1.28 \pm 0.02$ & $2.28 \pm 0.02$ \\
\hline
\end{tabular}
\caption[Zipf parameters: 1872-2022]{Estimated Zipf parameters obtained from linear regressions for each Brazilian census from 1872 to 2022. The fits follow $\ln[F(x)] = \alpha \ln(x) + \ln(\beta)$, where $\alpha$ is the slope and $\ln(\beta)$ the y-intercept. The corresponding power-law exponent is $a = 1 - \alpha$. Reported uncertainties correspond to the standard errors of the linear fits.}
\label{tab:zipf}
\end{table}

\begin{figure}[H]
    \centering
    \includegraphics[width=\linewidth]{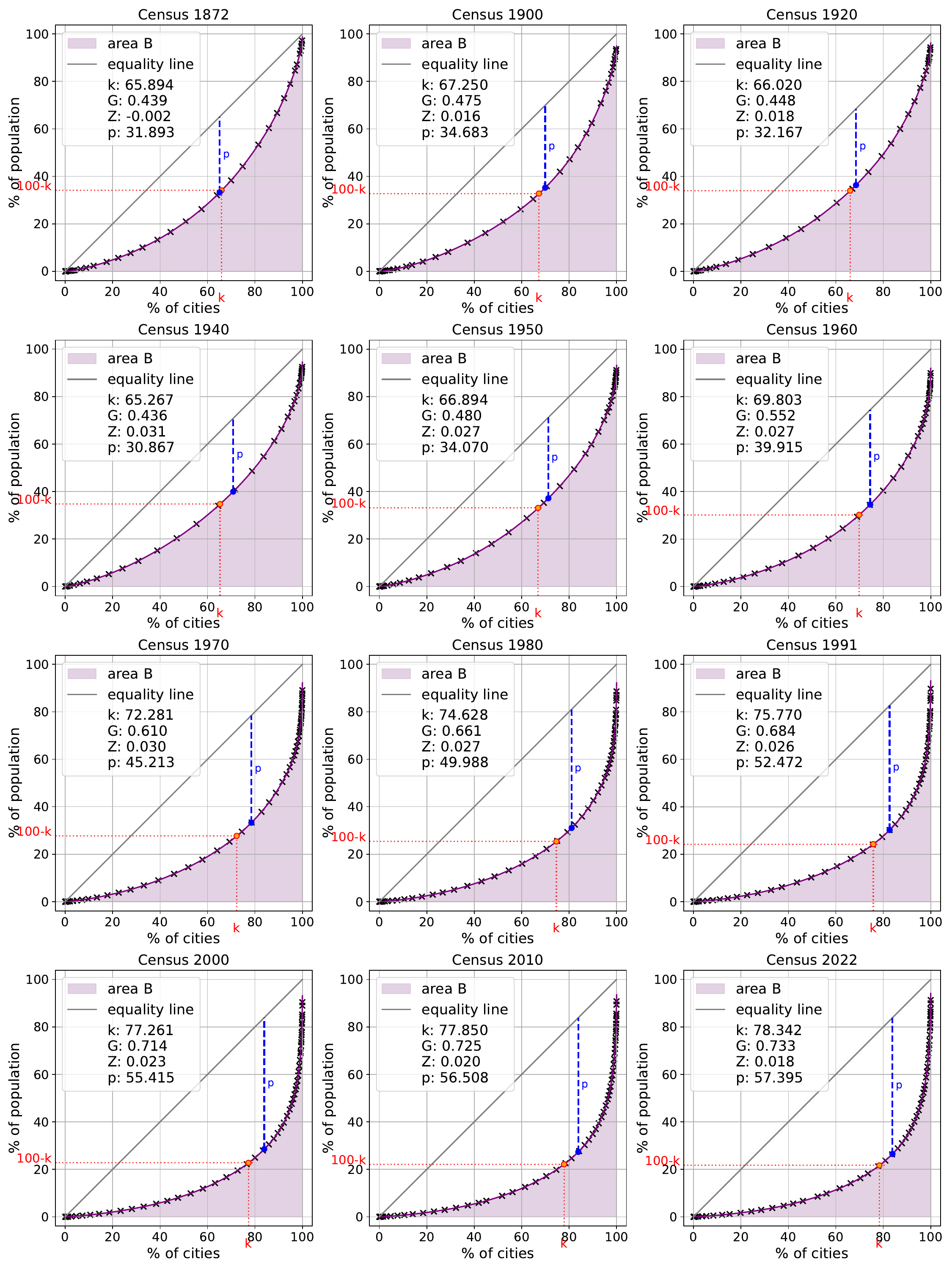}
    \caption[Lorenz curves: 1872-2022]{Lorenz curves for the Brazilian city size distribution across the censuses from 1872 to 2022. Each curve represents the cumulative share of population as a function of the cumulative share of cities, showing an increasing deviation of the interpolated Lorenz curves from the line of equality over time and thus the increase in concentration.}
    \label{fig:lorenz1872to2022}
\end{figure}

\begin{figure}[H]
    \centering
    \begin{minipage}{0.48\textwidth}
        \centering
        \includegraphics[width=\linewidth]{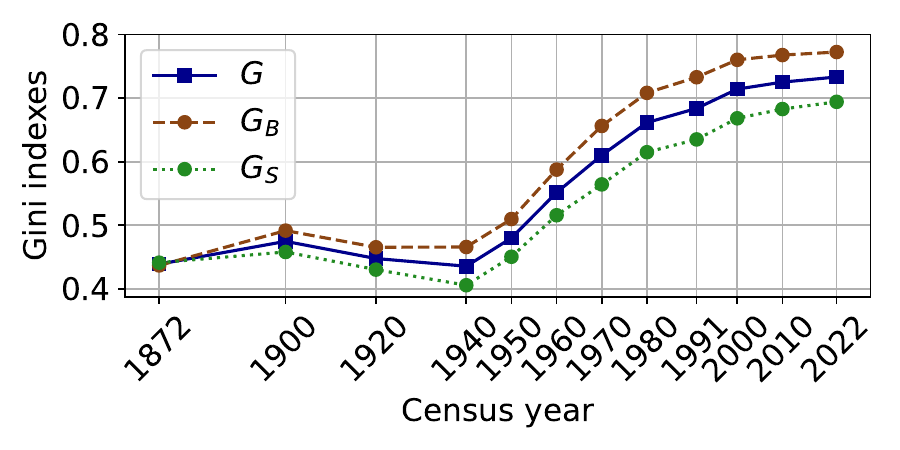}
        \caption[Gini indices $G$, $G_B$ and $G_S$: 1872-2022]{Temporal evolution of the Gini indices for the full distribution $G$, for the big-city segment $G_B$ and for the small-city segment $G_S$ from 1872 to 2022, showing a sustained increase in overall concentration throughout the period.}
        \label{fig:gini_1872to2022}
    \end{minipage}%
    \hfill
    \begin{minipage}{0.48\textwidth}
        \centering
        \includegraphics[width=\linewidth]{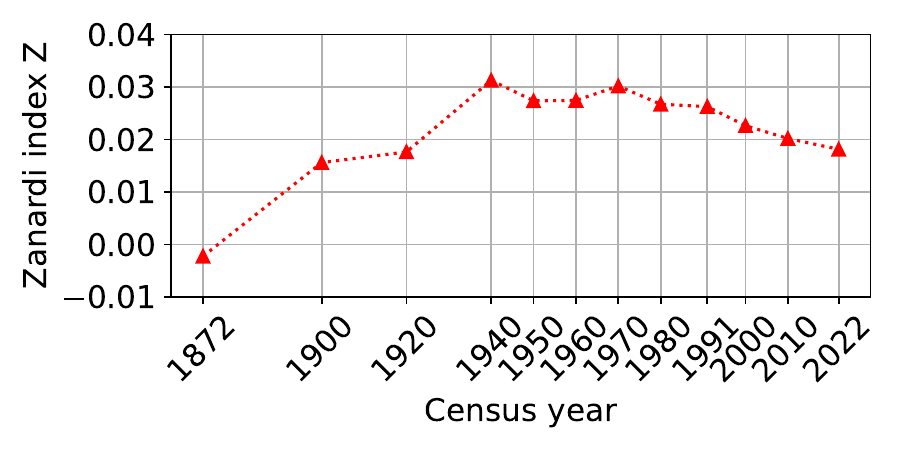}
        \caption[Zanardi indices $Z$: 1872-2022]{Temporal evolution of the Zanardi index $Z$ from 1872 to 2022, indicating that the asymmetry is slightly negative in 1872 and positive in all subsequent censuses, meaning that inequality becomes more concentrated in the big-city segment after 1900.}
        \label{fig:zanardi_1872to2022}
    \end{minipage}
\end{figure}

The Lorenz curves are presented in Fig. \ref{fig:lorenz1872to2022} and the inequality measures across censuses are summarized by the Gini indices in Fig. \ref{fig:gini_1872to2022} and by Zanardi indices in Fig. \ref{fig:zanardi_1872to2022}, offering a century-and-a-half view on concentration and asymmetry in the Brazilian city system. 
The Gini indices in Fig. \ref{fig:gini_1872to2022} show increasing population distribution inequality after 1940, even within the less-populated-city segment $G_S$, while the Zanardi index in Fig. \ref{fig:zanardi_1872to2022} is negative in 1872 and positive in all subsequent censuses, indicating that the asymmetry shifted from a slight predominance of inequality in the small-city segment in 1872 to a greater concentration of population in the group of more populated cities after 1900. 
The magnitude of $Z$ decreases after 1970, declining from $0.030$ in 1970 to $0.018$ in 2022.
This suggests that, although the asymmetry between the two segments remains predominantly positive in favor of the big-city segment after 1900, the rate at which this imbalance intensifies has become less pronounced over time, indicating a slower growth of population inequality, reflecting the fact that concentration remains higher in larger cities than in smaller ones within the city-size distribution.
Table \ref{tab:ginipietra} complements these results by reporting the values of the inequality measures for each census year. 
It shows that the concentration, measured by $G$, exhibits an overall increase over time, while the asymmetry measured by $Z$ changes from a slightly negative value in 1872 to positive values afterward, while the Kolkata index $k$ shifts towards higher values, indicating that an increasingly smaller fraction of cities accounts for most of the population.

An additional measurement associated to inequality can be obtained by the maximum vertical distance between the Lorenz curve and the line of equality, defined as the Pietra index $p$. 
The vertical dashed lines in Fig. \ref{fig:lorenz1872to2022} illustrate this definition. 
This index quantifies the maximum deviation between the observed distribution and the equality line.
Although this interpretation is mostly discussed in the context of income inequality, it is also meaningful for city size distributions, since excessive population concentration in large urban centers is associated with several social and infrastructural problems related to urban gigantism.
In this sense, the Pietra index provides a complementary measure of concentration, indicating the largest vertical distance between the empirical Lorenz curve and the equality line.
The values of the Pietra indices $p$ reported in Table \ref{tab:ginipietra} show that the distribution is far from a spatially balanced population arrangement, with $p$ ranging from approximately 31,9\% in 1872 to 57,4\% in 2022.
These results were calculated using monotonic cubic spline interpolation in Python. 
As shown in Figs. \ref{fig:numberofpoints1872to2022} and \ref{fig:lorenz1872to2022}, each Lorenz curve is represented by a finite set of empirical points, and the interpolation constructs a local cubic approximation between every pair of consecutive points. 
This procedure generates a continuous piecewise curve that preserves the monotonic increasing behavior expected for cumulative distributions without introducing artificial oscillations. 
The Pietra index is then obtained by determining the point of the complete interpolated Lorenz curve that is farthest from the equality line.

\begin{figure}[h!]
    \centering
    \includegraphics[width=0.6\textwidth]{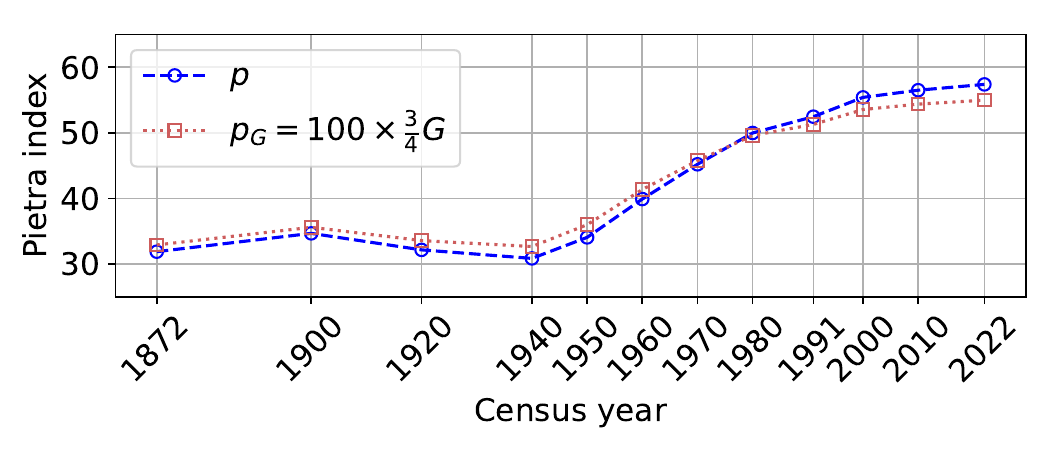}
    \caption[Comparison between the Pietra indices $p$ and $p_G$]{Comparison between the Pietra index $p$ obtained from a monotonic cubic spline interpolation of the Lorenz curve and the approximate Pietra index $p_G$ obtained from Eq. \ref{eq:pG} estimated from the Gini coefficient for all Brazilian censuses from 1872 to 2022.}
    \label{fig:pietra_comparison}
\end{figure}

Moreover, a comparative Pietra index, here defined as $p_G$, can be approximately estimated from the Gini coefficients with the expression given in Eq. (\ref{eq:pG}), derived from the quadratic approximation proposed in Ref. \cite{chakrabarti2025}. 
This expression provides an analytical connection with the Gini coefficient, allowing an approximate estimation of the Pietra index without requiring the explicit interpolation of the Lorenz curve.
Although the Gini and Zanardi coefficients already describe important aspects of concentration and asymmetry, the Pietra index adds an additional interpretation associated with redistribution and therefore complements these measures.
Fig. \ref{fig:pietra_comparison} compares the values of the Pietra indices $p$ and $p_G$ over time, showing that both measures follow similar trends throughout the analyzed period.
The values of $G$, $p$ and $p_G$ show an overall increasing trend over time, with fluctuations before 1950 and a more sustained increase afterward.
This pattern indicates increasing concentration in city population sizes over time.

\begin{figure}[h!]
    \centering
    \includegraphics[width=0.6\textwidth]{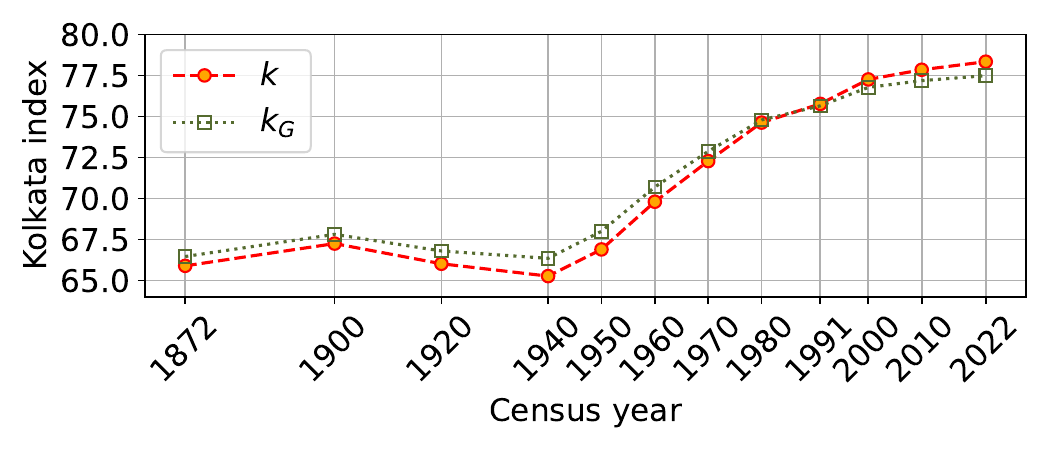}
    \caption[Comparison between the Kolkata indices $k$ and $k_G$]{Comparison between the Kolkata index $k$ obtained from the intersection of the Lorenz curve with the opposite diagonal $y = 100 - x$ and the approximate Kolkata index $k_G$ obtained from Eq. \ref{eq:kG} estimated from the Gini coefficient for all Brazilian censuses from 1872 to 2022.}
    \label{fig:kolkata_comparison}
\end{figure}

The Kolkata index $k$ obtained from the intersection of the Lorenz curve with the opposite diagonal $y = 100 - x$ increased systematically from approximately 66\% in 1872 to 78\% in 2022, as shown in Fig. \ref{fig:kolkata_comparison}, indicating that the population has become progressively concentrated in a smaller fraction of cities.
The index remained relatively stable between 1872 and 1950, but exhibited a sustained increase after 1950, reaching values above 70\% in 1960 and continuing to rise subsequently, which suggests a connection with the accelerated industrialization process that began in Brazil around 1950. 
The value of $k$ in Table \ref{tab:ginipietra} approaching 80\% in the most recent censuses is consistent with the Pareto principle, also known as the 80:20 rule, according to which approximately 80\% of the population would be concentrated in 20\% of the cities. 
This pattern was originally documented by the Italian economist Vilfredo Pareto (1848-1923) in his work \textit{Cours d'\'{e}conomie politique}, where he observed that roughly 80\% of the land in Italy was owned by 20\% of the population.

\begin{table}[H]
\centering
\begin{tabular}{c c c c c c c c c c c}
\hline
Year & $G$ & $G_B$ & $G_S$ & $p(\%)$ & $p_G(\%)$ & $p_G$ error & $Z$ & $k(\%)$ & $k_G(\%)$ & $k_G$ error \\
\hline
1872 & $0.439$ & $0.437$ & $0.441$ & $31.893$ & $32.925$ & $3.23\%$ & $-0.002$ & $65.894$ & $66.463$ & $0.86\%$ \\
1900 & $0.475$ & $0.492$ & $0.458$ & $34.683$ & $35.625$ & $2.72\%$ & $0.016$ & $67.250$ & $67.812$ & $0.84\%$ \\
1920 & $0.448$ & $0.465$ & $0.430$ & $32.167$ & $33.600$ & $4.46\%$ & $0.018$ & $66.020$ & $66.800$ & $1.18\%$ \\
1940 & $0.436$ & $0.466$ & $0.406$ & $30.867$ & $32.700$ & $5.94\%$ & $0.031$ & $65.267$ & $66.350$ & $1.66\%$ \\
1950 & $0.480$ & $0.510$ & $0.451$ & $34.070$ & $36.000$ & $5.67\%$ & $0.027$ & $66.894$ & $68.000$ & $1.65\%$ \\
1960 & $0.552$ & $0.588$ & $0.516$ & $39.915$ & $41.400$ & $3.72\%$ & $0.027$ & $69.803$ & $70.700$ & $1.28\%$ \\
1970 & $0.610$ & $0.656$ & $0.564$ & $45.213$ & $45.750$ & $1.19\%$ & $0.030$ & $72.281$ & $72.875$ & $0.82\%$ \\
1980 & $0.661$ & $0.708$ & $0.615$ & $49.988$ & $49.575$ & $0.83\%$ & $0.027$ & $74.628$ & $74.787$ & $0.21\%$ \\
1991 & $0.684$ & $0.733$ & $0.635$ & $52.472$ & $51.300$ & $2.23\%$ & $0.026$ & $75.770$ & $75.650$ & $0.16\%$ \\
2000 & $0.714$ & $0.760$ & $0.668$ & $55.415$ & $53.550$ & $3.37\%$ & $0.023$ & $77.261$ & $76.775$ & $0.63\%$ \\
2010 & $0.725$ & $0.768$ & $0.683$ & $56.508$ & $54.375$ & $3.77\%$ & $0.020$ & $77.850$ & $77.188$ & $0.85\%$ \\
2022 & $0.733$ & $0.772$ & $0.694$ & $57.395$ & $54.975$ & $4.22\%$ & $0.018$ & $78.342$ & $77.487$ & $1.09\%$ \\
\hline
\end{tabular}
\caption[Gini, Zanardi, Kolkata and Pietra indices]{Values of the Gini index $G$, the big-city and small-city Gini indices $G_B$ and $G_S$, the Pietra index $p$, the approximate Pietra index $p_G$ obtained from Eq. \ref{eq:pG}, the absolute percentage error of $p_G$ relative to $p$ computed as $|p_G - p|/p$, the Zanardi coefficient $Z$, the Kolkata index $k$, the approximate Kolkata index $k_G$ obtained from Eq. \ref{eq:kG}, and the absolute percentage error of $k_G$ relative to $k$ computed as $|k_G - k|/k$ for all censuses from 1872 to 2022. The mean errors of the estimations are approximately $3.4\%$ for $p_G$ and $0.9\%$ for $k_G$.}
\label{tab:ginipietra}
\end{table}

\section{Allometry in urban areas: 2022 census data}

\begin{figure}[H]
    \centering
    \includegraphics[width=\linewidth]{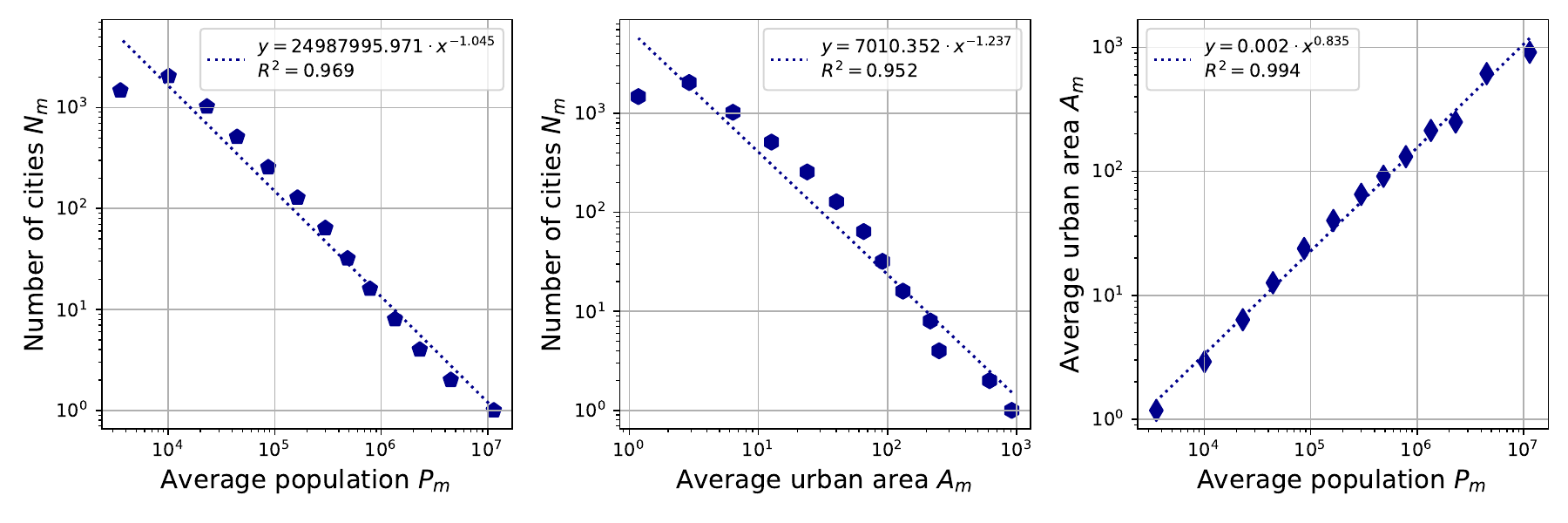}
    \caption[Urban city areas: 2022]{The scaling patterns for the hierarchy of the 5570 cities in Brazil: 2022. Left panel: scaling relation between urban population $P_m$ and city number $N_m$. Center panel: scaling relation between urban area $A_m$ and city number $N_m$. Right panel: allometric relation between urban area $A_m$ and population $P_m$.}
    \label{fig:urbanarea2022}
\end{figure}

The parameter $b$, as defined in Eq. (\ref{eq:allometry}) and associated with Fig. \ref{fig:urbanarea2022}, is the allometric scaling exponent. It is important in this analysis because it characterizes how urban area scales with population size. 
Allometry refers to the study of how specific characteristics of a living organism change in relation to its overall size in terms of growth or other quantities, like metabolic ones. In urban systems, allometry describes how spatial quantities change as cities grow, allowing the assessment of land use patterns.
The allometric scaling exponent $b$ is related to the fractal dimensions $D$ and $d$ by the following equation defined in Ref. \cite[p.~4]{chen2012}: 
\be
b = D/d. 
\label{eq:bdd}
\ee
Here, $D$ refers to the fractal dimension with respect to population and $d$ refers to the fractal dimension with respect to urban area.
The comparison between $D$ and $d$ indicates whether urban land area grows faster or slower than population.
The value $b$ can reflect whether urban land use is intensive or extensive. 
Values of $b<1$ imply that urban area increases at a slower rate than population, which is consistent with a more spatially compact configuration and increasing population density, as more people are accommodated per unit of urban area as cities grow \cite{chen2012}.

Figure \ref{fig:urbanarea2022} displays the data from Table \ref{tab:areas2022} (p. 43) through three scaling relations. 
The graph of $N_m$ versus $P_m$ illustrates the relationship between the number of cities and their average population size, following $N_m \propto P_m^{-D}$. 
The graph of $N_m$ versus $A_m$ shows the relationship between the number of cities and their average urban area, following $N_m \propto A_m^{-d}$.
Finally, the graph of $A_m$ versus $P_m$ represents the allometric relationship between urban area and population, following the power-law relation $A_m \propto P_m^{b}$.

From the data in Fig. \ref{fig:urbanarea2022}, the estimated values are $D = -1.045$, $d = -1.237$ and $b = 0.835$, obtained for all Brazilian cities using the 2022 census data. 
These results were acquired from a Python function that tested which of several linear fits had a coefficient of determination $R^2$ closer to 1.
The value $R^2$ is a statistical measure that represents the proportion of the variance in the dependent variable that can be predicted from the independent variables, and being close to 1 indicates a good fit of the model to the observed data.
Ref. \cite{chen2012} reported the allometric scaling exponent $b = 0.793$ for the USA (2000) and $b = 0.856$ for China (2005), as shown in Table \ref{tab:urban_results}. 
The USA adopts the urbanized area definition, which includes high-density peripheral regions and cities with more than 50,000 people. 
China uses the built-up area definition, focusing on urban infrastructure but not population density and encompassing cities with more than 100,000 inhabitants.
The value obtained in this study is $b = 0.835$, which lies in between these results.
The Brazilian dataset includes all cities with populations greater than 833 inhabitants, which is substantially lower than the population thresholds adopted for the USA and China in Ref. \cite{chen2012}.

\begin{table}[H]
    \centering
    \begin{tabular}{c c c c c}
        \hline
    Country & Year & $D$ & $d$ & $b$ \\
        \hline
        USA & 2000 & -0.974 & -1.213 & 0.793  \\
        China & 2005 & -1.262 & -1.435 & 0.856\\
        Brazil & 2022 & -1.045 & -1.237 & 0.835  \\
        \hline
    \end{tabular}
    \caption[Urban areas comparison]{Urban area results for the USA, China and Brazil. Results for the USA and China were reported in Ref. \cite{chen2012}.}
    \label{tab:urban_results}
\end{table}

The result $b=0.835$ suggests that the Brazilian urban system in 2022 follows a sublinear allometric regime since $b<1$, indicating that urban area grows more slowly than population. 
This behavior is consistent with increasingly dense urban configurations, where population growth is accommodated without a proportional expansion of land use.
From the perspective of fractal scaling, the relation in Eq. (\ref{eq:bdd}) indicates that the spatial distribution of population is more concentrated than that of urban land, which is reflected in $|d| > |D|$. 
This inequality implies that the hierarchical structure of urban areas decays faster than the hierarchical structure of population, suggesting that smaller cities occupy relatively larger areas compared to their population size, while larger cities concentrate population in smaller areas relative to their size.
The comparison with the USA and China in Ref. \cite{chen2012} shows that all three cases in Table \ref{tab:urban_results} exhibit $b<1$, suggesting that sublinear scaling may be a robust property of large urban systems despite differences in definitions of urban area.

\begin{figure}[H]
    \centering
    \includegraphics[width=\linewidth]{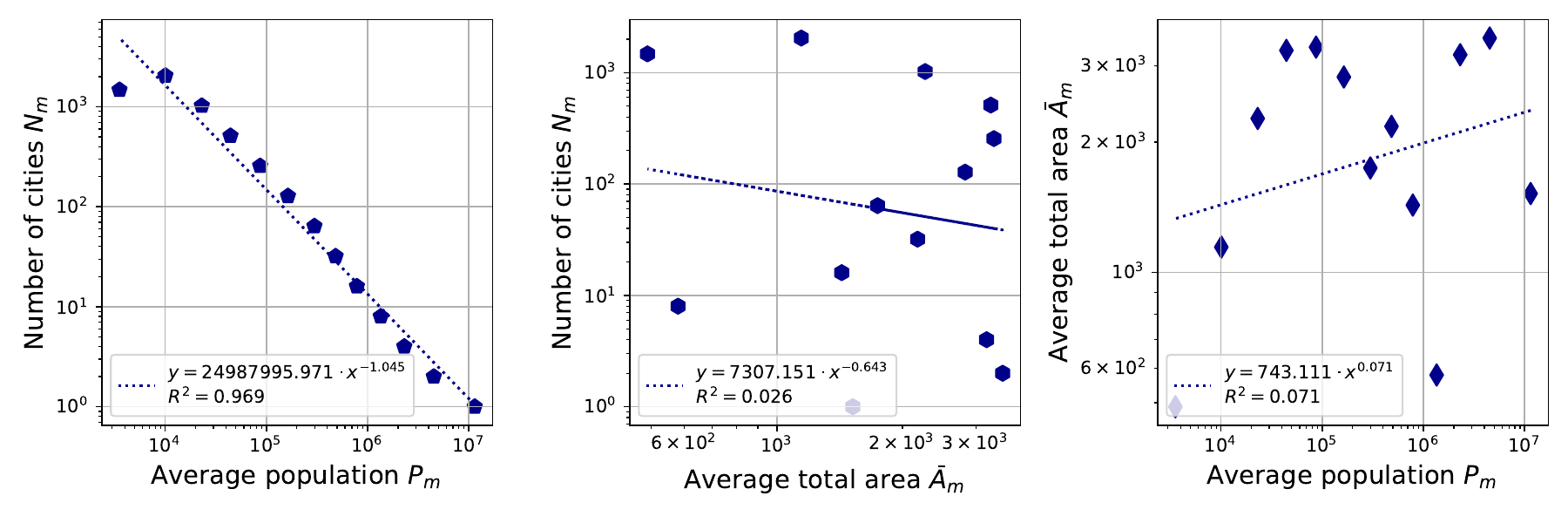}
    \caption[Total city areas: 2022]{The scaling patterns for the hierarchy of the 5570 cities in Brazil: 2022. Left panel: scaling relation between total population $P_m$ and city number $N_m$. Center panel: lack of scaling relation between total area $\bar{A}_m$ and city number $N_m$. Right panel: lack of allometric relation between total area $\bar{A}_m$ and population $P_m$.}
    \label{fig:totalarea2022}
\end{figure}

The same analysis was performed for city total area $\bar{A}_m$ instead of city urban area $A_m$. 
In this case, no clear scaling or allometric relation is observed between area and population, as indicated by the absence of a linear trend in the central and right plots of Fig. \ref{fig:totalarea2022} and by the $R^2$ values being much lower than 1.
These results highlight the distinction between administrative boundaries and functional urban space.
While urbanized area reflects the spatial extent of human occupation, total area includes large non-urban regions that introduce noise into the scaling behavior.
This contrast indicates that in the 2022 Brazilian census data, the allometric behavior is observed only when the spatial occupation variable is considered.

\clearpage
\chapter*{Conclusion}
\markboth{CONCLUSION}{CONCLUSION}
\addcontentsline{toc}{chapter}{Conclusion}
This dissertation explored the applicability of the Gompertz-Zipf distribution to describe Brazilian city size distributions in terms of population, utilizing data from the Brazilian Institute for Geography and Statistics censuses. 
Rather than focusing on a single dataset, this study analyzed twelve censuses spanning 150 years, from 1872 to 2022, providing a long-term perspective on the structure and evolution of the Brazilian city system.
The study aimed at addressing the ongoing debate in the literature regarding the validity and robustness of the Zipf law and power-law distributions in the city size context.
Our findings indicate that the Gompertz-Zipf distribution provides a statistically coherent and analytically simple representation of the city sizes, using Brazilian data as example. 
By dividing the distribution into two segments, a Gompertz distribution for less populated cities and a Zipf distribution for more populated cities, we were able to overcome some of the limitations associated with applying a single power-law model to the entire city size range, that requires assuming an arbitrary lower-limit population size, a practice which does not lead to a universal lower limit value and may cause sample bias.
Furthermore, the application of inequality measures to city size distributions provides additional information beyond the functional forms of the distribution. 
The Lorenz curve and the Gini coefficient are standard in income distribution studies, while the Zanardi index and the Gompertz function have not been systematically applied to city population distributions in the literature.
While the Gini coefficient quantifies the overall level of concentration, the Zanardi index captures the asymmetry of the distribution, indicating whether inequality is more pronounced in the lower or upper tail. 
The combined use of these measures allows a more complete characterization of the distribution.

A similar analysis made in Ref. \cite{moura2006} showed that the exponent of the Zipf law in Brazilian cities is $a = 2.22 \pm 0.34$ for 1970 and 1980 and $a = 2.26 \pm 0.11$ for 1991 and 2000. 
This is consistent with the present dissertation, which shows $a = 2.28 \pm 0.02$ for the 2022 Brazilian cities data.
The use of logarithmic binning with different bases demonstrated that the results are robust with respect to the binning choices for using a Gompertz-Zipf distribution to represent the Brazilian city system.
Both the Gompertz parameters and the Zipf exponent remain stable at different binning configurations, reinforcing that the observed patterns are intrinsic properties of the Brazilian city size distribution rather than a product of the analytical methodology.

The same Gompertz-Zipf population distribution analysis was applied to 150 years of data from Refs. \cite{IBGE2022census} and \cite{IBGEbook} for the Brazilian censuses of 1872, 1900, 1920, 1940, 1950, 1960, 1970, 1980, 1991, 2000, 2010 and 2022. 
The results show that the average Zipf exponent is $a = 2.63 \pm 0.53$, defined as the mean of all $a$ values, ranging from $a = 3.61 \pm 0.09$ in 1872 to $a = 2.28 \pm 0.02$ in 2022.
This reduction indicates an increasing presence of very populous cities from 1872 to 2022.
The relative variation of the Zipf exponent remains below $4\%$ after 1960. These results show that the Zipf parameters become stable from 1960 onward.
The parameter $B$ varies from $-2.39\times10^{-5}$ to $-1.49\times10^{-5}$, which corresponds to a relative variation of approximately $46\%$. 
Since $B$ is negative, larger absolute values of $B$ indicate a faster reduction in the number of cities as population increases, whereas smaller absolute values indicate a slower reduction. 
The oscillations of $B$ indicate periods with faster decreases and periods with slower decreases in the number of cities within the less populated segment as population increases.
With the chosen logarithmic binning base, the threshold separating these two segments exhibits good coefficients of determination ranging from $R^2=0.980$ to $R^2=0.999$ across all censuses, and varies between 20,216 and 102,857 people for the entire span of 150 years.
These results reinforce the need to determine, for each census, where the Zipf law begins to hold rather than adopting an arbitrary population cutoff.
Furthermore, the inequality analysis based on the Lorenz curve revealed a high level of concentration in the data.
The Gini coefficient increases from $G = 0.439$ in 1872 to $G = 0.733$ in 2022, while the Kolkata index increases from $k = 65.89$ to $k = 78.34$, indicating a long-term rise in population concentration. 
Similarly, the Zanardi index is slightly negative in 1872 and positive in all subsequent censuses, with its lowest value $Z=-0.002$ in 1872, the highest value $Z=0.031$ in 1940 and $Z = 0.018$ in 2022, showing that the asymmetry is oriented towards more populated cities. 
Although the magnitude of $Z$ remains relatively small, its positive sign after 1900 indicates a greater concentration of population in the group of more populated cities than in the group of less populated cities.

In addition, the analysis of urban allometry reinforces the broader relevance of scaling laws in the study of city growth. 
The allometric relation resulted in an exponent $b=0.835$ for the 2022 Brazilian data. Since $b<1$, population grows more rapidly than urban area, indicating a sublinear allometric regime with increasing population density as cities grow.
This sublinear regime means that urban area grows more slowly than population, leading to higher population density as cities grow.
This finding aligns with previous results for the United States $b=0.793$ and China $b=0.856$ as reported in Ref. \cite{chen2012}, and indicates that Brazilian cities follow similar spatial scaling patterns consistent with hierarchical and fractal urban structures. 
The absence of scaling relations when total area is considered instead of urban area indicates that the observed allometric behavior depends on spatial occupation.

In conclusion, this dissertation contributes to the literature by providing a comprehensive empirical analysis of the Brazilian city system over a long historical period. 
By combining theoretical models with extensive data, it suggests that the Gompertz-Zipf distribution offers a statistically coherent and analytically simple description of city size distributions.
Moreover, the results for over 150 years make the Brazilian case a relevant example within the broader study of complex systems and scaling phenomena.

A possible future analysis is to apply the methodology presented here to other countries or regions, allowing comparative studies of city size distributions and inequality patterns, or, for a more socio-economic analysis, to connect the results presented here with Brazilian historical events. 





\pagebreak
\phantomsection
\addcontentsline{toc}{chapter}{Bibliography}
\bibliography{references}
\renewcommand{\thechapter}{\arabic{chapter}}
\end{document}